\documentclass[12pt]{article}
\usepackage{amssymb,epsfig}

\renewcommand{\theequation}{\arabic{section}.\arabic{equation}}
\newcommand{\be}[1]{
\begin{equation}\label{#1}}
\newcommand{\ee}{\end{equation}}
\newcommand{\ba}[1]{
\begin{eqnarray}\label{#1}}
\newcommand{\ea}{\end{eqnarray}}
\newcommand{\baa}{\begin{eqnarray*}}
\newcommand{\btab}{\begin{tabular}}
\newcommand{\etab}{\end{tabular}}
\newcommand{\eaa}{\end{eqnarray*}}

\newcommand{\derleft}{\stackrel{\leftarrow}{D}\!}
\newcommand{\derright}{\stackrel{\rightarrow}{D}}

\def \labeltest #1 {\label{#1}}
\newcommand\re[1]{(\ref{#1})}

\newcommand\lr[1]{{\left({#1}\right)}}

\def\II{\hbox{{1}\kern-.25em\hbox{l}}}
\newcommand \widebar [1] {\overline{#1}}

\def \Tr {\mbox{Tr\,}}

\newcommand \vev [1] {\langle{#1}\rangle}

\newcommand \ket [1] {|{#1}\rangle}

\newcommand\bin[2]{\left({#1}\atop{#2}\right)}

\def \e {\mbox{e}}
\def \CO {{\cal O}}

\begin{document}

\begin{titlepage}
\begin{flushright}
\begin{tabular}{l}
LPT--Orsay--01--19\\
TPR--01--02\\
UB--ECM--PF--01/01\\
hep-ph/0102313
\end{tabular}
\end{flushright}
\vskip1cm
\begin{center}
  {\large \bf
     Evolution equation for the structure function $g_2(x,Q^2)$
  \\}
\vspace{1cm}
\def\thefootnote{\fnsymbol{footnote}}%
{\sc V.M.~Braun}${}^{1}$,
{\sc G.P.~Korchemsky}${}^2$
          and {\sc A.N.~Manashov}${}^{1,3}$\footnote{
Permanent address:\ Department of Theoretical Physics,  Sankt-Petersburg State
University, St.-Petersburg, Russia}
\\[0.5cm]
\vspace*{0.1cm} ${}^1${\it
   Institut f\"ur Theoretische Physik, Universit\"at
   Regensburg, \\ D-93040 Regensburg, Germany
                       } \\[0.2cm]
\vspace*{0.1cm} ${}^2$ {\it
Laboratoire de Physique Th\'eorique%
\footnote{Unite Mixte de Recherche du CNRS (UMR 8627)},
Universit\'e de Paris XI, \\
91405 Orsay C\'edex, France
                       } \\[0.2cm]
\vspace*{0.1cm} ${}^3$ {\it
Department d'ECM, Universitat de Barcelona,\\
08028 Barcelona, Spain} \\[1.0cm]

\vskip0.8cm
{\bf Abstract:\\[10pt]} \parbox[t]{\textwidth}{
  We perform an extensive study of the scale dependence of flavor-singlet
  contributions to the structure function $g_2(x,Q^2)$ in polarized
  deep-inelastic scattering. We find that the mixing between
  quark-antiquark-gluon and three-gluon twist-3 operators  only involves
  the three-gluon operator with the lowest anomalous dimension
  and is weak in other cases. This means, effectively, that only those
  three-gluon operators with the lowest anomalous dimension for each moment
  are important, and allows to formulate a simple two-component
  parton-like description of $g_2(x,Q^2)$ in analogy with the conventional
  description of twist-2 parton distributions.
  The similar simplification was observed earlier for the nonsinglet
  distributions, although the reason is in our case different.
}
\vskip1cm

\end{center}

\end{titlepage}

\tableofcontents

\newpage

\section{Introduction}
\setcounter{equation}{0}
Twist-three parton distributions in the nucleon are attracting
constant interest as unique probes of quark-gluon correlations
in hadrons.  Quantitative studies of twist-three effects are becoming possible
with the increasing precision of experimental data at SLAC and RHIC,
and can constitute an important part of the future spin physics program
on high-luminosity accelerators like ELFE, eRHIC, etc.

The structure function $g_2(x,Q^2)$ of polarized deep inelastic scattering
received most attention in the past. The  experimental studies at SLAC
\cite{E143,E154,E155} have confirmed theoretical expectations about the
shape of $g_2(x,Q^2)$ and provided first evidence on the
most interesting twist-3 contribution. On the theoretical side,
a lot of effort was invested to understand the physical interpretation
of twist-3 distributions (see e.g. \cite{Book,AELreview,KTreview} for the
review of various aspects) and
their scale dependence \cite{SV82,BKL84,BB89,ABH,Kodaira,BKM00}.
One-loop corrections to the coefficient functions have been calculated
\cite{Ji1,Ji2,I}.

In spite of the  significant progress that has been achieved,
understanding of the scale dependence of $g_2(x,Q^2)$ still poses an
outstanding theoretical problem. The difficulty is due to the well-known
fact \cite{Book,AELreview,KTreview} that the structure function $g_2(x,Q^2)$ presents by itself only
one special projection of a more general three-particle
quark-antiquark-gluon correlation function in the nucleon
that depends, generally, on two variables - the momentum fractions
carried by partons.

In deep inelastic scattering with a transversely polarized target, only
this special projection can be measured. On the other hand,
the scale dependence of the parent quark-antiquark-gluon correlation
function involves the ``full'' function in a notrivial way \cite{BKL84}
and the knowledge of one particular projection $g_2(x,Q_0^2)$
at a given value of $Q_0^2$ does not allow to predict $g_2(x,Q^2)$ at different
momentum transfers: a DGLAP-type evolution equation for $g_2(x,Q^2)$
in QCD does not exist or, at least, is not warranted.
The reason is simply that inclusive measurements
do not provide complete information on the relevant
three-particle parton correlation function.

{}From the phenomenological point of view this conclusion is not satisfactory
since it would mean that one cannot relate results of the measurements of
$g_2(x,Q^2)$ at different values of $Q^2$ to one another without model
assumptions. The theoretical challenge is, therefore, to find out whether
the complicated pattern of quark-gluon correlations can be reduced to a
few effective degrees of freedom.
In this case one will be able to find a meaningful approximation
%One has to look for meaningful approximations
to the scale dependence that introduces a minimum amount of
nonperturbative parameters.

Such an approximation is known  \cite{ABH} for the flavor-nonsiglet (NS)
contribution to the structure function.
To explain this result, it is convenient to
use the language of the Operator Product Expansion (OPE),
see Sect.~2 for more details.
The statement of the OPE is that odd moments $n=3,5,\ldots$
of the structure function $g_2(x,Q^2)$ can be expanded in contributions of
multiplicatively renormalized local quark-antiquark-gluon operators%
\footnote{
 We neglect the Wandzura-Wilczek twist-2 contributions throughout this paper.}
\ba{OPE}
 \int_0^1 \! dx\,x^{n-1} g_2^{\rm NS}(x,Q^2) &=&
   \sum_{k=0}^{n-3} C_{n-3}^k
   \left(\frac{\alpha_s(Q)}{\alpha_s(\mu)}\right)^{\gamma_{n-3}^k/b}
   \langle\!\langle O_{n-3}^k(\mu) \rangle\!\rangle ,
\ea where  $C_{n-3}^k$ are the coefficient functions and $\langle\!\langle
O_{n-3}^k(\mu) \rangle\!\rangle$ are reduced matrix elements normalized at the
scale $\mu$; $\gamma_{n-3}^k$ are the corresponding anomalous dimensions that we
assume are ordered with $k$: $\gamma^0_{n-3} < \gamma^1_{n-3} < \ldots
< \gamma^{n-3}_{n-3}$ for
each $n$, and $b=11/3 N_c -2/3 n_f$. Note that the number of contributing
operators rises linearly with $n$ in the r.h.s.\ of \re{OPE}. This should be
compared
%, in contrast
with the familiar case of leading twist-2 distributions. There,
%in which case
a single operator (flavor-nonsiglet) exists for each moment. A measurement of the moment
of $g_2(x,Q^2)$ cannot separate between contributions of different operators and
is, therefore, not sufficient to predict the scale dependence.

The situation simplifies drastically, however, in the large $N_c$ limit.
It turns out \cite{ABH} that
the tree-level coefficient functions, $C_{n-3}^k$,  of all
operators other than the one with the lowest anomalous dimension for each
$n$ are suppressed by powers of $1/N_c^2$ so that to this accuracy one
can approximate the sum in \re{OPE} by the first term $k=0$.
The corresponding anomalous dimension $\gamma_{n-3}^{k=0}$ can be
calculated analytically and result can be reformulated  as a DGLAP-type
evolution equation
\ba{AP:NS}
  Q^2 \frac{d}{d Q^2} \, g_2^{\rm NS}(x,Q^2) &=& \frac{\alpha_s}{4\pi}
     \int_x^1\! \frac{dz}{z} P^{\rm NS}(x/z) \,g_2^{\rm NS}(z,Q^2)\,,
\nonumber\\
   P^{\rm NS}(z) &=&
    \left[\frac{4C_F}{1-z}\right]_+
      +\delta(1-z)\left[C_F+\frac{1}{N_c}\left(2-\frac{\pi^2}{3}\right)\right]
      - 2C_F\,,
\end{eqnarray}
where $C_F=(N_c^2-1)/(2N_c)$. Here, we have also included the $1/N_c^2$ corrections
calculated in \cite{BKM00}.

The present paper is devoted to the extension of this analysis to the
flavor-singlet sector in which case twist-3 composite local three-gluon operators
have to be included. Coefficient functions of  three-gluon operators
vanish at tree level, so that gluon contributions appear entirely
through the evolution. The number of independent three-gluon operators
is, roughly speaking, half of the number of quark-gluon operators
and, similarly, is rising with the moment $n$. The subject of this work is to find out
whether the whole set of gluon operators contributes
significantly, or one can reduce the gluon contribution to a certain
single degree of freedom. We find that such a reduction is indeed
possible and formulate a two-channel DGLAP-type evolution
equation for the structure function $g_2(x,Q^2)$ that presents our
main result.

In physical terms, the approximation constructed in this paper
corresponds to the introduction of quark and gluon transverse spin
parton distributions which are identified with the particular components
of quark-antiquark-gluon and three-gluon parton correlation functions
that possess the lowest anomalous dimension.
%`trajectories'  with the lowest anomalous dimension.
We will find that, first, the structure function $g_2(x,Q^2)$ is
dominated by contributions of these two distributions at large scales
(including the ${\cal O}(\alpha_s)$ gluon contribution calculated in \cite{I}),
and, second, the `leakage' of transverse spin to genuine three-particle
degrees of freedom at lower scales  is small due to the specific pattern
of the QCD evolution.  We would like to emphasize that importance of
this result is not so much in the possibility to calculate the scale
dependence, but in the identification of important transverse spin
degrees of freedom that are preserved by  QCD interactions.

The outline of the paper is as follows. Sect.~2 is mainly introductory.
We introduce necessary notation and present a summary of the existing
calculations of the coefficient functions in the operator product expansion.
Sect.~3 is devoted to the general formalism of the
renormalization of twist-3 gluon operators.
We emphasize importance of the conformal symmetry and
introduce a convenient framework that allows to treat renormalization
as a quantum mechanical problem with hermitian Hamiltonian.
This section is necessarily rather technical and a reader who is only
interested in the applications may prefer to skip this discussion
and go over directly to Sect.~4 where we collect our results.
The main result of this paper is the generalization
of the DGLAP evolution equation \re{AP:NS} to the flavor-singlet
channel; it is repeated in Sec.~5 which also contains a summary
and conclusions.

In Appendix A we present a detailed calculation of the relevant anomalous
dimensions. In Appendix B we collect necessary formulae for the Racah 6j-symbols
of the $SL(2,R)$ group. The conformal basis representation of the QCD
evolution kernels is discussed in Appendix C.

\section{The Operator Product Expansion}
\setcounter{equation}{0}

The hadronic tensor which appears in the description
of deep inelastic  scattering of polarized leptons on polarized nucleons,
involves two structure functions%
\footnote{We define the nucleon spin vector  as
$s_\mu = \bar u(p,s)\gamma_\mu\gamma_5 u(p,s)$ where  $u(p,s)$ is the
nucleon spinor $\bar u(p,s) u(p,s) =2 M$, so that  $s^2=-4M^2$.}
\ba{Wa}
  W^{(A)}_{\mu\nu} &=&\frac{1}{p\cdot q}
   \varepsilon_{\mu\nu\alpha\beta} q^\beta
\left\{ s^\beta g_1(x_B,Q^2) +\left[s^\beta- \frac{s\cdot q}{p\cdot q}p^\beta
\right]g_2(x_B,Q^2)\right\}
\ea
and is related to the antisymmetric part of the Fourier-transform of the
T-product of two electromagnetic currents:
\ba{T-product}
  W^{(A)}_{\mu\nu} &=& \frac{1}{\pi}\,{\rm Im}\, T^{(A)}_{\mu\nu}\,,
\nonumber\\
  i T^{(A)}_{\mu\nu} &=& \frac{i}{2} \!\int \!d^4x\,e^{iqx}
\langle p,s| T\{j_\mu(x/2)j_\nu(-x/2)-j_\nu(x/2)j_\mu(-x/2)\} |p,s\rangle\,.
\ea
The light-cone expansion of (\ref{T-product})
at $x^2\to 0$ goes in terms of nonlocal light-cone
operators of increasing twist, schematically
\ba{twistexpansion}
&&\frac{i}2 T\{j_\mu(x)j_\nu(-x)-j_\nu(x)j_\mu(-x) \} \stackrel{x^2\to 0}{=}
\nonumber
\\
&&
\qquad\qquad
\frac{i\varepsilon_{\mu\nu\alpha\beta}}{16\pi^2}
        \frac{\partial}{\partial x_\alpha}\left\{[C\ast O_\beta]^{\rm tw-2}
   + [C\ast O_\beta]^{\rm tw-3} +({\rm higher~twists})\right\},
\ea
where $C\ast O_\beta$ stands for the product (convolution) of the coefficient
functions and operators of the corresponding twist. This expression is explicitly
$U(1)$-gauge invariant, i.e. $\partial/\partial x_\mu T\{j_\mu(x)j_\nu(-x)\} =0$.

\subsection{Twist-2}
For the sake of completeness and in order to facilitate the
comparison to twist-3, we collect here the relevant portion
of the results for the leading-twist.

Retaining the tree-level quark contribution and the leading
gluon correction that starts at order ${\cal O}(\alpha_s)$
one obtains \cite{I}
\ba{OPE1}
 [C\ast O_\beta]^{\rm tw-2} &=& \frac{x_\beta}{x^4}
       \sum_{q=u,d,s,\ldots}\!\!\!e^2_q\,\,
\int_0^1\!\! du\,\, \Bigg\{
\Big[\bar q(ux)\!\not\!x\gamma_5 q(-ux)+(x\to -x)\Big]_{\mu^2_{\overline{\rm MS}}}
\\
&+&\frac{4\alpha_s}{\pi}
\Big( \ln(-x^2\mu^2_{_{\overline{\rm MS}}})+2\gamma_E\Big)
(u\ln u +u(1-u)) \Tr\{ G_{x\xi}(ux)\widetilde G_{x\xi}(-ux)\}\Bigg\}\,,
\nonumber
\ea
where $G_{x\xi}= x_\eta G^a_{\eta\xi}t^a$ and
the subscript $[\ldots]_{\mu^2}$ indicates the normalization scale of the
operator. Here and in what follows it is implied that the gauge invariance
of nonlocal light-cone operators is restored by including the gauge factors
$P\exp(i\int dz_\mu A^\mu(z)$.
Note simplicity of the answer \re{OPE1}: the entire gluon contribution
can be eliminated by choosing the proper scale of the quark operator
$\mu^2_{\overline{\rm MS}} = 1/(-x^2e^{2\gamma_E})$. This property
is lost in the momentum space since after the Fourier transformation
contributions of different light-cone separations get mixed.

Going over to the matrix elements, one introduces the quark and
gluon  helicity distributions
\ba{Deltag}
\vev{p,s|\bar q(x)\not\!x\gamma_5 q(-x)|p,s}
&=&(sx)\int_{-1}^1\!d\xi\, \e^{2i\xi px}\,\Delta q(\xi,\mu^2)\,,
\nonumber\\
\vev{p,s|\Tr\Big\{G_{x\alpha}(x)\widetilde G_{x\alpha}(-x)\Big\}|p,s}
&=&\frac{i}{4}(sx)(px)
\int_{-1}^1\!d\xi\, \e^{2i\xi px}\,\xi \Delta g(\xi,\mu^2)\,.
\ea
In the first case, positive (negative) $\xi$ correspond to the
contribution of quarks (antiquarks)
$\Delta q(x_B) = q^\uparrow(x_B)-q^\downarrow(x_B)$,
$\Delta q(-x_B)
\equiv \Delta \bar q(x_B) = \bar q^\uparrow(x_B)-\bar q^\downarrow(x_B)$,
respectively. For gluons, $\Delta g(\xi)=\Delta g(-\xi)$.

Ignoring the gluon contribution for a moment,
making a Fourier transformation of (\ref{OPE1}), taking imaginary part and
comparing with the definition of structure functions in (\ref{Wa}) one obtains
\be{g11}
    g_1(x_B,Q^2) = \frac{1}{2}\sum_q e^2_q\,
    [\Delta q(x_B,\mu^2 = Q^2)+\Delta q(-x_B,\mu^2=Q^2)]\,,
\ee
where $x_B= Q^2/(2pq)$ is the Bjorken variable, and twist-2 contribution
to $g_2(x_B,Q^2)$
\be{gWW}
   [g_2(x_B,Q^2)]^{\rm tw-2} = g_2^{WW}(x_B,Q^2) =
  -g_1(x_B,Q^2) +
   \int_{x_B}^1 \frac{dy}{y}\, g_1(y,Q^2)\,.
\ee
Equation \re{g11} tells that the structure function $g_1(x,Q^2)$ is a
measure of the quark helicity distribution in the nucleon, as well known.
Equation \re{gWW} is the familiar Wandzura-Wilczek relation \cite{WW}.
We would like to stress that despite the fact that the relation in \re{g11}
is affected by higher order perturbative radiative corrections, the
Wandzura-Wilczek relation between the structure functions, Eq.~\re{gWW},
is exact to all orders of perturbation theory and is only modified by twist-3
contributions to $g_2(x_B,Q^2)$ that are subject of this paper.
%
%It is worthwhile to note that the relation in \re{g11}
%is affected by perturbative radiative corrections, whereas the
%Wandzura-Wilczek relation between the structure functions is exact to
%all orders of perturbation theory and is only modified by twist-3
%contributions that are subject of this paper.
%
The reason for this is that the relation \re{gWW} follows from particular
(and unique) form of the Lorentz structure for the antisymmetric part of the T-product
in \re{twistexpansion} which in turn is dictated by the $U(1)$ gauge invariance.
%
%The reason for this is
%that $U(1)$ gauge invariance only allows for a single Lorentz structure
%for the antisymmetric part of the T-product in \re{twistexpansion}.
Although the coefficient function in front of the twist-2 operator in
\re{twistexpansion}
has a nontrivial perturbative expansion, it affects both structure
functions $g_1(x,Q^2)$ and $g_2(x,Q^2)$ simultaneusly and in the same way.
Hence the Wandzura-Wilczek relation holds true.

For most of the subsequent discussion it will be convenient to
go over to the moments space:
\be{moms}
f(n,Q^2) = \int_0^1 \!dx_B\, x_B^{n-1}\, f(x_B,Q^2)
\,,
\ee
for any function $f$.
In particular, restoring the gluon contribution we get \cite{I}
\ba{mom-g1}
 &&g_1(n,Q^2) = \frac12
\sum_{q=u,d,s,\ldots}\!\!\!e^2_q\,
\Bigg\{\Delta q(n,\mu^2_{\overline{\rm MS}}) +
\Delta \bar q(n,\mu^2_{\overline{\rm MS}})
\nonumber\\
&&
\qquad \qquad\qquad+
\frac{\alpha_s}{2\pi}
\frac{n-1}{n(n+1)}\Delta g(n,\mu^2)\left[
\ln\frac{Q^2}{{\mu^2_{\overline{\rm MS}}}}-\psi(n)-1-\gamma_E\right]
\Bigg\},
\nonumber\\
&&[g_2(n,Q^2)]^{\rm tw-2}=-\frac{n-1}n\,g_1(n,Q^2)\,.
\ea

\subsection{Twist-3}
The twist-3 contribution to the T-product in \re{twistexpansion} is more
complicated. One obtains the following expression\cite{I}%
\footnote{To avoid misunderstanding, note that this result does not
 include the ${\cal O}(\alpha_s)$ correction to the coefficient function
 of quark-antiquark-gluon operators. This correction has been calculated
  recently in \cite{Ji1,Ji2} using a different operator basis}
\ba{OPE-tw3}
&& \hspace*{-5mm}[C\ast O_\beta]^{\rm tw-3}=
\nonumber\\
&& \frac{i}{2x^2}
        \sum_{q=u,d,s,\ldots}\!\!\!e^2_q\,\,
\int_0^1\!\! du\!\int^u_{-u} \!\!dv\,\Bigg\{
(u+v)\,S_\beta(u,v,-u)
+ (u-v)  S_\beta(-u,v, u)
\nonumber\\
&& + \bar u^2 \frac{\alpha_s}{\pi}
  \Big[(u+v)\widetilde O_{\beta}(v,u,-u)+2u \widetilde O_{\beta}(u,v,-u)
+ (u-v)\widetilde O_{\beta}(u,-u,v)\Big]
\\
&& +\frac{4\alpha_s}{\pi}
\left(\ln(-x^2\mu^2_{\overline{\rm MS}})+2\gamma_E+1\right)\Big[
 (\bar u u+\mbox{$\frac14$} \bar u^2 +u\ln u)
\left[v\, \widetilde O_{\beta}(v,u,-u)+ u\,\widetilde O_{\beta}(u,v,-u)\right.
\nonumber\\&&
\left.
- v\, \widetilde O_{\beta}(u,-u,v)\right]
 -\frac{1}{12}\bar u^2(u+2)\left[\widetilde O_{\beta}(u,-u,v)+\widetilde O_{\beta}(u,v,-u)+
\widetilde O_{\beta}(v,u,-u)\right]\Big]\Big\}_{\mu^2_{\overline{\rm MS}}},
\nonumber
\ea
where $\bar u=1-u$ and we have introduced the C-even nonlocal quark-gluon operator
\ba{Spm}
 S_\beta(u,v,-u) &=& S_\beta^{+}(u,v,-u)+S_\beta^{-}(-u,v,u)\,,\nonumber\\
   S_\beta^\pm (a,b,c) &=& \frac12\bar q(ax)[ig\widetilde G_{\beta x}(bx)\pm
   gG_{\beta x}(bx)\gamma_5]\!\not\!xq(cx)\,
\ea
and the nonlocal three-gluon operator
\be{Odual}
  \widetilde O_{\beta}(u,v,w)
 = \frac{ig}{2} f^{abc}G^a_{x\alpha}(ux)
   \widetilde G^b_{x\beta}(vx)G^c_{x\alpha}(wx)\,.
\ee
The relatively complicated expression in the last two lines in \re{OPE-tw3}
reflects quark-gluon mixing and reproduces the corresponding term in
the renormalization group equation for the twist-3 operator
$S_\beta(u,v,-u)$~\cite{BB89,Muller}
\ba{QGrenorm}
\lefteqn{
 [ S_\beta(u,v,-u)]_{\mu_2^2} \,\,{=}\,\,
[S_\beta(u,v,-u)]_{\mu_1^2}
\,\,-\,\frac{\alpha_s}{2\pi}
\ln\frac{\mu_2^2}{\mu_1^2}
\! \int_{-u}^u\!\!ds\!\int_{-u}^s\!\! dt\, (2u)^{-3}}
\nonumber\\&\times&
 \!\!\Big\{[2u(s-t)+4(u-s)(t+s)]\widetilde O_{\beta}(s,v,t)
- 2|u|(s-t)[\widetilde O_{\beta}(s,t,v)-\widetilde O_{\beta}(v,s,t)]\Big\}.
\mbox{\hspace*{0.8cm}}
\ea
Similar to the twist-2 contribution in Eq.~\re{OPE1}
this  contribution
can be eliminated by appropriate choice of the normalization scale of the
quark operator. Finally, the expression in the second line in \re{OPE-tw3}
is not affected by the scale choice and defines a `genuine' twist-3 gluon
coefficient function.
%
%In addition, the expression in the second line in \re{OPE-tw3} defines a
%`genuine' twist-3 gluon coefficient function that cannot be eliminated by the
%scale choice in the quark operator.

Nucleon matrix elements of the nonlocal operators in \re{Spm}, \re{Odual}
define parton correlation functions
\be{Dq}
\vev{p,s| S_\mu^\pm(u,v,-u)|p,s}
= 2i(px)\left[s_\mu (px) - p_\mu (sx)\right] \!\int_{-1}^1\!\! {\cal
D}\xi\,
\e^{ipx[\xi_1 u + \xi_2 v - \xi_3 u]} D_q^\pm(\xi_1,\xi_2,\xi_3)
\ee
and
\be{Dg}
\vev{p,s| \widetilde O_{\mu}(u,v,-u)|p,s}
= -2(px)^2\left[s_\mu (px) - p_\mu (sx)\right] \!\int_{-1}^1\!\! {\cal D}\xi\,
\e^{ipx[\xi_1 u + \xi_2 v - \xi_3 u]} D_g(\xi_1,\xi_2,\xi_3)
\ee
where the integration measure is given by
\be{Dxi}
 \int_{-1}^1\!\! {\cal D}\xi \equiv \int_{-1}^1\!\!d\xi_1d\xi_2d\xi_3
 \,\delta(\xi_1+\xi_2+\xi_3)\,.
\ee
The quark correlation functions $D_q^\pm(\xi_i)$ have the following
symmetry property:
\be{bose}
  D_q^\pm(\xi_1,\xi_2,\xi_3) = (D_q^\mp)^\ast(-\xi_3,-\xi_2,-\xi_1)\,.
\ee
They are in general complex functions, but the imaginary parts do not
contribute to the structure functions and can be omitted \cite{Jaffe83}.
In turn, the gluon correlation function $D_g(\xi_i)$
is real and antisymmetric to the interchange of the first and
the third argument:
\be{bose-g}
D_g(\xi_1,\xi_2,\xi_3)=- D_g(\xi_3,\xi_2,\xi_1)\,.
\ee
Substituting \re{OPE-tw3} into \re{twistexpansion} and taking the
Fourier transform one obtains the moments of the structure function
for positive {\it odd\/} $n$ %$n=1,3,\ldots$
\ba{otvet}
 [g_2(n,Q^2)]^{\rm tw-3} &=& \frac12  \sum_{q=u,d,s,\ldots}\!\!\!e^2_q\,\,
\frac{4}{n}\int_{-1}^1\!{\cal D}\xi\,
%\Bigg\{ D_q(\xi_i,{\mu_{\overline{\rm MS}}^2})\, \Phi_n^{q}(\xi_1,\xi_3) \,
\Bigg\{ D_q(\xi_i,{\mu_{\overline{\rm MS}}^2})\,
\Phi_n^{q}(\xi_1,\xi_3) \,
\\[3mm]
&&{}\hspace*{-0.5cm}
+ \frac{\alpha_s}{4\pi} \frac{D_g(\xi_i,{\mu_{\overline{\rm MS}}^2})}{n+1}
\Bigg[ \Phi_n^{g}(\xi_i)+\Omega_n^{qg}(\xi_i)
\Bigg(\ln\frac{Q^2}{\mu_{\overline{\rm MS}}^2}
-\psi\left(n\right)-\!\gamma_E\!-1\!\Bigg) \Bigg]\Bigg\}\,.
\nonumber\ea
Here $D_q(\xi_i)$ is
the distribution function corresponding to the $C-$even
combination of quark-gluon operators \re{Spm}
\be{Dqs}
\vev{p,s| S_\mu(u,v,-u)|p,s}
= 4i(px)\left[s_\mu (px) - p_\mu (sx)\right] \!\int_{-1}^1\!\! {\cal
D}\xi\,
\e^{ipx[\xi_1 u + \xi_2 v - \xi_3 u]} D_q(\xi_1,\xi_2,\xi_3)
\ee
so that
\be{Dsym=ReD}
D_q(\xi_1,\xi_2,\xi_3) =
{\mbox{\rm Re}}\,D^+_q(\xi_1,\xi_2,\xi_3) =
{\mbox{\rm Re}}\,D^-_q(-\xi_3,-\xi_2,-\xi_1)\,,
\ee
the quark coefficient function is defined as %
\footnote{In difference to Ref.~\cite{BKM00} we prefer to define
the quark contribution in terms of ${\mbox{\rm Re}}\,D^+_q(\xi_i)$
instead of ${\mbox{\rm Re}}\,D^-_q(\xi_i)$. Because of this, the quark
coefficient function in \re{psi-n} differs from the one given in \cite{BKM00}
by the replacement $\xi_3 \leftrightarrow -\xi_1$. The gluon coefficients
are symmetric under this transformation and are not affected.}
\ba{psi-n}
\Phi_n^{q}(\xi_1,\xi_3)&=&%\Phi_n^{q,-}(-\xi_3,-\xi_1)
=-\frac{\partial}{\partial \xi_3}
            \frac{\xi_1^{n-1}-(-\xi_3)^{n-1}}{\xi_1+\xi_3}
\ea
and the gluon coefficient functions can be expressed in terms of
$\Phi_n^{q}$ as
%
%\ba{CD}
%&&\Phi_n^g(\xi_i) =
%\left[\Phi_{n-1}^{q}(\xi_1,\xi_3)+\Phi_{n-1}^{q}(\xi_1,-\xi_1-\xi_3)\right]
%+(\xi_1\leftrightarrow -\xi_3)\,,
%\\[3mm]
%&&\Omega_n^{qg}(\xi_i) =
%\left(1+\frac{2}{n(n-2)}\right) \Phi_n^g(\xi_i)
% +\frac{2(n-1)}{n(n-2)}\left[\Phi_{n-1}^{q}(-\xi_1-\xi_3,\xi_3)
% +\Phi_{n-1}^q(\xi_1+\xi_3,-\xi_1)\right]\!.
%\nonumber
%\ea
%
%
\ba{CD}
&&\Phi_n^g(\xi_i) =
\left[\Phi_{n-1}^{q}(\xi_1,\xi_3)+\Phi_{n-1}^{q}(-\xi_1-\xi_3,\xi_3)\right]
+(\xi_1\leftrightarrow -\xi_3)\,,
\\[3mm]
&&\Omega_n^{qg}(\xi_i) =
\left(1+\frac{2}{n(n\!-\!2)}\right) \Phi_n^g(\xi_i)
 +\frac{2(n\!-\!1)}{n(n\!-\!2)}\left[\Phi_{n-1}^{q}(\xi_1,-\xi_1-\xi_3)
 +\Phi_{n-1}^q(-\xi_3,\xi_1+\xi_3)\right]\!.
\nonumber
\ea
Explicit expressions for $\Phi^q_n$ and $\Phi^g_n$ for the lowest
moments $n=3,5,7$ can be found in \cite{I}.

Equivalently, one may choose to expand moments of the structure function
in contributions of local operators, for example
\ba{localop}
{}[ S^\pm_\mu ]^k_N &=& \frac12\bar q (\derleft\cdot x)^k
    [ig\widetilde G_{\mu\nu}\pm  gG_{\mu\nu}\gamma_5]\not\! x\,x^\nu
\,(\derright\cdot x)^{N-k}q\,,
\nonumber\\
{}[G_\mu]_N^k &=& \frac{i}2 g f^{abc} G_{x\alpha}^a\,(\derleft\cdot x)^k\,
\widetilde G_{x\mu}^b\,(\derright\cdot x)^{N-1-k}\,G_{x\alpha}^c\,.
\ea
The reduced matrix elements $\langle\!\langle\ldots
\rangle\!\rangle$ of local operators
\ba{vev1}
\vev{p,s| [S_\mu^\pm]^k_N|p,s}
= 2(ipx)^{N+1}\left[s_\mu (px) - p_\mu (sx)\right]
\langle\!\langle [S^\pm]^k_N \rangle\!\rangle\,,
\nonumber\\
\vev{p,s| [G_\mu]^k_N|p,s}
= 2(ipx)^{N+1}\left[s_\mu (px) - p_\mu (sx)\right]
\langle\!\langle [G]^k_N \rangle\!\rangle\,
\ea
are equal to moments of the three-particle distributions%
\footnote{The distinction between `plus' and `minus' distributions
is delicate since it is affected by the convention used to define the
$\gamma_5$ matrix. We use $\gamma_5=i\gamma_0\gamma_1\gamma_2\gamma_3 =
-i \gamma^0\gamma^1\gamma^2\gamma^3$ and $\epsilon^{0123}=-\epsilon_{0123}=1$
\cite{Okun}. A sign change in the definition of the $\gamma_5$ matrix
results in the replacement $S^+ \leftrightarrow S^-$ and, for matrix
elements $\langle\!\langle [S]^k_N\rangle\!\rangle \to
(-1)^N \langle\!\langle [S]^{N-k}_N\rangle\!\rangle$.
The coefficient functions in the OPE are changed accordingly, so that the
results for the physical observables remain intact.}
\be{Dg-moments}
  \langle\!\langle [S^\pm]^k_N \rangle\!\rangle =
\int_{-1}^1\!\! {\cal D}\xi\,
\xi_1^k \xi_3^{N-k} D_q^\pm(\xi_i)\,,
\qquad
  \langle\!\langle [G]^k_N \rangle\!\rangle =
\int_{-1}^1\!\! {\cal D}\xi\,
\xi_1^k \xi_3^{N-1-k} D_g(\xi_i)\,.
\ee
The symmetry relations \re{bose} and \re{bose-g} imply that
$
  \langle\!\langle [S^\pm]^k_N \rangle\!\rangle^* = (-1)^N
 \langle\!\langle [S^\mp]^{N-k}_N \rangle\!\rangle
$
and
$
\langle\!\langle [G]^k_N\rangle\!\rangle=-\langle\!\langle [G]^{N-1-k}_N
\rangle\!\rangle $. Therefore, the number of independent quark and gluon matrix
elements contributiong to a given $n=(N+3)$th moment is equal to
$\ell_q=N+1$ and
$\ell_g=[N/2]$, respectively. Finally, we define the reduced matrix
elements corresponding to the C-even quark-gluon operator \re{Spm} as
\be{S-red}
\langle\!\langle [S]^k_N\rangle\!\rangle
={\rm Re}\,\langle\!\langle [S^+]^{k}_N \rangle\!\rangle
=(-1)^N{\rm Re}\,\langle\!\langle [S^-]^{N-k}_N \rangle\!\rangle
=\int_{-1}^1\!\! {\cal D}\xi\,
\xi_1^k \xi_3^{N-k} D_q(\xi_i)\,.
\ee

\subsection{Transverse spin parton densities: why and why not}

In order to pursue a parton model-like interpretation, one can introduce
transverse spin distributions as the specific projections of general
quark-antiquark-gluon and three-gluon operators. Their definition
is suggested by the explicit form of the contribution of these operators
to the operator product expansion \re{OPE-tw3}
\ba{Tq}
 \lefteqn{\hspace*{-1cm}\int^u_{-u} \!dv\,\vev{p,s|
\left[(u+v)\,S^+_\mu(u,v,-u)+(u-v)S^-_\mu(u,v,-u)\right]|p,s}=
\hspace*{5cm}{}}
\nonumber\\
&&\hspace*{3.5cm}{}=-i\left[s_\mu - p_\mu\frac{ (sx)}{(px)}\right]
\!\int_{-1}^1\!
 \!d\xi\, \e^{2i\xi upx}\,\, \Delta q_T(\xi,\mu^2)\,
\ea
and
\ba{Tg}
 \lefteqn{\hspace*{-2cm}\int^u_{-u} \!dv\,\vev{p,s|
\Big[(u+v)\widetilde O_{\mu}(v,u,-u)+2u\, \widetilde O_{\mu}(u,v,-u)
+ (u-v)\widetilde O_{\beta}(u,-u,v)\Big]
|p,s}=
\hspace*{5cm}{}}
\nonumber\\
&&\hspace*{3.5cm}{}=2\left[s_\mu (px) - p_\mu (sx)\right]
\!\int_{-1}^1\!
 \!d\xi\, \e^{2i\xi upx}\,\,\xi\, \Delta g_T(\xi,\mu^2)\,,
\ea
so that
\ba{partonmoments}
\int_{-1}^1\! d\xi \, \xi^{n-1} \Delta q_T(\xi) &=&
{4} \int_{-1}^1\!\! {\cal D}\xi\,\Phi_n^q(\xi_i)\,
         D_q(\xi_i)\,,
\nonumber\\
   \int_{-1}^1\! d\xi \, \xi^{n-1} \Delta g_T(\xi) &=&
  2 \int_{-1}^1\!\! {\cal D}\xi\,
     \,\Phi_n^g(\xi_i)\, D_g(\xi_i)\,.
\ea
The functions $\Delta q_T(x_B,Q^2)$ and $\Delta g_T(x_B,Q^2)$
describe the momentum fraction distribution of the
transverse spin of the proton and have the same support property
as the parton distribution in (\ref{Deltag}).
Note that $\Delta g_T(\xi)=\Delta g_T(-\xi)$.
However, in contrast with \re{Deltag}, they do not have
any probabilistic interpretation but rather can be expressed through the more
general three parton correlation functions
$D_q(\xi_1,\xi_2,\xi_3)$ and $D_g(\xi_1,\xi_2,\xi_3)$ integrating
out the dependence on one  momentum fraction. The explicit
expressions for the lowest moments \re{partonmoments} look as follows
\ba{Explicit-q}
&&
\int_{0}^1\! dx \, \lr{\Delta q_T(x)+\Delta q_T(-x)} =0,
\\
&&
\int_{0}^1\! dx \,x^2 \lr{\Delta q_T(x)+\Delta q_T(-x)} =
{4} \int_{-1}^1\!\! {\cal D}x\,D_q(x_i)
=4 \langle\!\langle [S]^0_0 \rangle\!\rangle,
\\
&&
\int_{0}^1\! dx \,x^4 \lr{\Delta q_T(x)+\Delta q_T(-x)} =
{4} \int_{-1}^1\!\! {\cal D}x\,(x_1^2-2x_1x_3+3x_3^2)D_q(x_i)
\\
&&\hspace*{54.5mm}
=4\left[\langle\!\langle [S]^2_2 \rangle\!\rangle
- 2\langle\!\langle [S]^1_2 \rangle\!\rangle
+ 3\langle\!\langle [S]^0_2 \rangle\!\rangle
\right]
\nonumber
\ea
for the quark distribution and
\ba{Explicit-g}
&&
\int_{0}^1\! dx \,\Delta g_T(x) = \int_{0}^1\! dx \,x^2 \Delta g_T(x) = 0,
\\
&&
\int_{0}^1\! dx \,x^4 \Delta g_T(x) = 10 \int_{0}^1\!\! {\cal D}x\,x_1D_g(x_i)
= - 10 \langle\!\langle [G]^0_2 \rangle\!\rangle
\ea
for the gluon distribution.

At tree level, neglecting the ${\cal O}(\alpha_s)-$correction to \re{otvet}, one
obtains using \re{partonmoments}
\be{BCform}
    [g_2^{\rm Born}(x_B,Q^2)]^{\rm tw-3} =  \frac{1}{2}\sum_q e^2_q
      \int_{x_B}^1 \frac{dy}{y}
  [\Delta q_T(y)+\Delta q_T(-y)]
\ee
that looks very similar to the leading twist expression \re{g11} and \re{gWW}.
The two contributions in the square brackets can be interpreted as the
contributions of quarks and antiquarks, respectively. Following the
analogy with the leading twist, it is convenient to introduce the
combinations of definite signature
\be{signature}
   \Delta q_T^\pm (y, Q^2) = \Delta q_T(y, Q^2) \pm \Delta q_T(-y, Q^2)\,.
\ee
The distribution $\Delta q_T^+ (y, Q^2)$ corresponds to the even signature
and can be obtained by the analytic continuation from even moments $N$
of the OPE. The distribution $\Delta q_T^-(y, Q^2)$
can be obtained by the analytic continuation from odd moments $N$ and
defines the valence quark contribution. The gluon contribution enters
into \re{otvet} through ${\cal O}(\alpha_s)$ corrections and, according
to \re{partonmoments}, its `genuine' twist-3 part is parameterized by the
gluon distribution $\Delta g_T(x)$. This suggests that similar to the
leading twist expressions \re{g11} and \re{gWW}, the twist-3 structure
function $g_2^{\rm tw-3}(x)$ can be described in terms of the
quark and gluon distributions, $\Delta q_T^+(x)$ and $\Delta g_T(x)$, respectively.

A deficiency of this interpretation is, however, that it does not
go through beyond the leading order. This is seen explicitly
on the gluon contribution in Eqs.~\re{otvet} and \re{CD}: The coefficient function
$\Omega^{qg}_n(\xi_i)$ that is responsible for the mixing with quark-gluon
operators does not coincide with $\Phi^{g}_n(\xi_i)$ and, therefore,
this mixing brings in gluon contributions that are not expressed entirely in terms
of $\Delta g_T(\xi)$ defined in \re{partonmoments}. Another reason is that the scale dependence of the
distributions introduced in \re{Tq}, \re{Tg} involves the full
three-particle functions $D_q$ and $D_g$ in a nontrivial way and, again,
brings in additional degrees of freedom.

Aim of this paper is to analyse the effects of QCD evolution in some detail. We
will find that although the above mentioned difficulties do exist, their
numerical impact is likely to be minimal. We will then be able to write an
approximate effective two-channel evolution equation involving the two
distributions $\Delta q^+_T(y,\mu^2)$ and $\Delta g_T(y,\mu^2)$ in full analogy
with the flavor-singlet DGLAP evolution equations in the leading twist.

\section{Hamiltonian approach to the three-particle evolution equations}
\setcounter{equation}{0}

Choosing ${\mu_{\overline{\rm MS}}^2}=Q^2$ one can
eliminate large logarithmic corrections to the gluon coefficient function
in \re{otvet}. To the leading logarithmic accuracy $(LL)$ and retaining
the flavor-singlet (S) contribution to the structure function $g_2(x)$
we write
\be{otvet-LL}
g_{2}^{{\rm LL}}(n,Q^2) = \langle e^2_q \rangle \,\frac{2}{n}
\int_{-1}^1\!{\cal D}\xi\,  \Phi_n^{q}(\xi_1,\xi_3)\,
   D_q^{\rm S}(\xi_i,Q^2)\,,
\ee
where
\be{singlet}
 \langle e^2_q\rangle =  \frac{1}{n_f}\sum_{q=u,d,s,\ldots}\!\!\!e^2_q\,,
\qquad
 D_q^{\rm S}(\xi_i) = D_u(\xi_i)+D_d(\xi_i)+D_s(\xi_i)+\ldots\,.
\ee
The gluon distribution is not present explicitly (to this accuracy)
but reappears through the evolution of the quark distribution to
lower scales. To see how this happens, expand \re{otvet-LL} in contributions
of flavor-singlet local operators
$\langle\!\langle S^k_N(Q^2) \rangle\!\rangle$
defined in \re{S-red}.
Using \re{psi-n} and \re{Dg-moments} one finds
\be{otvet-LL-1}
g_{2}^{{\rm LL}}(n,Q^2)=
    \langle e^2_q \rangle\,\frac{2}{n} \sum_{k=0}^N (-1)^{N-k}
      (N-k+1) \langle\!\langle S^k_N(Q^2)\rangle\!\rangle\,.
\nonumber
\ee

The scale dependence of the reduced matrix elements is described by the
system of coupled evolution equations
\ba{EQ-local}
Q^2 \frac{d}{d Q^2}\langle\!\langle S^k_N(Q^2) \rangle\!\rangle &=&
-\frac{\alpha_s}{4\pi}\left(
[V^{qq}_N]_{kk'}\langle\!\langle S^{k'}_N(Q^2) \rangle\!\rangle +
[V^{qg}_N]_{km'}\langle\!\langle G^{m'}_N(Q^2) \rangle\!\rangle
\right),
\nonumber\\
Q^2 \frac{d}{d Q^2}\langle\!\langle G^m_N(Q^2) \rangle\!\rangle &=&
-\frac{\alpha_s}{4\pi}\left(
[V_N^{gq}]_{mk'}\langle\!\langle S^{k'}_N(Q^2) \rangle\!\rangle +
[V_N^{gg}]_{mm'}\langle\!\langle G^{m'}_N(Q^2) \rangle\!\rangle
\right)
\ea
with $[V^{AB}_N]$ being the known matrices of anomalous
dimensions \cite{BKL84}, $k\,,k'=0,...,N$ and $m\,,m'=0,...,[N/2]-1$. (Here
$N=n-3$ is the number of derivatives in the quark-gluon operator,
$[\ldots]$ stands for an integer part.) Solving these equations one
defines $[3N/2]+1$ linear combinations of the matrix elements
\be{conf-ops}
\langle\!\langle{\cal O}_{N,\,\alpha}\rangle\!\rangle  =
\sum_{0\le k\le N} C_{\alpha k}^q (N)\,
\langle\!\langle[S]^k_N\rangle\!\rangle +
\sum_{0\le m\le [N/2]-1} C_{\alpha m}^g (N)\,
\langle\!\langle[G]^m_N\rangle\!\rangle\,,
\qquad \alpha=0,...,[3N/2]
\ee
that are renormalized multiplicatively and obtain the moments  of the
structure function in the standard form \re{OPE}. The
corresponding anomalous dimensions $\gamma_N^\alpha$ can be found by
diagonalizing the full matrix of the anomalous dimensions entering \re{EQ-local}
\be{full}
(C_{\alpha}^q,C_{\alpha}^g)\bigg[ {\cal V}_N - \gamma_N^\alpha\II\bigg]=0\,,
\qquad
{\cal V}_N = \left(\begin{array}{cc} V_N^{qq} & V_N^{qg}
                                 \\  V_N^{gq} & V_N^{gg}
                   \end{array}\right)\,.
\ee
The left eigenstates of the mixing matrix define the vector of
the coefficient functions $(C_{\alpha k}^q(N),C_{\alpha m}^g (N))$ entering \re{conf-ops}.

For lowest values of the moments the mixing matrix ${\cal V}_N$ looks
as follows. For $N=0$ the matrix consists of only one element
\be{N=1}
{\cal V}_{0}=\frac{17}6N_c+\frac16\frac1{N_c}+\frac23n_f\,,
\ee
while for $N=2$ it has the following form
%\be{N=3}
%{\cal V}_{2}= \left [\begin {array}{cccc} {\frac {287}{60}}\,N_c - {\frac
%{37}{60}}\,\frac1{N_c} + \frac15\,{n_f} & \frac35\,{N_c}+{\frac
%{23}{20}}\,\frac1{N_c}+\frac25\,{n_f} & -{\frac {3}{20}}\,{N_c}+{\frac
%{3}{20}}\,\frac1{N_c}+\frac15\,{n_f} &
%%%\frac1{40}\,n_f
%\frac1{10}\,n_f
%\\
%\noalign{\medskip}
%%
%\frac13\,{N_c}+\frac1{12}\,\frac1{N_c}+\frac2{15}\,{n_f}
%&
%{\frac {59}{12}}\,{N_c}-\frac32\,\frac1{N_c}+{\frac {4}{15}}\,{n_f}
%&
%\frac12\,{N_c}+{\frac {7}{12}}\,\frac1{N_c}+\frac2{15}\,{n_f}
%&
%%%{\frac {7}{120}}\,{n_f}
%{\frac {7}{30}}\,{n_f}
%\\
%\noalign{\medskip}
%%
%-\frac1{12}\,{N_c}+\frac12\,\frac1{N_c}+\frac15\,{n_f}
%&
%\frac12\,{N_c}+\frac74\,\frac1{N_c}+\frac25\,{n_f}
%&
%{\frac {17}{4}}\,{N_c}-\frac76\,\frac1{N_c}+\frac15\,{n_f}
%&
%%%-{\frac {17}{120}}\,{n_f}
%-{\frac {17}{30}}\,{n_f}
%\\
%\noalign{\medskip}
%%
%{\frac {23}{120}}\,{N_c}
%&
%{\frac {7}{40}}\,{N_c}
%&
%-{\frac {37}{120}}\,{N_c}
%&
%\frac23\,{n_f}+{\frac {307}{60}}\,{N_c}
%\end {array}\right]
%\ee
%
%
\be{N=3}
{\cal V}_{2}= \left [\begin {array}{cccc}
   \frac {17}{4}\,N_c-\frac76\,\frac1{N_c}+\frac15\,n_f
 & \frac12\,{N_c}+\frac74\,\frac1{N_c}+\frac25\,{n_f}
 & -\frac1{12}\,{N_c}+\frac12\,\frac1{N_c}+\frac15\,{n_f}
 & -\frac {17}{30}\,n_f
\\
\noalign{\medskip}
  \frac12\,{N_c}+{\frac {7}{12}}\,\frac1{N_c}+\frac2{15}\,{n_f}
&{\frac {59}{12}}\,{N_c}-\frac32\,\frac1{N_c}+{\frac {4}{15}}\,{n_f}
&\frac13\,{N_c}+\frac1{12}\,\frac1{N_c}+\frac2{15}\,{n_f}
&{\frac {7}{30}}\,{n_f}
\\
\noalign{\medskip}
  -{\frac {3}{20}}\,{N_c}+{\frac {3}{20}}\,\frac1{N_c}+\frac15\,{n_f}
&\frac35\,{N_c}+{\frac {23}{20}}\,\frac1{N_c}+\frac25\,{n_f}
&\frac {287}{60}\,N_c - {\frac{37}{60}}\,\frac1{N_c} + \frac15\,{n_f}
&\frac1{10}\,n_f
\\
\noalign{\medskip}
  -{\frac {37}{120}}\,{N_c}
& {\frac {7}{40}}\,{N_c}
& {\frac {23}{120}}\,{N_c}
& \frac23\,{n_f}+{\frac {307}{60}}\,{N_c}
\end {array}\right]
\ee
Going through an explicit calculation of \re{conf-ops} and \re{full} and
putting $N_c=n_f=3$ one finds for the two lowest moments \re{otvet-LL-1}
\ba{N=0,2-sing}
\frac{3}{2}\langle e^2_q\rangle^{-1} g_2^{\rm LL}(3,Q^2) &=& L^{\gamma_0^0/b}
  \langle\!\langle S^0_0\rangle\!\rangle\,,
\nonumber\\
\frac{5}{2}\langle e^2_q\rangle^{-1} g_2^{\rm LL}(5,Q^2) &=&
   L^{\gamma_2^0/b}
  \Big[
    0.415\langle\!\langle S^2_2 \rangle\!\rangle
     - 2.558\langle\!\langle S^1_2 \rangle\!\rangle
  + 2.776\langle\!\langle S^0_2 \rangle\!\rangle
  + % 2.732
    0.966\langle\!\langle G^0_2 \rangle\!\rangle
 \Big]
\nonumber\\
&+&
 L^{\gamma_2^1/b}
  \Big[
   0.340 \langle\!\langle S^2_2 \rangle\!\rangle
    -0.152 \langle\!\langle S^1_2 \rangle\!\rangle
   -0.261 \langle\!\langle S^0_2 \rangle\!\rangle
  %  -0.324
  % -0.029
  -0.114\langle\!\langle G^0_2 \rangle\!\rangle
 \Big]
\nonumber\\
&+&
 L^{\gamma_2^2/b}
  \Big[
    0.134 \langle\!\langle S^2_2 \rangle\!\rangle
  + 0.496 \langle\!\langle S^1_2 \rangle\!\rangle
   +0.404 \langle\!\langle S^0_2 \rangle\!\rangle
   %-2.570
   %- 0.227
  -0.909 \langle\!\langle G^0_2 \rangle\!\rangle
 \Big]
\nonumber\\
&+&
 L^{\gamma_2^3/b}
  \Big[
    0.111\langle\!\langle S^2_2 \rangle\!\rangle
  + 0.214\langle\!\langle S^1_2 \rangle\!\rangle
  + 0.080\langle\!\langle S^0_2 \rangle\!\rangle
  % -0.161
  %+ 0.014
  +  0.057\langle\!\langle G^0_2 \rangle\!\rangle
 \Big]
\ea
where $L = \alpha_s(Q^2)/\alpha_s(\mu^2)$,
all reduced matrix elements on the r.h.s. are
normalized at the scale $\mu^2$ and the flavor-singlet anomalous dimensions are
equal to
\be{lowdim-sing}
 \gamma_0^0 = 10.5556\,,\quad
 \gamma_2^0 =10.7393 \,,\quad
 \gamma_2^1 = 13.5155\,,\quad
 \gamma_2^2 =  17.6794\,,\quad
 \gamma_2^3 = 18.1714\,.
\ee
If $\mu^2=Q^2$, then $L=1$ and the coefficients in front of
$\langle\!\langle S^2_2 \rangle\!\rangle$,
$\langle\!\langle S^1_2 \rangle\!\rangle$,
$\langle\!\langle S^0_2 \rangle\!\rangle$,
$\langle\!\langle G^0_2 \rangle\!\rangle$ coincide with their
tree-level values 1, -2, 3 and 0, respectively, as expected from \re{otvet-LL-1}.
Note that the largest coefficients occur in the contribution of the
operator with the lowest anomalous dimension, that is  similar
to flavor-nonsinglet case \cite{ABH}, and the most important correction
is apparently associated with the operator with the {\em second-largest}
anomalous dimension. Aim of this work is to explain this structure
and understand how it extends
 for arbitrary moment $n$. To this end, we develop
a new %technical
framework for solving the three-particle evolution
equations, dubbed Hamiltonian approach in what follows.
A short account of the same technique is presented in our letter \cite{BKM00},
where it was used to calculate $1/N_c^2$ correction to the evolution in
flavor-nonsiglet sector.

The basic idea of our approach can be explained as follows. As it follows
from \re{N=3}, the mixing matrices ${\cal V}_N$ is \re{EQ-local} do not have
any obvious symmetry and, in general, are quite complicated. In particular, they
are not hermitian and their eigenvectors are not orthogonal to each other. On
the other hand, by a numerical diagonalization, Eq.~\re{full}, one finds
that all eigenvalues of these matrices (anomalous dimensions)
are real for arbitrary $N$. This property is not obvious and allows to
suspect some hidden symmetry of the problem, which is not manifest
in the particular representation of the evolution equations \re{EQ-local}
involving only forward matrix elements of the operators.
We will argue that this symmetry
indeed exists and is nothing else as the familiar conformal symmetry of the
QCD Lagrangian.

The conformal symmetry manifest itself through
%which is translated into
 the $SL(2,\mathbb{R})$ invariance
of the renormalization group equations describing the evolution of
the local twist-3 operators including operators with the total derivatives.
The $SL(2,\mathbb{R})$ symmetry of these evolution equations is obscured by the
restriction to forward matrix elements of the operators in \re{EQ-local}
(or, equivalently, the condition that momentum fractions of the partons sum to
 zero in \re{otvet-LL}, cf. \re{Dxi}, $\xi_1+\xi_2+\xi_3=0$).
Since the operators containing total derivatives have vanishing forward matrix
elements, it seems natural to neglect their mixing with the twist-3 operators
\re{localop}  in the discussion of deep inelastic scattering. But it is this
 reduction that complicates the structure of the evolution equations
if it is imposed from the beginning.

Our approach relies on the conformal symmetry of the evolution equations and
 can be illustrated by the following scheme indicating a chain of
 transformations on the matrix of the evolution kernels ${\cal V}_N$:
\baa%{picture}
&&
{}
\nonumber
\\[-3mm]
&&
\underbrace{
\left(
\begin{array}{c}
\mbox{\rm forward}\\
\mbox{\rm non-hermitian kernels}
\end{array}
\right)}_{\mbox{$\ell_q+\ell_g$}}
\Rightarrow
\underbrace{\left(
\begin{array}{c}
\mbox{\rm non-forward}\\
\mbox{\rm  hermitian kernels}\\
\end{array}
\right)}_{\mbox{$\frac12[\ell_q(\ell_q-1)+\ell_g(\ell_g-1)]$}}
\Rightarrow
\underbrace{\left(
\begin{array}{c}
{\rm  hermitian~kernels}\\
{\rm in~conformal~basis}
\end{array}
\right)}_{\mbox{$\ell_q+\ell_g$}}
%\\&&
%\hspace*{1.1cm}N+1\hspace*{2.2cm} N(N+1)/2 \hspace*{2.3cm} N+1\nonumber
\eaa
Instead of dealing with the non-hermitian ``forward'' mixing matrices ${\cal V}_N$
of dimension $\ell_q+\ell_g=(N+1)+[N/2]$, with $\ell_q$ and $\ell_g$ being
the total number of quark-antiquark-gluon and three-gluon forward matrix
elements, we choose to consider much bigger matrices
(but hermitian with respect to the so-called conformal scalar product) of dimension  $[\ell_q(\ell_q-1)+\ell_g(\ell_g-1)]/2$
that take into account the mixing with the
operators containing total derivatives. Diagonalizing the thus defined
``non-forward'' evolution kernels we expand its eigenstates over the
basis  of ``spherical harmonics'' of the conformal $SL(2,\mathbb{R})$ group
and obtain much simpler matrix equation of the coefficients in this expansion
(see Appendix C).
Thanks to the conformal invariance, the corresponding matrix, defining the
non-forward evolution kernels in the conformal basis representation,
has smaller dimension, $\ell_q+\ell_g$, and is now {\it hermitian}.
We then make a forward projection at the very end.

\subsection{Coefficient functions of local operators}

An arbitrary local three-particle
 operator ${\cal O}_{N,\alpha}$ is defined by the set of coefficients
in the expansion over the standard basis of operators built from the elementary
fields $\Phi_k$ and covariant derivatives (cf. \re{conf-ops}), schematically
\be{cf1}
{\cal O}_{N,\alpha} = \sum_{k_1+k_2+k_3=N} w_{k_1k_2k_3}\,
 (D^{k_1} \Phi_1)(D^{k_2} \Phi_2)(D^{k_3} \Phi_3)\,,
\ee
or, equivalently, by a characteristic homogenous polynomial of three
variables
\be{def:Psi}
   \Psi_{N,\alpha}(x_1,x_2,x_3) = \sum_{k_1+k_2+k_3=N} w_{k_1k_2k_3}^{N,\alpha}\,
  x_1^{k_1}x_2^{k_1}x_3^{k_1}\,,
\ee
which we shall refer to as the {\it coefficient function of the operator\/} (not to be
mixed with the coefficient functions in the OPE).
The rationale for the name is that with the help of the coefficient
function the local operator can be ``projected out'' of the corresponding
nonlocal operator, in our case
\ba{direct}
{\cal O}_{N,\alpha} &=&
\left[2N_c
\Psi_{N,\alpha}^{q}
(\partial_{z_1},\partial_{z_2},\partial_{z_3}) S_\mu(z_1,z_2,z_3) +
n_f\Psi_{N-1,\alpha}^{g}(\partial_{z_1},\partial_{z_2},\partial_{z_3})
\widetilde O_\mu(z_1,z_2,z_3)\right]\Bigg|_{z_i=0}
\nonumber\\&=&
\Big(2N_c\,\Psi_{N,\alpha}^{q}(\partial_{z_i}),
n_f\,\Psi_{N-1,\alpha}^{g}(\partial_{z_i})\Big)
\left (
\begin{array}{c}
S_\mu(z_i)\\
\widetilde O_\mu(z_i)
\end{array} \right)\Bigg|_{z_i=0}\,,
\ea
where $\Psi_N^{q}(x_i)$ and $\Psi_{N-1}^{g}(x_i)$ are
homogenous polynomials of degree $N_{\bar qgq}=N$ and
$N_{g\widetilde g g}=N-1$,
respectively.%
\footnote{The difference in the number of derivatives
is compensated by the  different dimensions of the quark and gluon field.}
The construction of the multiplicatively renormalizable operators
\re{direct} is equivalent to finding of the appropriate coefficient
functions. The normalization color factors have been included in
\re{direct} for later convenience (see Eq.~\re{sc-prod} below).

To expose the symmetries of the problem, it is convenient to start with
the evolution equations for the corresponding nonlocal operators
$S_\mu(z_i)$ and $\widetilde O_\mu(z_i)$ defined in \re{Spm} and
\re{Odual}, respectively. They have the following general form
\ba{EQ-nonlocal}
Q^2 \frac{d}{d Q^2} S_\mu(z_1,z_2,z_3) &=& -\frac{\alpha_s}{4\pi}\left[
\widehat H_{qq}\, S_\mu\,(z_1,z_2,z_3)+
\widehat H_{qg}\,\widetilde O_\mu\,(z_1,z_2,z_3)\right],
\nonumber
\\
Q^2 \frac{d}{d Q^2}\widetilde O_\mu(z_1,z_2,z_3)
&=& -\frac{\alpha_s}{4\pi}\left[
\widehat H_{gq}\,S_\mu\, (z_1,z_2,z_3)+
\widehat H_{gg}\,\widetilde O_\mu\,(z_1,z_2,z_3)\right],
\ea
where $z_i$ stand for the  light-cone coordinates of the quarks and gluons,
and $\widehat H_{ab}$ $(a,b=q,g)$ are  integral operators
acting on quark (gluon) coordinates and describing the
interaction between quarks and gluons on the light-cone,
$\widehat H_{qq}\,S_\mu\,(z_i)= \int d z'_k
\widehat H_{qq}(z_i|z'_k)S_\mu(z'_k)$
and similar for the other kernels. Note that
the short-distance expansion of the
nonlocal operators $S^\pm_\mu$ and $\widetilde O_\mu$ gives rise to the local
twist-3 operators \re{localop} as well as operators
with the total
derivatives. In difference to the previous discussion
we do not assume the translation invariance, $S_\mu\,(z_1,z_2,z_3)\neq
S_\mu\,(z_1+\delta,z_2+\delta,z_3+\delta)$, so that the mixing with
operators containing total derivatives is included in \re{EQ-nonlocal}.
The explicit expressions for the kernels $\widehat H_{ab}$
can be found in \cite{BB89}. They will not be needed for what follows.

The evolution equation \re{EQ-nonlocal} has the form of the Schr\"odinger
equation with the $2\times 2$ matrix of the evolution kernels
$\widehat H_{ab}$ playing the r\^ole of the Hamiltonian.
Let $\widehat\Psi_{N,\alpha}^{q}(z_i)$ and $\widehat\Psi_{N-1,\alpha}^{g}(z_i)$
be homogenous polynomials in the light-cone
coordinates of quarks and gluons of degree $N$ and $N-1$,
respectively, satisfying the Schr\"odinger equation
\be{Sch-dual}
\left(
\begin{array}{cc}
\widehat H_{qq} & \widehat H_{qg} \\ \widehat H_{gq} & \widehat H_{gg}
\end{array}
\right) \left(\widehat\Psi_{N,\alpha}^{q}(z_i) \atop
\widehat\Psi_{N-1,\alpha}^{g}(z_i)\right) = {\cal E}_{N,\alpha}
\left(\widehat\Psi_{N,\alpha}^{q}(z_i)\atop
\widehat\Psi_{N-1,\alpha}^{g}(z_i)\right)
\ee
with the same evolution kernels  as in \re{EQ-nonlocal}.
The nonlocal operators $S_\mu(z_i)$ and $\widetilde O_\mu(z_i)$
can be expanded in terms of these functions with certain operator-valued
coefficients
\be{short-dist}
\left (
\begin{array}{c}
S_\mu(z_i)\\
\widetilde O_\mu(z_i)
\end{array} \right)_{\!Q^2} =\sum_{N,\alpha}
\left (
\begin{array}{c}
\widehat\Psi_{N,\alpha}^{q}(z_i)\\
\widehat\Psi_{N-1,\alpha}^{g}(z_i)
\end{array} \right)\,  {\cal O}_{N,\alpha}(Q^2)\,,
\ee
where the subscript in the l.h.s.\ stands for the normalization scale.
It follows from the evolution equation that the local operators
${\cal O}_{N,\alpha}$ that appear in this expansion
are renormalized multiplicatively and their anomalous dimensions are
determined by the corresponding energy eigenvalues ~${\cal E}_{N,\alpha}$
\be{anom-dim}
\gamma_N^\alpha = {\cal E}_{N\!,\,\alpha}
\ee
with the subscript $\alpha=0,1,...,[3N/2]$ enumerating different solutions
to \re{Sch-dual}. The scale dependence of  ${\cal O}_{N,\alpha}$
takes, therefore, the standard form
\be{scale-dep}
{\cal O}_{N,\alpha}(Q^2) = {\cal O}_{N,\alpha}(\mu^2) \left(
\frac{\alpha_s(Q^2)}{\alpha_s(\mu^2)}\right)^{\gamma_{N}^\alpha/b}.
\ee

So far we have introduced two different sets of polynomials:
$(\widehat\Psi_N^{q}(z_i),\widehat\Psi_{N-1}^{g}(z_i))$
and $(\Psi_N^{q}(x_i),\Psi_{N-1}^{g}(x_i))$. The former set
defines the expansion \re{short-dist} of nonlocal operators $S_\mu(z_i)$ and
$\widetilde O_\mu(z_i)$ over the complete set of local multiplicatively
renormalizable operators ${\cal O}_{N,\alpha}$,
while the latter determines the particular form of the
multiplicatively renormalizable operators
${\cal O}_{N,\alpha}$, cf.~\re{cf1}, \re{def:Psi}.
It follows from the definition \re{direct} and \re{short-dist} that
\be{psi-phi}
\left[2N_c\,
\Psi_{N,\alpha}^{q}(\partial_{z_i})
\widehat\Psi_{N',\alpha'}^{q} (z_i) +
n_f\,
\Psi_{N-1,\alpha}^{g}(\partial_{z_i})
\widehat\Psi_{N'-1,\alpha'}^{g}(z_i)\right]\Bigg|_{z_i=0}=
\delta_{NN'}\delta_{\alpha\alpha'},
\ee
and in this sense the polynomials $\widehat\Psi$
are dual to the coefficient functions $\Psi$. In what
follows we shall refer to $\widehat\Psi-$functions as coefficient functions
of a local operator in the dual representation.
One can determine the
functions $\Psi^{q}_{N,\alpha}$ and $\Psi^{g}_{N-1,\alpha}$ from the
orthogonality condition \re{psi-phi} provided the complete set of
eigenfunctions $\widehat\Psi^{q}_{N,\alpha}$ and $\widehat\Psi^{g}_{N-1,\alpha}$
of the
Schr\"odinger equation \re{short-dist} is given.
This task seems complicated, but in fact is not.
Using conformal symmetry, we will be able to find an explicit
expression connecting the ``direct'' $\Psi(x_i)$ and dual
$\widehat\Psi(z_i)$ coefficient functions
of a multiplicatively renormalizable operator. The answer is
given in Eq.~\re{exp-dual} below.

\subsection{Conformal symmetry}

A remarkable property of the evolution kernels in \re{EQ-nonlocal} is that
they are invariant under the projective transformations of the light-cone
coordinates of quark and gluons, $z_k$,
\be{SL2}
z_k \to \frac{az_k+b}{cz_k+d}\,,\qquad ad-bc=1\,.
\ee
This invariance has its roots in the conformal symmetry of the QCD
Lagrangian and the transformations \re{SL2} form the
$SL(2,\mathbb{R})$ (collinear) subgroup of the full conformal group
acting on the fields
``living'' on the light-cone. As well known, the conformal symmetry of QCD
is broken by quantum corrections. However,
since the leading-order renormalization group equations are driven by
tree-level counterterms, they have to respect the symmetry of the QCD Lagrangian.

The action of the $SL(2,\mathbb{R})$ transformations \re{SL2}
on (quantum) fields
$\Phi_a(z)$,  where $a=\bar q\,,q\,,g\,,\widetilde g$ corresponds to
$\bar q(z),q(z), G_{x\perp}(z),\widetilde G_{x\perp}(z)$, respectively,
is defined as
\be{fields}
\Phi_a(z)\to  (cz+d)^{-2j} \Phi_a\lr{\frac{az+b}{cz+d}}
\ee
and is described by three generators
$\widehat L_+$, $\widehat L_-$ and $\widehat L_0$ that
can be realized as first-order differential operators acting on the field
coordinates:
\ba{L's}
\widehat L^-_{a}\Phi_a(z)&=&-\partial_z \Phi_a(z)\,,
\nonumber
\\
\widehat L^+_{a}\Phi_a(z)&=&(z^2\partial_z+2j_a z) \,\Phi_a(z)\,,
\\
\widehat L^0_{a}\Phi_a(z)&=&(z\partial_z+j_a)\, \Phi_a(z)\,.
\nonumber
\ea
Here $j_a=(l_a+s_a)/2$ is the conformal spin of the field
$\Phi_a(z)$, with $l_a$ being a
canonical dimension (3/2 for quarks and 2 for gluons)  and $s_k$
the spin projection  on
the light-cone direction, $\Sigma_{pz} \Phi_a = i s_a \Phi_a$.
In the case at hand the spin projections have their maximum values
$s_{\bar q}=s_q=1/2$, $s_g=s_{\widetilde g}=1$, leading to
\be{j's}
j_{\bar q}=j_q=1\,,\qquad j_g=j_{\widetilde g}=3/2\,.
\ee
In order to unify the notation,
we introduce two $2\times 2$ matrices of the evolution
kernels $\widehat {\cal H}$ and the generators of conformal transformations
$\widehat {\cal L}_k$ $(k=\pm,0)$ in the
quark-antiquark-gluon and three-gluon channels
\be{matrices}
\widehat {\cal H}=
\left(\begin{array}{cc}
  \widehat H_{qq} & \widehat H_{qg} \\
  \widehat H_{gq} & \widehat H_{gg}
\end{array}\right)\,,\qquad
\widehat {\cal L}_k=
\left(\begin{array}{cc}
  \widehat L_{\bar q g q}^k & 0 \\
  0 & \widehat L_{g\widetilde g g}^k
\end{array}\right)\,,
\ee
where $\widehat L_{\bar q g q}^k$ and
$\widehat L_{g\widetilde g g}^k$ $(k=\pm,0)$ are the total
three-particle $SL(2,\mathbb{R})$ generators acting on
antiquark-quark-gluon and three-gluon coordinates, respectively:
\be{Three-L}
\widehat L_{\bar q g q}^k = \widehat L_{\bar q}^k + \widehat L_{q}^k +
\widehat L_{g}^k
\,,\qquad \widehat L_{g\widetilde g g}^k=\widehat L_{g}^k +
\widehat L_{\widetilde g}^k + \widehat L_{g}^k\,.
\ee
The conformal invariance of the evolution equation is stated as
\be{conf-sym}
[\widehat {\cal H}, \widehat {\cal L}_k] = [\widehat {\cal H},
\widehat {\cal L}^2] =[\widehat {\cal L}^2, \widehat {\cal L}_k] = 0
\ee
where
\be{L2}
\widehat {\cal L}^2 = \frac12(\widehat {\cal L}_+\widehat {\cal L}_-+\widehat {\cal L}_-\widehat {\cal L}_+) + \widehat {\cal L}_0^2
\ee
is the three-particle quadratic Casimir operator.

Thanks to the conformal invariance \re{conf-sym} the solutions to \re{Sch-dual}
can be classified according to the representations of the $SL(2,\mathbb{R})$
group. Namely, we can impose additional constraints on the eigenfunctions
\be{constraints}
\left[
\widehat {\cal L}_0-J \right]
\left(\widehat\Psi_{N}^{q}(z_i)\atop
\widehat\Psi_{N-1}^{g}(z_i)\right) =
\widehat {\cal L}_{-} \left(\widehat\Psi_{N}^{q}(z_i)\atop\widehat
\Psi_{N-1}^{g}(z_i)\right) =
\left[
\widehat {\cal L}^2-J(J-1)\right]
\left(\widehat\Psi_{N}^{q}(z_i)\atop
\widehat\Psi_{N-1}^{g}(z_i)\right) =0\,,
\ee
where the $SL(2,\mathbb{R})$ generators are given by the same expressions
as in \re{L's}. Since the eigenfunctions are homogenous polynomials in the
light-cone coordinates $z_k$, the first equation in \re{constraints} is
automatically
satisfied provided
\be{J}
J=\frac72 + N_{\bar q q g} = \frac92 + N_{g\widetilde g g}\,.
\ee
where  $N_{\bar q q g}$ and $N_{g\widetilde g g}$ count the total number of
covariant derivatives in the corresponding local operator,
$N_{\bar q q g}=N_{g\widetilde g g}+1=N$. Notice that for $J=7/2$, or
equivalently $N_{\bar q q g}=0$ there exists no  three-gluon contribution.
The second condition ensures that $\widehat\Psi_{N}^{q}$ and
$\widehat\Psi_{N-1}^{g}$ are invariant under translations of the light-cone
coordinates. This requirement  defines the so-called highest weight of the
%$SL(2,\mathbb{R})$
discrete series representation of the
$SL(2,\mathbb{R})$ group, labeled by the conformal
spin $J=N+7/2$. An infinite tower of solutions to \re{Sch-dual} can be
obtained from the highest weight by the repeated application of  the
`step-up' operator $\widehat {\cal L}_+$:
\be{desc}
\left({}^n\widehat\Psi_{N}^{q}(z_i)\atop
{}^n\widehat \Psi_{N-1}^{g}(z_i)\right)=
\widehat {\cal L}_{+}^n \left(\widehat\Psi_{N}^{q}(z_i)\atop\widehat
\Psi_{N-1}^{g}(z_i)\right).
\ee
Since $\widehat {\cal L}_+$ commutes with the Hamiltonian, Eq.~\re{conf-sym}, all
states ${}^n\widehat \Psi_{N}$, with different
$n=0,1,2,\ldots$ have the same energy.
As we will show in a moment
(see Eq.~\re{desc-dual}), the corresponding operators
${}^n{\cal O}_{N}$
are just those obtained from the highest weight state operator by
adding the $n$-th power of the total derivative and, therefore,
they do not survive
upon taking a forward matrix element.
Note that the expansion in \re{short-dist} formally includes operators
with arbitrary powers of total derivatives,
but  we can  ignore their contribution and concentrate
on studying the properties of the highest weights \re{constraints} only.

Going over from the dual coefficient functions $\widehat \Psi(z_i)$ to
the coefficient functions $\Psi(x_i)$ defined in \re{direct}
corresponds to going over to a different (non-standard)
representation of the conformal group.
Using the relation \re{psi-phi} and requiring
\be{L-dual}
\left[L_k^{\pm,0} \Psi(\partial_{z_1},\partial_{z_2},
\partial_{z_3})\right] \widehat\Psi(z_1,z_2,z_3)\bigg|_{z_i=0}
=\Psi(\partial_{z_1},\partial_{z_2},\partial_{z_3})\left[\widehat L_k^{\mp,0}
\widehat  \Psi(z_1,z_2,z_3)\right]\bigg|_{z_i=0}
\ee
one finds the following representation of the $SL(2,\mathbb{R})$ generators on
the space of the coefficient functions $\Psi(x_1,x_2,x_3)$
\ba{dual-L}
L_k^0\, \Psi(x_k) &=& (x_k\partial_{x_k} + j_k) \Psi(x_k),
\nonumber\\
L_k^+\, \Psi(x_k) &=& -x_k \Psi(x_k),
\\
L_k^-\, \Psi(x_k) &=& (x_k\partial_{x_k}^2 +
2j_k\partial_{x_k})
\Psi(x_k)
\nonumber
\ea
with the conformal spins $j_k$ defined in \re{j's}. Note that these
expressions are more complicated compared to the standard expressions
\re{L's}. Eqs.~\re{dual-L} and \re{L's} define the $SL(2)$ generators
in two different representations and, as such, they are related to each
other through a transformation
\be{T}
 L_k^{\pm}=-\mathbb{T}^{-1}
\widehat L_k^{\pm}\mathbb{T}\,,\qquad
 L_k^{0}=\mathbb{T}^{-1}
\widehat L_k^{0}\mathbb{T}\,,\qquad
\widehat \Psi(z_i)=[\mathbb{T} \Psi](z_i)
\ee
that maps into each other the coefficient functions in two different
representations. The explicit form of the $\mathbb{T}-$transformation
is given by
\be{map-ex}
\widehat \Psi(z_i)
%=[\mathbb{T} \Psi](z_i)
=\Psi(\partial_{x_i})\prod_{i=1}^3 (1-x_i\,z_i)^{-2j_i}\Big|_{x_i=0}
=
\prod_{i=1}^3\int_0^\infty \, \frac{dt_i\,t^{2j_i-1}}{\Gamma(2j_i)} e^{-t_i}
\,\Psi(z_i t_i)\,.
\ee
To verify this relation, use \re{L's} and \re{dual-L} to check
that $\widehat L_k^{\pm}\widehat \Psi(z_i)=
-[\mathbb{T} (L_k^{\pm}\Psi)](z_i)$ and
$\widehat L_k^{0}\widehat \Psi(z_i)=
[\mathbb{T} (L_k^{0}\Psi)](z_i)$.
The conformal constraints \re{constraints} on the coefficient functions
corresponding to the highest weight look exactly as before:
\be{constraints-dual}
\left[
{\cal L}_0-J \right]
\left(\Psi_{N}^{q}(x_i)\atop
\Psi_{N-1}^{g}(x_i)\right) =
{\cal L}_{-} \left(\Psi_{N}^{q}(x_i)\atop
\Psi_{N-1}^{g}(x_i)\right) =
\left[
{\cal L}^2-J(J-1)\right]
\left(\Psi_{N}^{q}(x_i)\atop
\Psi_{N-1}^{g}(x_i)\right) =0\,,
\ee
with the $SL(2)$ generators defined in the representation \re{dual-L}.
Note that the ``step-up'' operator $L_+$ has become very simple
and its action consists of adding the sum of derivatives acting
on each of the three fields, as we anticipated:
\be{desc-dual}
\left({}^n\Psi_{N}^{q}(x_i)\atop
{}^n\Psi_{N-1}^{g}(x_i)\right)= \widehat{\cal L}_{+}^n
\left(\Psi_{N}^{q}(x_i)\atop
\Psi_{N-1}^{g}(x_i)\right)
=(x_1+x_2+x_3)^n\left(\Psi_{N}^{q}(x_i)\atop
\Psi_{N-1}^{g}(x_i)\right)\,.
\ee
Finally, applying the transformation \re{map-ex} to the coefficient
function \re{def:Psi} we obtain
the following expression for the dual coefficient function
\be{exp-dual}
\widehat\Psi(z_i)=\sum_{k_1+k_2+k_3=N} w_{k_1 k_2 k_3} \,z_1^{k_1} z_2^{k_2}
z_3^{k_3}\,
\frac{\Gamma(k_1+2j_1)}{\Gamma(2j_1)}
\frac{\Gamma(k_2+2j_2)}{\Gamma(2j_2)}\frac{\Gamma(k_3+2j_3)}{\Gamma(2j_3)}\,.
\ee
This relation establishes the one-to-one correspondence between the coefficient
functions \re{def:Psi} and their dual counter-parts \re{exp-dual}.

Much of the following discussion is based on the fact that
the coefficient functions of multiplicatively
renormalizable operators $\Psi_{N,\alpha}(x_i)$ satisfying the
highest weight condition \re{constraints-dual} are orthogonal with
respect to the so-called conformal scalar product. This property becomes
crucial in establishing the hermiticity of the evolution Hamiltonian in
\re{Sch-dual}. The hermiticity property will be quite helpful in the
further analysis and as we argue below is a direct consequence of the
conformal symmetry.

Eq.~\re{psi-phi} suggests to define the following scalar product on the
space of  coefficient functions given by \re{def:Psi}
\be{dif-rep}
\vev{\Psi_{N,\alpha}|\Psi_{N,\beta}}\sim\Psi_{N,\alpha}^{q}(\partial_{z_i})
\widehat\Psi_{N,\beta}^{q} (z_i)\bigg|_{z_i=0}=
\Psi_{N,\alpha}(\partial_i)
[\mathbb{T} \Psi]_{N,\beta}(z_i)\bigg|_{z_i=0}.
\ee
For the coefficient functions of local operators without total
derivatives that satisfy
the constraints~(\ref{constraints}) (i.e. those that we are interested in)
one can equivalently  rewrite the definition in \re{dif-rep}
in a more familiar integral form:
\be{int-rep}
\vev{\Psi_{N,\alpha}|\Psi_{N,\beta}}=
\frac{\Gamma(2j_1+2j_2+2j_3+2N)}
{\Gamma(2j_1)\Gamma(2j_2)\Gamma(2j_3)}
\int_0^1 [dx]\,x_1^{2j_1-1} x_2^{2j_2-1} x_3^{2j_3-1}\,
\Psi_{N,\alpha}(x_i)\Psi_{N,\beta}(x_i)\,,
\ee
where $j_k$ are the conformal spins of the operators entering \re{cf1}.
% and an overall normalization factor is chosen for later convenience.
Here, the integration goes over the region $0\le x_k\le 1$, $x_1+x_2+x_3=1$
and the integration measure is defined as
\be{int-measure}
[dx]=dx_1 dx_2 dx_3\,\delta(x_1+x_2+x_3-1)\,.
\ee
The $SL(2)$
generators \re{dual-L} are (anti)self-adjoint operators on the space of
the coefficient functions endowed with the scalar product \re{dif-rep}.
%,\re{int-rep}.
Indeed using Eqs.~(\ref{L-dual}), \re{dual-L} and (\ref{T})
 it is straightforward to verify  that
\be{herm}
\vev{\Psi_{N,\alpha}| L_k^{\pm} \Psi_{N,\beta}}
=-\vev{ L_k^{\mp} \Psi_{N,\alpha}|\Psi_{N,\beta}}\,, \qquad
\vev{\Psi_{N,\alpha}| L_k^{0} \Psi_{N,\beta}}
=\vev{ L_k^{0} \Psi_{N,\alpha}|\Psi_{N,\beta}}\,.
\ee
As a consequence, the two-particle Casimir operators
${L}^2_{ik} = ({L}_i+{L}_i)^2$ defined as \re{L2} are the self-adjoint
operators.

\subsection{Helicity basis}

As we will see in the next section, the evolution Hamiltonians \re{matrices}
can be written in terms of the two-particle Casimir operators
$\widehat{L}^2_{ik} = (\widehat{L}_i+\widehat{L}_i)^2$ in the dual
representation.
This property ensures that the Hamiltonians inherit hermiticity
properties of the generators and are self-adjoint operators as well.
As a consequence, their eigenvalues alias  the anomalous dimensions
\re{anom-dim} are real and the corresponding eigenfunctions are
orthogonal to each other
\be{hint}
\vev{\Psi_{N,\alpha}|\Psi_{N,\beta}}\sim \delta_{\alpha\beta}\,.
\ee
In fact, for three-gluon operators there is a complication that
the solution to the Schr\"o\-din\-ger equation \re{Sch-dual}
has to be found on the subspace of functions
$\widehat\Psi^g_{N,\alpha}(z_i)$
that are antisymmetric under the exchange of the first and third
gluon $z_1\leftrightarrow z_3$. The permutation operator $P_{13}$
that projects onto the states with correct symmetry is not a
self-adjoint operator so that one has to be careful%
\footnote{The situation is in fact very similar
to the  Hartree-Fock construction of the completely antisymmetric
fermionic wave functions in quantum mechanics.}.
In physical terms, the problem arises because, as we will see
in a minute, the three-gluon operator $\widetilde O_\mu$ contains
a sum of contributions of gluons with opposite helicity,
antisymmetrized because of the crossing symmetry. The way out
\cite{BFLK} is therefore to write down the evolution equation for
the helicity eigenstates and restore the crossing symmetry at the
end.

The construction of the helicity basis is based on the decomposition
of the quark and gluon fields entering the definition of the nonlocal
operators $S_\mu(z_i)$ and $\widetilde O_\mu(z_i)$ into the components
of different chirality
\be{chir}
q^\pm(z) = \frac{1\mp\gamma_5}2 q(z)\,,\qquad
\bar q^\pm(z) = \bar q(z)\frac{1\mp\gamma_5}2 \,,\qquad
G_{\mu}^{\pm}(z) = \frac12\left[G_{\mu,n}\pm i \epsilon^\perp_{\mu\nu}
G_{\nu,n}\right],
\ee
where $\epsilon_{\mu\nu}^\perp=\epsilon_{\mu\nu\alpha\beta}
p_\alpha n_\beta/(pn)$. The  fields defined in this way
satisfy the conditions
\be{hel}
\gamma_5 q^\pm(z)=\pm q^\pm(z)\,,\qquad
i \epsilon^\perp_{\mu\nu} G_{\nu}^{\pm}(z) = \pm  G_{\mu}^{\pm}(z)
\ee
and describe quark, antiquark and gluon of a definite helicity.
Making this decomposition one obtains the
expressions for the nonlocal operators $S_\mu(z_i)$ and $\widetilde O_\mu(z_i)$
in terms of the chiral fields. We remind that $\mu =1,2$ is a transverse index
and in order to construct the three-particle states with definite overall
helicity one has to take particular linear combinations, projecting
onto the two complex vectors in the transverse plane:
\be{comp}
   w_\mu = e_{\perp,\mu}^{(1)}+i e_{\perp,\mu}^{(2)}\,,\qquad
   \widebar w_\mu = e_{\perp,\mu}^{(1)}-i e_{\perp,\mu}^{(2)}\,.
\ee

We find for the quark-antiquark-gluon operator
\ba{S-w}
S_w(z_1,z_2,z_3) &=& \bar q^{-}(z_1) G_w^{+}(z_2)\!\not\!n q^{+}(z_3)
+ \bar q^{+}(z_3) G_w^{+}(z_2)\!\not\!n q^{-}(z_1)\,,
\nonumber\\
S_{\widebar w}(z_1,z_2,z_3) &=& \bar q^{+}(z_1)
 G_{\widebar w}^{-}(z_2)\!\not\!n q^{-}(z_3)
+ \bar q^{-}(z_3) G_{\widebar w}^{-}(z_2)\!\not\!n q^{+}(z_1)\,.
\ea
Notice that the  quark and the antiquark have opposite helicity
in both cases, and helicity of the gluon $\pm 1$ concides with the
total helicity of the system. (We tacitly imply that the momenta of
the three partons are aligned along the same
light-cone direction defined by the  proton momemtum $p$.)
The similar decomposition of the three-gluon operator looks as follows
\ba{O-w}
\widetilde O_w(z_1,z_2,z_3) &=& \frac12\left[
T_w(z_1,z_2,z_3)- T_w(z_3,z_2,z_1)\right],
\nonumber\\
\widetilde {O}_{\widebar w}(z_1,z_2,z_3) &=& \frac12\left[
T_{\widebar w}(z_1,z_2,z_3)- T_{\widebar w}(z_3,z_2,z_1)\right],
\ea
where the notation was introduced
\ba{O-h}
T_w(z_1,z_2,z_3)&=&\frac{ig}2 f^{abc} G^{a,-}_{\bar w}(z_1)G^{b,+}_{w}(z_2)
G^{c,+}_{w}(z_3)\,,
\nonumber\\
T_{\widebar w}(z_1,z_2,z_3)&=&\frac{ig}2 f^{abc}
G^{a,+}_{w}(z_1)G^{b,-}_{\widebar w}(z_2)
G^{c,-}_{\widebar w}(z_3)\,.
\ea
The operators $T_w(z_i)$ and $T_{\widebar w}(z_i)$ describe the state
of three gluons with the total helicity $\pm 1$, respectively.

It follows from \re{O-w} and \re{O-h} that the operators
$\widetilde O_w(z_1,z_2,z_3)$ and $T_w(z_1,z_2,z_3)$ are
antisymmetric under the interchange of gluons with the same and
the opposite helicity, respectively:
\ba{sym-h}
\widetilde O_w(z_1,z_2,z_3) &=& - \widetilde O_w(z_3,z_2,z_1)
=  -P_{31} \widetilde O_w(z_1,z_2,z_3)\,,
\nonumber\\
T_w(z_1,z_2,z_3) &=& - T_w(z_1,z_3,z_2) =  -P_{23} T_w(z_1,z_2,z_3)
\ea
with $P_{ik}$ being the permutation operators.
Using this, we can invert \re{O-w} to get
\ba{T-O}
T_w(z_1,z_2,z_3)&=&\widetilde O_w(z_1,z_2,z_3)+\widetilde
O_w(z_2,z_3,z_1)-\widetilde O_w(z_3,z_1,z_2)
\nonumber\\
&=&(1+P_{12} P_{31}-P_{23} P_{31})\,\widetilde O_w(z_1,z_2,z_3)\,.
\ea
The same relation holds between the operators $T_{\widebar w}$ and
$\widetilde O_{\widebar w}$.

The relations \re{O-w} and \re{T-O} allow one to rewrite the evolution
equation \re{EQ-nonlocal} for the three-gluon operator $\widetilde O(z_i)$
in terms of $T(z_i)$ with definite helicity.
 The difference  amounts to the following redefinition of
the evolution kernels \re{matrices}
$ \widehat {\cal H}\to \widehat {\cal H}^h$
\be{H-h}
{\cal \widehat H}^h=\left(\begin{array}{cc}
  \widehat H_{qq} & \widehat H_{qh} \\
  \widehat H_{hq} & \widehat H_{hh}
 \end{array}\right)
 =\left(\begin{array}{cc}
  1 & 0 \\
  0 & (1+ P_{12}-P_{23})
\end{array}\right)\left(\begin{array}{cc}
  \widehat H_{qq} & \widehat H_{qg} \\
  \widehat H_{gq} & \widehat H_{gg}
\end{array}\right)\left(\begin{array}{cc}
  1 & 0 \\
  0 & \frac12(1-P_{31})
\end{array}\right).
\ee
Note that the kernel $\widehat H_{qq}$ is not affected by this transformation.
Despite the fact that the two evolution equations are obviously equivalent,
the premium  in dealing with helicity operators $T_w(z_i)$
(or $T_{\widebar w}(z_i)$)
is that, as we will show below, the evolution kernel $\widehat {\cal H}^h$
becomes hermitian on the space of the coefficient functions.
The original kernel in \re{matrices} is not hermitian due to the presence
of additional permutation operators in \re{H-h}. The explicit  expressions for
the operator $\widehat {\cal H}^h$ will be given below.

According to \re{O-w}, going over from
$\widetilde{O}(z_i)$ to
the helicity states $T(z_i)$ is equivalent to the following ansatz for
the dual coefficient function of the three-gluon operator:
\be{Psi-hel}
\widehat\Psi^{g}_{N-1}(z_1,z_2,z_3) = \frac12
\left[\widehat\Psi^h_{N-1}(z_1,z_2,z_3)
- \widehat\Psi^h_{N-1}(z_3,z_2,z_1)\right]\,,
\ee
where  the coefficient functions $\widehat\Psi^h_{N-1}(z_i)$ are defined by
the short distance expansion of $T_w(z_i)$.
Applying the equivalence relation \re{T-O} one finds
\be{equiv}
\widehat\Psi^h_{N-1}(z_1,z_2,z_3)=\widehat\Psi^{g}_{N-1}(z_1,z_2,z_3)
+\widehat\Psi^{g}_{N-1}(z_2,z_3,z_1)
-\widehat\Psi^{g}_{N-1}(z_3,z_1,z_2)\,.
\ee

Substituting \re{O-w} into \re{direct}, one rewrites the multiplicatively
renormalizable operators as
\be{Addition}
{\cal O}_{N,\alpha} =\left.\left[
2N_c\,\Psi_{N,\alpha}^{q}(\partial_{z_i})S_\mu(z_i)
+n_f\,\Psi_{N-1,\alpha}^{h}(\partial_{z_i}) T_\mu(z_i)\right]
\right|_{z_i=0}\,,
\ee
and obtains similar relations between thus defined
coefficient functions $\Psi_{N-1,\alpha}^h(x_i)$
and $\Psi_{N-1,\alpha}^g(x_i)$.
We find that the local three-gluon operators are projected out of the
 $T(z_i)$ by the following coefficient function
\be{Psi-h}
\Psi_{N-1,\alpha}^h(x_1,x_2,x_3)=\frac12\left[\Psi_{N-1,\alpha}^g(x_1,x_2,x_3)
-\Psi_{N-1,\alpha}^g(x_1,x_3,x_2)\right]
\ee
that has the same symmetry \re{sym-h} as the helicity operator itself:
\be{Psi-h1}
\Psi^h_{N-1}(x_1,x_3,x_2)=P_{23}\, \Psi^h_{N-1}(x_1,x_2,x_3)=
- \Psi^h_{N-1}(x_1,x_2,x_3)\,.
\ee
The inverse relation looks as
\be{rel-Psi}
\Psi_{N-1,\alpha}^g(x_1,x_2,x_3) = \Psi_{N-1,\alpha}^h(x_1,x_2,x_3)+
\Psi_{N-1,\alpha}^h(x_3,x_1,x_2)-
\Psi_{N-1,\alpha}^h(x_2,x_3,x_1)\,.
\ee
Notice that the relations between the two sets of the coefficient functions
are different in the direct and dual representations. Nevertheless,
it follows from \re{Psi-hel} and \re{rel-Psi} that the coefficient functions
$\widehat\Psi^h_{N-1}(z_i)$ and $\Psi^h_{N-1}(x_i)$ satisfy the same conformal
constraints \re{constraints} and \re{constraints-dual} as
$\widehat\Psi^g_{N-1}(z_i)$ and $\Psi^g_{N-1}(x_i)$, respectively.

Last but not least, the original Schr\"o\-din\-ger equation
(\ref{Sch-dual}) can be transformed into a Schr\"o\-din\-ger equation
for the coefficient functions $\Psi^q_N(x_i)$ and $\Psi^h_{N-1}(x_i)$
\be{Sch-eq-h}
\left(
\begin{array}{cc}
H_{qq} & H_{qh} \\ H_{hq} & H_{hh}
\end{array}
\right) \left(\Psi_N^{q}(x_i)\atop \Psi_{N-1}^{h}(x_i)\right)
= {\cal E}_N \left( \Psi_N^{q}(x_i)\atop \Psi_{N-1}^{h}(x_i)\right),
\ee
where the evolution kernels ``without a hat'' in the helicity basis,
$H_{ab}^h$, are obtained from the corresponding kernels
$\widehat H_{ab}^h$ \re{H-h} through the $\mathbb{T}-$transformation
\re{T}
\be{H-T}
 H_{ab} \equiv \mathbb{T}_a^{-1}\widehat H_{ab}\mathbb{T}_b\,.
%= \widehat H_{ab}(L_k^{\pm,0})\,,
\ee
%i.e. by changing the representation of the $SL(2)$ generators.

\subsection{Conformal basis}

Solution of the evolution equations \re{Sch-eq-h} can be simplified
significantly by expanding the coefficient functions $\Psi_N^{q}(x_i)$ and
$\Psi_{N-1}^{h}(x_i)$ over the basis of ``spherical harmonics''
consistent with conformal symmetry constraints \re{constraints}.

\subsubsection{General construction}

For a generic three-particle operator, \re{cf1} and \re{def:Psi},
such  a `conformal basis' can be constructed as follows \cite{BDKM}.
Conformal symmetry allows to fix the total three-particle conformal
spin $J=j_1+j_2+j_3+N$ of the state that translates to the condition
that the corresponding coefficient function satisfies Eqs~\re{constraints}.
We define the set of functions ${Y}_{Jj}^{(31)2}$ by requiring
that they obey Eq.~\re{constraints} and, in addition,
have a definite value of the conformal spin
in one of the two-particle channels, $(31)$ for definiteness.
The latter condition leads to
\be{conf-bas}
{L}^2_{31} {Y}_{Jj}^{(31)2}(x_i) = j(j-1){Y}_{Jj}^{(31)2}(x_i)\,,
\ee
where $j=j_1+j_3+n$ and $n=0,...,N$.

Taken together, Eqs.~(\ref{conf-bas}), \re{constraints} determine the
functions ${Y}_{Jj}^{(31)2}$ uniquely and have the following
solution:
\ba{Y-q}
Y_{Jj}^{(31)2}(x_i) &=&r_{Jj}(x_1+x_2+x_3)^{J-j-j_2}
(x_1+x_3)^{j-j_1-j_3}\nonumber\\
&&\times\> P_{J-j-j_2}^{(2j_2-1,2j-1)}\lr{\frac{x_1-x_2+x_3}{x_1+x_2+x_3}}
P_{j-j_1-j_3}^{(2j_3-1,2j_1-1)}\lr{\frac{x_1-x_3}{x_1+x_3}}\,.
\ea
Here $P^{(\alpha,\beta)}_n(x)$ is the Jacobi polynomial and
$r_{Jj}$ an arbitrary normalization factor.
%Applying the ${\mathbb T}$-transformation (\ref{map-ex}) to the
%functions in (\ref{Y-q}) one finds the corresponding dual basis functions
%$\widehat Y_{Jj}^{(31)2}(z_i)$
%\be{Y-q-dual}
%{\widehat Y}^{(31)2}_{Jj}(z_i)=\widehat r_{Jj}
%\,(z_1-z_3)^{j-j_1-j_3}\,(z_1-z_2)^{J-j-j_2}
%\,{}_2 F_1
%\left(j-J+j_2,j+j_3-j_1,2j;\frac{z_1-z_3}{z_2-z_1}
%\right)
%\ee
%where the normalization factor $\widehat r_{Jj}$ is related to
%$r_{Jj}$ as
%\be{r-factor}
%\widehat r_{Jj}=r_{Jj}
%\frac{\Gamma(J-j+j_2)\,\Gamma(J+j-j_2)\,\Gamma(j+j_1-j_3)\,\Gamma(j-j_1+j_3)
%}{\Gamma(j-j_1-j_3+1)\,\Gamma(J-j-j_2+1)\,\Gamma(2j)}
%\prod_{i=1}^3\frac{1}{\Gamma(2j_i)}\,.
%\ee
The basis functions $Y_{Jj}^{(31)2}(x_i)$ are orthogonal with respect to the
scalar product (\ref{int-rep}). Requiring that
${\|{Y}_{Jj}^{(31)2}\|^2}=1$ we fix the normalization to be
\ba{norm}
%{\|{Y}_{Jj}^{(31)2}\|^2}
r_{Jj}^{-2}
&=&
%\Gamma(2J)\prod_{i=1}^3\frac{1}{\Gamma(2j_i)}\,
%\frac{\Gamma(2j_1+2j_2+2j_3)}
\frac{\Gamma(2J)}
{\Gamma(2j_1)\Gamma(2j_2)\Gamma(2j_3)}
\frac{\Gamma(j+j_{1}-j_{3})\Gamma(j-j_{1}+j_{3})}
{\Gamma(j-j_1-j_3+1)\Gamma(j+j_1+j_3-1) (2j-1)}\,\times
\nonumber\\
&&\times\frac{\Gamma(J-j+j_2)\Gamma(J+j-j_2)}
{\Gamma(J-j-j_2+1)\Gamma(J+j+j_2-1)(2J-1)}\,.
\ea

The above construction of the conformal basis involves an obvious
ambiguity in which order the spins of partons are coupled to the total
spin $J$. Choosing in \re{conf-bas} a different two-particle channel
one obtains a  different  conformal basis related to the original one
through the  matrix $\Omega$ of  Racah 6j-symbols
of the $SL(2,\mathbb{R})$ group
\be{tr-matr}
{Y}_{Jj}^{(31)2}(x_i)=\sum_{j_1+j_2\le j'\le J-j_3}\Omega_{jj'}(J)\,
{Y}_{Jj'}^{(12)3}(x_i)\,.
\ee
Properties of the Racah 6j-symbols as well as their
explicit expressions  in terms of the generalized hypergeometric series
${}_4F_3(1)$ are summarized  in Appendix~A.

\subsubsection{Quark-gluon conformal basis}

We now specify the above general construction to the particular cases of
quark-antiquark-gluon and three-gluon operators and define
\ba{basis}
&&Y^q_{N,k}(x_1,x_2,x_3)=
\ Y^{(31)2}_{N+7/2,k+2}(x_1,x_2,x_3)
\bigg|_{j_1=j_3=1\,,j_2=3/2}
\nonumber\\
&&Y^h_{N-1,k-1}(x_1,x_2,x_3)=
Y^{(31)2}_{N+7/2,n+2}(x_1,x_2,x_3)
\bigg|_{j_1=j_2=j_3=3/2}\,,
\ea
where the prefactors have been chosen for later convenience. The
three-particle conformal spin is given by $J=N+7/2=j_{\bar q}
+j_{\bar g}+j_q+N = j_g+j_{\tilde g}+j_g+N-1$ in both cases,
and the conformal spin in the $(31)-$subchannel is equal to
$j=k+2=j_{\bar q}+j_q+k=j_g+j_g+k-1$. Notice
that $j\ge 3$  and, therefore, the basis functions
$Y^h_{N,k-1}(x_i)$ in the gluon-gluon subchannel
are well-defined for $k-1\ge 0$. In what follows we
assume that $Y^h_{N-1,-1}(x_i)=0$.

Each of the two sets of
functions in \re{basis} forms an orthonormal basis with respect to
the conformal scalar product \re{int-rep} whose explicit form depends
on the conformal spin of the particles and, therefore, is different for
the quark-antiquark-gluon and three-gluon systems.
This suggests to define the scalar product on the space
of two-dimensional vectors of coefficient functions
in the following form
\be{sc-prod}
\vev{\Psi_1|\Psi_2}=2N_c\vev{\Psi_1^q|\Psi_2^q}+n_f\vev{\Psi_1^h|\Psi_2^h}
\,,\qquad
\Psi_a=\left({\Psi_a^q\atop\Psi_a^h}\right)\,,
\ee
where the scalar product in each sector, $\vev{\Psi_1^q|\Psi_2^q}$ and
$\vev{\Psi_1^h|\Psi_2^h}$, is obtained from \re{int-rep} by
substituting $j_1=j_3=1$, $j_2=3/2$ and $j_1=j_2=j_3=3/2$, respectively.
As we show below such choice of the scalar product ensures the hermiticity
of the evolution kernels.

We shall look for the solutions to the evolution equation \re{Sch-eq-h}
in the following form
\be{ex-exp}
\bin{\Psi^q_N(x_i)}{\Psi^h_{N-1}(x_i)}=
\sum_{k=0}^{N}\lr{
\begin{array}{l}
{\frac{u^q_{N,k}}{\sqrt{2N_c}}\ Y^q_{N,k}(x_i)} \\[3mm]
{\frac{u^h_{N,k}}{\sqrt{n_f}}\ Y^h_{N-1,k-1}(x_i)}
\end{array}}
\ee
with $u^q_{N,k}$ and $u^h_{N,k}$ being the expansion coefficients.
In this way, the evolution kernels finally become symmetric and real matrices
acting on the vector of the expansion coefficients. We fix the normalization
of the coefficients by requiring the eigenstates \re{ex-exp} to have the
unit norm
\be{norms}
\|\Psi\|^2=2N_c \|\Psi^q_{N,\alpha}\|^2+ n_f\|\Psi^h_{N-1,\alpha}\|^2
=\sum_{k=0}^N |u^q_{N,k}|^2 + |u^h_{N,k}|^2=1
\ee
where $u^h_{N,0}=0$.

\subsection{QCD evolution kernels}

To one-loop accuracy, the evolution kernels are given~\cite{BFLK} by
the sum of two-particle Hamiltonians describing the pair-wise interaction
between quarks and gluons on the light-cone:
$\widehat H = \widehat H_{12}+ \widehat H_{23}+ \widehat H_{31}$, or,
equivalently $H= H_{12}+ H_{23}+ H_{31}$.
Conformal invariance \re{conf-sym} then implies that each pair-wise
Hamiltonian $H_{ik}$ only depends on the sum of conformal spins of
the interacting particles, e.g. in the ``direct'' representation
 $H_{ik}=H(J_{ik})$, where
\be{L-ik}
L_{ik}^2 = (L_i + L_k)^2 = J_{ik}(J_{ik}-1)
\ee
and the $SL(2)$ generators $L_i^\alpha$ $(\alpha=\pm,0)$ are given  in
Eq.~\re{L's}.
The explicit form of this dependence can most easily be obtained by
comparing the eigenvalues. To this end it is sufficient to calculate the
one-gluon exchange diagrams  in a simplified situation when momenta of the
contributing partons sum up to zero. Alternatively, one can
start with  the known integral representation for the evolution
Hamiltonian~\cite{BB89} and project it onto the conformal basis \re{basis}.

\subsubsection{Diagonal evolution kernels}

The diagonal quark evolution kernel is given by
\be{on-q}
H_{qq}\ket{Y^{q}_{N,n}}=\left[H_{S^+} + \frac{2n_f}3\delta_{J_{13},2}\right]
\ket{Y^{q}_{N,n}}\,,
\ee
where $b=11/3 N_c +2/3n_f $ is the lowest-order coefficient of the
QCD $\beta-$function. The first term in brackets stands for the
flavour-nonsinglet evolution kernel (see below) and the
second term proportional to $\delta_{J_{13},2}$ comes
from the additional Feynman diagram in which the quark and the antiquark
annihilate to produce a gluon that splits again into the $q\bar q$-pair.
Since the quark and the antiquark are produced in this way in one spatial
point, contribution of this diagram is different from zero only for
$J_{13}=j_q+j_{\bar q}=2$. The flavour-nonsinglet
Hamiltonian $H_{S^+}$ can be represented as~\cite{BFLK,BDM98,Belitsky99,DKM99}:
\be{HSp}
 H_{S^+}=N_c\,H^{(0)}-\frac{2}{N_c}\, H^{(1)},
\ee
where
\ba{H-NS}
 { H}^{(0)} &=& V_{qg}^{(0)}(J_{12}) +
 U_{qg}^{(0)}(J_{23})
 \\
 { H}^{(1)} &=& V_{qg}^{(1)}(J_{12}) + U_{qg}^{(1)}(J_{23})
 + U_{qq}^{(1)}(J_{13})\,.
\ea
Here, the notation was introduced for the two-particle quark-quark and
quark-gluon kernels
\ba{nonplanar}
V_{qg}^{(0)}(J)&=&\psi(J+3/2)+\psi(J-3/2)-2\psi(1)-3/4\,,
\\
U_{qg}^{(0)}(J)&=&\psi(J+1/2)+\psi(J-1/2)-2\psi(1)-3/4\,,
\nonumber\\
V_{qg}^{(1)}(J)&=&\frac{(-1)^{J-5/2}}{(J-3/2)(J-1/2)(J+1/2)}\,,
\quad
U_{qg}^{(1)}(J)=-\frac{(-1)^{J-5/2}}{2(J-1/2)}\,,
\nonumber\\
%V_{qq}^{(1)}(J)&=&\psi(J)-\psi(1)-3/4\,,
%\quad
U_{qq}^{(1)}(J)&=&\frac12\left[\psi(J-1)+\psi(J+1)\right]-\psi(1)-3/4\,,
\nonumber
\ea
where $\psi(x)=d\ln\Gamma(x)/dx$.

The diagonal gluon kernel is defined as
\be{on-g}
H_{hh}\ket{Y^{h}_{N-1,k-1}} =
N_c\left[H_{3/2}-V_{3/2}\right]\ket{Y^{h}_{N-1,k-1}}
\ee
where
\ba{h32}
H_{3/2}&=& 2 \left[ U_{gg}(J_{12}) + U_{gg}(J_{23}) + U_{gg}(J_{31})\right] -  b/N_c\,,\\
V_{3/2}&=&V_{gg}(J_{12})+V_{gg}(J_{31})
%\frac{1-P_{23}}2
\ea
and
\ba{U,V}
U_{gg}(j) &=& \psi(j)-\psi(1)\,,
\\
V_{gg}(j) &=& \frac2{j(j-1)}+\frac{3(1+(-1)^j)}{(j-2)(j-1)j(j+1)}\,.
\ea
We notice that the kernel $H_{hh}$ is invariant
under permutations of the two gluons of the same helicity
\be{sym-diag}
 [H_{hh},P_{23}]=0\,.
\ee
Note that the  pair-wise
diagonal Hamiltonians $\widehat H_{ik}$ and $H_{ik}$
have the same functional dependence on the  Casimir operators,
i.e.  $\widehat H_{ik}=h(\widehat J_{ik})$ and
$ H_{ik}=h(J_{ik})$ with the same function $h$.

\subsubsection{Off-diagonal evolution kernels}

The off-diagonal kernels describe the mixing between quark-antiquark-gluon
and three-gluon states. The conformal symmetry implies that the conformal
spin of both states should be the same, and this appplies {\em both}
to the total conformal spin of the three-parton system and the
conformal spin of the parton pair involved in the mixing.
It follows that, therefore, acting on the basis
function in the quark sector, $\ket{Y^{q}_{N,k}}$, the evolution kernel
$ H_{hq}$ transforms it into the basis function in the gluon sector,
$\ket{Y^{h}_{N-1,k-1}}$, with the {\em same} $N,k$.
Similarly, the evolution kernel $ H_{qh}$ maps gluon
conformal basis into the quark one. As a consequence, the off-diagonal kernels
can be written down in the following form
\be{W-qh}
H_{qh}=W_{qh}(J_{31})\,  \frac{1\!-\!P_{23}}{2}\,,\qquad
H_{hq}=\frac{1\!-\!P_{23}}{2}\,W_{hq}(J_{31}) \,,
\ee
where the operators $W_{qh}$ and $W_{hq}$ is defined as
\ba{off-diag-ex}
&&\hspace*{-1cm}
% H_{qh}
W_{qh}
 \ket{ Y^{h}_{N-1,k-1}}= -n_f
\left[\frac{J_{13}^2-J_{13}+2(-1)^{J_{13}}}{J_{13}(J_{13}-1)
\sqrt{(J_{13}+1)(J_{13}-2)}}
\right]
%\left(\frac{1\!-\!P_{23}}{2}\right)
\ket{ Y^{q}_{N,k}},
\nonumber\\
&&\hspace*{-1cm}
% H_{hq}
W_{hq}
 \ket{ Y^{q}_{N,k}}= -2N_c
%\left(\frac{1\!-\!P_{23}}{2}\right)
\left[\frac{J_{13}^2-J_{13}+2(-1)^{J_{13}}}
{J_{13}(J_{13}-1)
\sqrt{(J_{13}+1)(J_{13}-2)}}
\right]
\ket{ Y^{h}_{N-1,k-1}}.
\ea
The following comments are in order.
The off-diagonal kernels $H_{qh}$ and $H_{hq}$
originate from the
Feynman diagrams in which quark and antiquark annihilate into two gluons
of opposite helicity. As a consequence, these kernels depend on the
conformal spin in the quark-antiquark subchannel, $J_{13}$, and the
additional projector $(1-P_{23})/2$ takes into account antisymmetry
of the three-gluon state under the interchange of gluons with the same
helicity, Eq.~\re{sym-h}.

It follows from \re{off-diag-ex} that the off-diagonal kernels are adjoint to
each other, $ H_{qg}= H_{gq}^\dagger$, with respect to the scalar product
\re{sc-prod}. Indeed, calculating the matrix elements of the
off-diagonal kernels between the states $\Psi^q_N$ and $\Psi^g_{N-1}$ defined
 in \re{ex-exp} and
taking into account their symmetry properties \re{Psi-h1} one finds
\ba{herm-off}
\vev{\Psi^q_{N}| H_{qh}\Psi^h_{N-1}}&=&
     \vev{ H_{hq}\Psi^q_N|\Psi^h_{N-1}}
\nonumber\\&=&
 -\sqrt{2 N_c n_f}
\sum_{k=1}^N
\frac{(k+1)(k+2)+2(-1)^k}{(k+1)(k+2)\sqrt{k(k+3)}}
\lr{u^q_{N,k}}^* u^h_{N-1,k-1}\,,
\ea
where we have substituted $J_{31}=k+2$. Repeating the similar calculation for
the matrix elements of the diagonal kernels,
$\vev{\Psi^q_{N}| H_{qq}\Psi^q_{N}}$ and
$\vev{\Psi^h_{N-1}| H_{hh}\Psi^h_{N-1}}$, one makes sure that
$ H_{qq}^\dagger= H_{qq}$ and $ H_{hh}^\dagger=
 H_{hh}$.
We conclude that the matrix of the evolution kernels entering the
Schr\"odinger equation \re{Sch-eq-h} is a hermitian operator
on the space of the coefficient functions endowed with
the scalar product \re{sc-prod}. As a consequence,
its eigenvalues ${\cal E}_N$ are real and the corresponding
eigenfunctions are orthogonal to each other.

\subsection{Expansion in conformal operators}
Once the Schr\"odinger equation in the helicity basis \re{Sch-eq-h} is solved,
we can easily restore the gluon coefficient function $\Psi^g_{N,\alpha}$
using the symmetry relation \re{rel-Psi}
and reconstruct the multiplicatively
renormalizable operators \re{direct}.
Their reduced forward matrix elements can be expressed in terms of the
multiparton distributions as follows:
\be{Red-O}
\langle\!\langle \,{\cal O}_{N\!,\,\alpha}(Q^2)\, \rangle\!\rangle =
\frac{1}{||\Psi_{N\!,\,\alpha}||^2}
\int_{-1}^1 {\cal D}\xi \left[4N_c\,\Psi_{N\!,\,\alpha}^q(\xi_i) D_q(\xi_i,Q^2)
+ n_f\,\Psi_{N-1,\,\alpha}^g(\xi_i) D_g (\xi_i,Q^2)\right]\,.
\ee
Notice that in contrast with \re{int-rep} the integration goes here over the
region $\xi_1+\xi_2+\xi_3=0$. Comparing this expression with the expansion is
the standard operator basis \re{conf-ops}
and assuming that the eigenvectors $\Psi_{N\!,\,\alpha}$
is normalized to unity
one gets the equivalence relations
\ba{C's}
4N_c\, \Psi_{N\!,\,\alpha}^q(\xi_i)\Big|_{\sum \xi_i=0} \
&=& \sum_{0\le k\le N}
C_{\alpha k}^q (N)\ \xi_1^n \xi_3^{N-k}\,,
\nonumber\\
n_f\, \Psi_{N-1,\,\alpha}^g(\xi_i)\Big|_{\sum \xi_i=0} &=&
\sum_{0\le m\le [N/2]-1} C_{\alpha m}^g (N)\
\left(\xi_1^m \xi_3^{N-1-m} - \xi_1^{N-1-m} \xi_3^m\right)\,.
\ea

Next, expanding the OPE coefficient function $\Phi_n^q(\xi_1,\xi_3)$
\re{otvet-LL} over the eigenstates of the evolution kernels
\be{Phi-eig}
\bin{\Phi_{N+3}^q(\xi_1,-\xi_1-\xi_3,\xi_3)}0 = \sum_\alpha w_{N,\alpha}\left.
\bin{2 N_c\,\Psi_{N\!,\,\alpha}^q(\xi_i)}{n_f\,\Psi_{N-1,\,\alpha}^g(\xi_i)}
\right|_{\xi_1+\xi_2+\xi_3=0}
\ee
we can decompose  moments of the structure function $g_2(x,Q^2)$
\re{otvet-LL-1} in multiplicatively renormalizable contributions \re{Red-O} as
\be{otvet-LL-reex}
g_2^{\rm LL}(N+3,Q^2) =\frac12 \langle e^2_q\rangle \,\frac{4}{N+3}
\sum_{\alpha=0}^{[3N/2]} w_{N,\alpha}
\langle\!\langle \,{\cal O}_{N\!,\,\alpha}(Q^2)\, \rangle\!\rangle\,.
\ee
The expansion coefficients $w_{N,\alpha}$ can be expressed in terms
of  $C_{\alpha k}^q (N)$ and $C_{\alpha m}^g (N)$ by comparing the
coefficients in front of different powers of $\xi_1$ and $\xi_3$ in
the both sides of \re{Phi-eig}.

\subsubsection{Quark coefficient function}
There exists, however, a more efficient way of finding the same coefficients.
To this end, observe  that the OPE coefficient function
$\Phi_{N+3}^q(\xi_1,-\xi_1-\xi_3,\xi_3)$ can uniquely be continued from
the hyperplane $\xi_1+\xi_2+\xi_3=0$ to arbitrary values of $\xi_i$
by requiring that the thus defined function
$\Phi_{N+3}^q(\xi_1,\xi_2,\xi_3)$ satisfies the
conformal constraints \re{constraints-dual} (and coincides with
$\Phi_{N+3}^q(\xi_1,-\xi_1-\xi_3,\xi_3)$ at $\xi_2=-\xi_1-\xi_3$).
To find an explicit expression, let us expand
$\Phi_{N+3}^q(\xi_1,\xi_2,\xi_3)$ over the conformal basis
\re{basis}
\be{Phi-exp}
\Phi_{N+3}^q(x_1,x_2,x_3)=\sum_{k=0}^N \phi^q_{N,k}\,Y^q_{N,k}(x_i)
\ee
with $\phi^q_{N,k}=\vev{\Phi_{N+3}^q|Y^q_{N,k}}$. Projecting out the
hyperplane $\sum_i x_i=0$, we find using
\re{Y-q} that the basis functions are reduced to
\be{basis-red}
Y^q_{N,k}(x_i)\bigg|_{\sum_i x_i=0}\sim (x_1+x_3)^N P^{(1,1)}_{k}
\lr{\frac{x_1-x_3}{x_1+x_3}}
\ee
and form an orthogonal basis on the subspace $x_1+x_3=1$ and
$0\le x_1,x_3\le 1$. This property allows us to calculate the expansion
coefficients $\phi_{N,k}$ as
\be{Geg}
\phi^q_{N,k}\sim \int_{0}^1 dx_1dx_3\,\delta(1-x_1-x_3)\,x_1x_3 P^{(1,1)}_{k}
\lr{\frac{x_1-x_3}{x_1+x_3}}
\Phi_{N+3}^q(x_1,-x_1-x_3,x_3)\,.
\ee
Substituting the actual expression for $\Phi_{N+3}(\xi_i)$ \re{psi-n}
and performing the integration one arrives at
\ba{Phi-coef}
\phi^q_{N,k}&=&
\lr{ \left[(-1)^{N-k}+1\right]\frac{N+k+5}{N-k+1}
+\left[(-1)^{N-k}-1\right]\frac{N+k+4}{N-k+2}}
\nonumber\\
&\times&
\lr{\frac{(k+1)(k+2)(2k+3)(N-k+1)(N-k+2)}{8(N+3)(N+k+4)(N+k+5)}
\frac{\Gamma^4(N+3)}{{\Gamma(2N+6)}}}^{1/2}\,.
\ea
Using these coefficients one can easily calculate the norm
of the quark coefficient function
\be{norm-q-cf}
\| \Phi^q_{N+3}\|^2 = %\frac1{N_c}
\sum_{k=0}^{N}(\phi^q_{N,k})^2
=\frac{\Gamma^4(n)}{\Gamma(2n)}\frac{n^4}{4}\left[1+\frac1{n^2}
\lr{1-4\psi(n)-4\gamma_E} -\frac2{n^3}
\right],
\ee
%\be{norm-Phi}
%\|\Phi^q_{N+3}\|=\left[\sum_{k=0}^N \lr{\phi_{N,k}^q}^2\right]^{1/2}
%= \frac{\Gamma^2(n)}{\Gamma(2n)}
%\frac{n^{3/2}}2\left[1+\frac1{2n^2}\lr{1-4\ln n-4\gamma_E}+{\cal O}(1/n^3)
%\right]\,,
%\ee
where in the r.h.s. $n=N+3$.

Finally, using \re{Phi-exp} and \re{Phi-coef} one finds the
expansion coefficients for moments of the structure function
\be{w-sc}
\omega_{N,\alpha}={\vev{\Phi_{N+3}^q|\Psi^q_{N,\alpha}}}
=\frac{1}{\sqrt{2N_c}}{\sum_{k=0}^N \phi^q_{N,k} u^q_{N,k}}\,.
\ee
The coefficients $u^q_{N,k}$ stand for the
expansion of a multiplicatively renormalizable operator in the
conformal basis, cf. Eqs.~\re{ex-exp} and \re{norms}.
Their dependence on the particular operator (index $\alpha$)
is tacitly assumed.

\subsubsection{Gluon coefficient functions}
The similar procedure can be used to calculate the expansion coefficients
of the gluon coefficient functions $\Phi_k^g$ and $\Omega_k^{qg}$ defined in
\re{otvet} and \re{CD}.
The corresponding conformal harmonics in the helicity basis are given
in the forward direction by
\be{h-forward}
Y^h_{N-1,k}(x_i)\bigg|_{\sum_i x_i=0} \sim
%\frac1{\sqrt{2n_f}}
%\frac{\Gamma(2N+6)}{\Gamma(N+k+7)\Gamma(N-k)}
(x_1+x_3)^{N-1} P^{(2,2)}_k\lr{\frac{x_1-x_3}{x_1+x_3}}.
\ee
In order to decompose the gluon coefficient function \re{CD}
over this basis we first construct the same coefficient function in the
helicity representation
\be{g-cf-h}
\Phi^h_{N+3}(x_i) = \frac12\left[\Phi^g_{N+3}(x_1,x_2,x_3)-
\Phi^g_{N+3}(x_1,x_3,x_2) \right]
\ee
so that $P_{23}\Phi^h_{N+3}(x_i)=-\Phi^h_{N+3}(x_i)$ and then
define the expansion coefficients $\phi^h_{N-1,k}$ as
\be{g-cf}
\Phi^h_{N+3}(x_i)=\sum_{k=0}^{N-1} \phi^h_{N-1,k}Y^h_{N-1,k}(x_i)\,.
\ee
Using orthogonality of the Jacobi polynomials we can calculate these
coefficients as
\be{cf-J}
\phi^h_{N-1,k} \sim \int_0^1 dx_1dx_3\delta(1-x_1-x_3)\,
 (x_1x_3)^2 P_k^{(2,2)}\lr{\frac{x_1-x_3}{x_1+x_3}}
\Phi^h_{N+3}(x_1,-x_1-x_3,x_3)\,,
\ee
Substituting the gluon coefficient function
\re{CD} into \re{g-cf-h} and \re{cf-J} and going through
the calculation one finds the following explicit expressions for the expansion
coefficients for {\it even} conformal spins $N$
\be{coeff-even}
\phi^h_{N-1,k}=%\sqrt{2n_f}\,
-{\frac {(N+k+5)(N+7+5/2\,k+1/2\,{k}^{2})}{(k+1 ) (N-k+1 )}}\, h_{N,k}
\ee
for even $k$ and
\be{coeff-odd}
\phi^h_{N-1,k}=%\sqrt{2n_f}\,
{\frac {(k+4) (N+k+6)}{2(N-k)}}\, h_{N,k}
\ee
for odd $n$. Here, the normalization constant $h_{N,k}$ is given by
\be{g-norm}
h_{N,k}^2={\frac { (k+1 ) (2\,k+5 ) (N-k )
 (N-k+1 ) (N+3 )^2}{8 (k+3 ) (k+4
 ) (k+2 ) (N+5+k ) (N+k+6 )}}\frac{\Gamma^4(N+3)}{\Gamma(2N+6)}\,.
\ee
Using \re{g-cf} we can calculate the norm of the gluon coefficient function
\be{norm-g-cf}
\| \Phi^h_{N+3}\|^2 = %\frac1{2n_f}
\sum_{k=0}^{N-1}(\phi^h_{N-1,k})^2
=\frac{\Gamma^4(n)}{\Gamma(2n)}\frac{n^3(n\!-\!1)}{16}
\left[1+\frac{2(n-4)(n+1)}{(n-1)^2n}(\psi(n)+\gamma_E)\right],
\ee
where $n=N+3$, as above.

The second gluon coefficient function, $\Omega_n^{qg}(\xi_i)$, appears in the
expression for the moments of the structure function \re{otvet} due to
mixing between quark-antiquark-gluon and three-gluon distribution amplitudes.
Similar to \re{g-cf-h} and \re{g-cf}, one can transform this function into the
helicity representation and decompose it over the conformal basis
\be{Omega-h}
\Omega_{N+3}^{qh}(x_i) = \frac12\left[\Omega^{qg}_{N+3}(x_1,x_2,x_3)-
\Omega^{qg}_{N+3}(x_1,x_3,x_2) \right]
= \sum_{k=0}^{N-1} \varpi^h_{N-1,k} Y^h_{N-1,k}(x_i)\,.
\ee
Substituting $\Omega_n^{qg}(\xi_i)$ by its explicit expression \re{CD} it
becomes straightforward to calculate the coefficients $\varpi^h_{N-1,k}$ using
\re{cf-J}. The resulting explicit expression is cumbersome
and will not be displayed here.
Comparing the expansions \re{g-cf} and \re{Omega-h} one
finds, however, \cite{BKM00}  that the two coefficient functions
$\Phi_n^g(\xi_i)$ and $\Omega_n^{qg}(\xi_i)$
are to a good accuracy proportional to each other
\be{close}
\Omega_n^{qg}(\xi_i)\approx c(n)\,\Phi_n^g(\xi_i)\,.
\ee
Applying the transformation \re{g-cf-h} and \re{Omega-h} to the both
sides of this relation and using the orthogonality of the conformal basis
we can calculate the coefficient $c(n)$ as
\be{c(n)}
c(n) = {\vev{\Omega_n^{qh}|\Phi_n^h}}/{\|\Phi_n^h\|^2}
= \sum_{k=0}^{N-1}\phi^h_{N-1,k} \varpi^h_{N-1,k}/{\|\Phi_n^h\|^2}
\ee
with the norm ${\|\Phi_n^h\|}$ given in \re{norm-g-cf}.
Going through the calculation one finds that $c(n)$ is given at large
$n$ by the following expansion:
\be{c-large}
c(n) = 1+ \frac1{n^2}\left[4\ln n +4\gamma_E -6 \right]+ {\cal O}(1/n^3)\,.
\ee

\setcounter{equation}{0}
\section{Results}

In this section we present detailed results on the solution of
the Schr\"odinger equation in Eq.~\re{Sch-eq-h}.
The main advantage of the Hamiltonian approach described
in the previous section is that it allows to understand qualitative
features of the solutions using the intuition
and a wealth of analytical tools well known from quantum mechanics.
For this analysis, we decompose the full Hamiltonian in \re{Sch-eq-h} in
two parts:
\be{split}
{\cal H}^h={\cal H}_0 + {\cal V}\,,\qquad
{\cal H}_0 =\lr{
\begin{array}{cc}
H_{qq} & 0 \\ 0 & H_{hh}
\end{array}}
\,,\qquad
{\cal V} =\lr{
\begin{array}{cc}
 0 & H_{qh} \\ H_{hq} & 0
\end{array}}
\,.
\ee
The Hamiltonian ${\cal H}_0$ governs the scale-dependence separately in the
`quark' and `gluon' sectors, whereas the off-diagonal kernel ${\cal V}$
describes the mixing between the two sectors. Our strategy throughout
this section will be first to solve the Schr\"odinger equation for
${\cal H}_0$ and then examine the deformation of the spectrum induced by
${\cal V}$ that we, somewhat imprecisely, refer to as the quark-gluon mixing.
The rationale for such a two-step procedure is that properties of the
individual Hamiltonians $H_{qq}$ and $H_{hh}$ have been studied in detail in
recent works \cite{BDM98,Belitsky99,DKM99} using their relation to completely
integrable models. We will, therefore, be able to use these results.

\subsection{Diagonal evolution kernels}
%
%%%%%%%%%%%%%%%%%%     FIGURE 1          %%%%%%%%%%%%%%%%%%%%%%%%%%%%
\begin{figure}[t]
\centerline{\epsfxsize10.0cm\epsffile{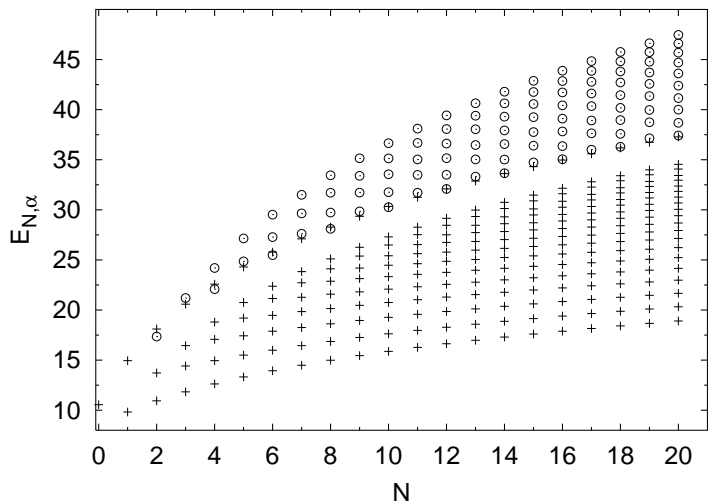}}
\caption[]{\small The spectrum of anomalous dimensions of flavor-singlet
twist-3 operators with the mixing between quark-antiquark-gluon (crosses)
and three-gluon operators (open circles) switched off, cf. \re{split}. }
\label{sp-diag}
\end{figure}
%%%%%%%%%%%%%%%%%%%%%%%%%%%%%%%%%%%%%%%%%%%%%%%%%%%%%%%%%%%%%%%%%%%%%%
%
The spectrum of eigenvalues of ${\cal H}_0$ is obviously given
by the superposition of the two independent spectra of quark-antiquark-gluon
and three-gluon operators
\be{LO}
H_{qq} \Psi^{q,(0)}_{N,\alpha} = {\cal E}^{q}_{N,\alpha} \Psi^{q,(0)}_{N,\alpha}\,,
\qquad
H_{hh} \Psi^{h,(0)}_{N-1,\beta} =
{\cal E}^{g}_{N,\beta} \Psi^{h,(0)}_{N-1,\beta}\,
\ee
and the corresponding eigenfunctions are given by
\be{LO-f}
\Psi_{N,\alpha}^{(0)}(x_i)=\bin{\Psi^{q,(0)}_{N,\alpha}(x_i)}{0}\,,\qquad
\Psi_{N,\beta+N+1}^{(0)}(x_i)=\bin{0}{\Psi^{h,(0)}_{N-1,\beta}(x_i)}\,.
\ee
The superscript $(0)$ stands to remind that the off-diagonal mixing terms are
omitted and the subscripts $\alpha$ and $\beta$ enumerate different `quark' and
`gluon' eigenstates corresponding to multiplicatively renormalizable operators
with $N$ (quark) or $N-1$ (gluon) covariant derivatives with the same canonical
dimension (and the same conformal spin $J=N+7/2$). In addition, we require that
the `gluon' eigenfunctions satisfy the symmetry property \re{Psi-h1}.
For a given $N$ the total number of the `quark' and `gluon' eigenstates is
equal to $\ell_q=N+1$ and $\ell_g=[N/2]$, respectively,
 so that $\alpha=0,\ldots,N$ and $\beta=0,\ldots,[N/2]-1$.

The results of the numerical calculation of the spectrum for
$N<20$ and $n_f=3$ are shown in Fig.~\ref{sp-diag}.
Note that the gluon eigenstates (open circles) and
the quark eigenstates (crosses) occupy two bands that lie on the
top of each other. For large $N$ the eigenvalues (anomalous dimensions)
rise logarithmically
\ba{band}
& 4C_F \ln N \le {\cal E}^{q}_{N,\alpha} \le 4 N_c \ln N\,, &
\nonumber
\\
&
4 N_c \ln N \le {\cal E}^{g}_{N,\beta} \le 6 N_c \ln N\,
&
\ea
and the coefficients in front of $\ln N$ are related to the color charges
of the corresponding (classical) parton configurations.
Since $\sim {\cal{O}}(N)$ levels have to fit within the band-width
$\sim {\cal{O}}(\ln N)$ \re{band}, the distance between the neighboring
levels in general goes to zero. The analysis of the `fine structure'
of the spectra \cite{Belitsky99,DKM99} reveals, however,  that in the limit
$N\to\infty$ three levels remain separated from the rest of the spectrum
by a finite gap.
These three special levels are: the lowest quark level, the highest quark
level and the lowest gluon level; they will play a decisive r\^ole in what
follows.

\subsubsection{The highest quark-antiquark-gluon state}
%
%%%%%%%%%%%%%%%%%%     FIGURE 2          %%%%%%%%%%%%%%%%%%%%%%%%%%%%
\begin{figure}[t]
\centerline{\epsfxsize10.0cm\epsffile{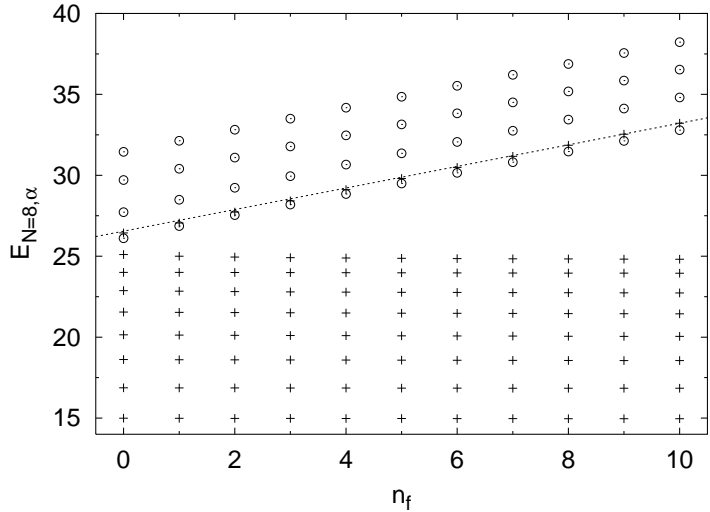}}
\caption[]{\small The dependence of the anomalous dimensions
of flavor-singlet  quark-antiquark-gluon (crosses)
and three-gluon operators (open circles) on the number of light quark
flavors $n_f$. The dotted line corresponds to the linear dependence
 $\sim (2/3) n_f$, see text. The quark-gluon mixing is switched off,
cf. \re{split}. }
\label{nfflow}
\end{figure}
%%%%%%%%%%%%%%%%%%%%%%%%%%%%%%%%%%%%%%%%%%%%%%%%%%%%%%%%%%%%%%%%%%%%%%
%
The eigenvalues ${\cal E}^{q}_{N,\alpha}$ and ${\cal E}^{g}_{N,\beta}$
depend on the number of light quark flavors $n_f$.
This dependence is shown in Fig.~\ref{nfflow}  for $N=8$ and
$0 \le n_f \le 10$ and reveals the following remarkable pattern:
In the gluon sector the $n_f-$dependence of all energy levels is linear,
 ${\cal E}^{g}_{N,\beta}\sim 2n_f/3$, and it can be traced to the additive
$b-$correction to \re{h32}. In contrast to this, in the quark sector all energy
levels except the highest one vary very slowly with $n_f$. At the same time, the
$n_f-$dependence  of the highest quark level is almost identical
 to that of the gluon levels, ${\cal E}^{q}_{N,N}\sim 2n_f/3$.
To understand this property, notice that the $n_f-$dependence of the
diagonal quark kernel in Eq.~\re{on-q} comes entirely from the annihilation
term $(2n_f/3)\delta_{J_{13},2}$. At large $n_f$ this term dominates
and the corresponding eigenstate is given by
\be{highest}
\Psi^{q,(0)}_{N,N}(x_i) = Y^q_{N,k=0}(x_i)+{\cal O}(1/n_f)\,.
\ee
%
%%%%%%%%%%%%%%%%%%     FIGURE 3          %%%%%%%%%%%%%%%%%%%%%%%%%%%%
\begin{figure}[t]
\centerline{\epsfxsize13.0cm\epsffile{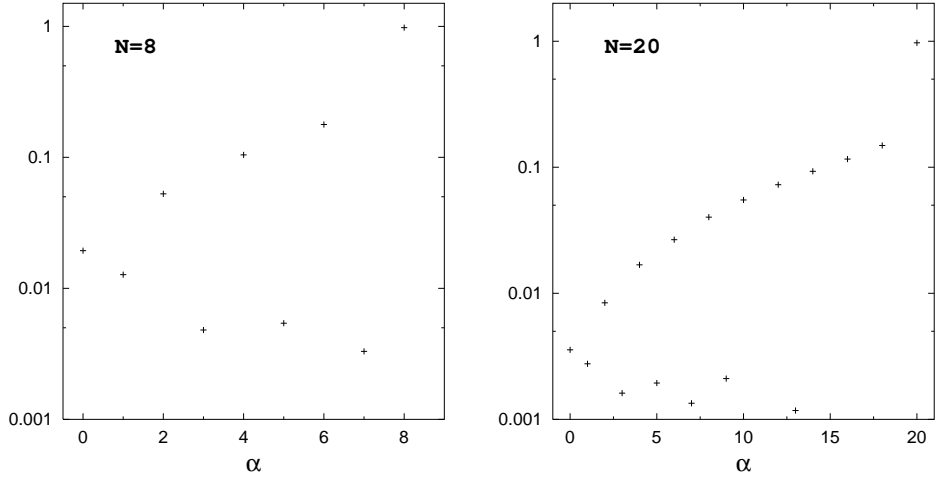}}
\caption[]{\small The coefficients of the expansion \re{exp1}
of $ Y^q_{N,k=0}(x_i)$ over the eigenstates
of the quark Hamiltonian $H_{qq}$.
$N=8$ and $N=20$ for the left and the right panel,
respectively.}
\label{Qhigh}
\end{figure}
%%%%%%%%%%%%%%%%%%%%%%%%%%%%%%%%%%%%%%%%%%%%%%%%%%%%%%%%%%%%%%%%%%%%%%
%
Remarkably enough,
this relation holds true with high accuracy for
small values of $n_f$ as well, including $n_f\to0$. To illustrate this,
we show in Fig.~\ref{Qhigh} the coefficients in the expansion of
$ Y^q_{N,k=0}(x_i)$ over the complete set of the quark eigenstates:
\be{exp1}
   Y^q_{N,k=0}(x_i) = \sum_{\alpha=0}^N
    \langle Y^q_{N,k=0}|\Psi^{q,(0)}_{N,\alpha}\rangle
   \,\Psi^{q,(0)}_{N,\alpha}(x_i)
\ee
for $n_f=3$ and two different values of $N$. Note that the normalization
is such that $\|\Psi_{N,\alpha}^{q,(0)}\|=1$ and, as a consequence,
the sum of the coefficients squared is equal to one. It is seen that
the overlap with the exact wave function of the highest `quark' state
$\alpha=N$ is very close to unity.

As a yet another test of the approximation for the wave function
\re{highest} one evaluates the corresponding energy (see Appendix A
for details)
\ba{quark-high}
{\cal E}^{q}_{N,N} \simeq \vev{Y^q_{N,k=0} |H_{qq}| Y^q_{N,k=0}}
 &=& N_c \left[2\Psi(N+3)+2\Psi(N+4)+4\gamma_E-\frac6{(N+3)^2}-\frac{19}6\right]
\nonumber\\&&{}
+ \frac{1}{N_c}\frac{6}{(N+3)^2(N+4)} + \frac23 n_f\,.
\ea
The few first terms of the expansion of the matrix element at large $N$ are
\be{estim}
{\cal E}^{q}_{N,N} \simeq N_c \left[
    4 \ln(N+3) +4 \gamma_E - \frac{19}{6} - \frac{19}3\frac1{(N+3)^2}\right]
+ \frac23 n_f + {\cal O}\lr{\frac1{(N+3)^3}}\,.
\ee
The approximation \re{quark-high} is compared with the exact numerical
calculations of the energy in Table~\ref{tab:energy} (two middle columns).
The accuracy turns out to be very good, better than 1\% for all $N$.

\begin{table}[t]
\begin{center}
\begin{tabular}{|c||c|c||c|c||c|c|}
\hline
  &  ${\cal E}^q_{N,0}$ &  $\vev{\Phi^q |H_{qq}| \Phi^q}$ &
     ${\cal E}^q_{N,N}$ &  $\vev{Y_{k=0} |H_{qq}| Y_{k=0}}$ &
     ${\cal E}^g_{N,0}$ &  $\vev{\Phi^h |H_{hh}| \Phi^h}$
\\
\hline
${N=2}$ & 10.933 & 10.987 & 18.107  & 17.993 &  17.350 & 17.350
\\
${N=4}$ & 12.629 & 12.644 & 22.587  & 22.395 &  22.087 & 22.160
\\
${N=6}$ & 13.938 & 13.946 & 25.785  & 25.561 &  25.474 & 25.541
\\
${N=8}$ & 14.994 & 14.999 & 28.285  & 28.046 &  28.107 & 28.162
\\
${N=10}$ & 15.876 & 15.881 & 30.343  & 30.094 & 30.260 & 30.304
\\
${N=30}$ & 20.824 & 20.828 & 41.634  & 41.367 & 41.847 & 41.878
\\
\hline
\end{tabular}
\end{center}
\caption{
Exact numerical results for the energy of the three `special' levels
(see text) compared with the calculation using the
approximations for the corresponding eigenfunctions.
The results for the lowest quark-antiquark-gluon,
the highest quark-antiquark-gluon
and the lowest three-gluon eigenstates are shown in the two left, two middle
and two right columns, respectively.
 }
\label{tab:energy}
\end{table}

Last but not least, we note that according to \re{basis-red} the wave function
entering \re{highest},
$Y^q_{N,0}(x_i) \sim (x_1+x_3)^N = (-x_2)^N$, depends on a single
(gluon) momentum fraction only and, therefore, it defines a local
operator \re{cf1} and \re{def:Psi} that contains derivatives acting
on the gluon but not on the quark fields.
All such operators can be obtained from a Tailor expansion
of the nonlocal operator \re{Spm} in which the quark and the antiquark
are located at the same space-time point and which can be rewritten as a
two-gluon operator using the equations of motion:
\be{EOM2}
 S_\beta(0,v,0) = i\bar q(0)g \widetilde G_{\beta x}(vx)\!\not\!xq(0)\,
  = -i \widetilde G^a_{\beta x}(vx) D^{ab}_\xi G^b_{\xi x}(0)\,.
\ee
Here $a,b$ are color indices and $D^{ab}$ is the covariant
derivative in the adjoint representation.
Thus, to a good accuracy, the quark-antiquark-gluon three-particle operator
with the highest anomalous dimension is in fact a two-particle
gluon operator!

\subsubsection{The lowest quark-antiquark-gluon state}

Since eigenfunctions of the hermitian Hamiltonian $H_{qq}$ are orthogonal
to each other and since the eigenfunction corresponding
to the highest anomalous dimension turns out to be very close to the
eigenfunction of the $n_f$-dependent contribution of the annihilation diagram
that gives rise to the second term in the Hamiltonian \re{on-q}, it
follows that
the wave functions of all other levels have a negligible overlap with this term
and, therefore, are to a good accuracy $n_f$-independent, in agreement with
Fig.~\ref{nfflow}.

This means that within the accuracy of \re{highest}, the
eigenfunctions and eigenvalues of all quark levels
other than the highest one coincide with those
of the flavor-nonsinglet quark Hamiltonian $H_{S^+}$ that was studied
in \cite{BDKM,Belitsky99,DKM99}.
In particular, it has been shown that
the wave function corresponding to the lowest anomalous dimension
coincides, in the large-$N_c$ limit, with the tree-level coefficient
function in the OPE \re{psi-n} continued from the hyperplane
$\xi_1+\xi_2+\xi_3=0$ to  arbitrary values of $\xi_i$ using the
conformal symmetry, cf. \re{Phi-exp}:
\be{iden}
\Psi^{q,(0)}_{N,\alpha=0} = \frac{\Phi^q_{N+3}(x_i)}{\|\Phi^q_{N+3}\|}+
{\cal O}(1/N_c^2)\,.
\ee
The additional factor in the r.h.s.\ takes into account the different
normalization of the states.
%
%%%%%%%%%%%%%%%%%%     FIGURE 4          %%%%%%%%%%%%%%%%%%%%%%%%%%%%
\begin{figure}[t]
\centerline{\epsfxsize13.0cm\epsffile{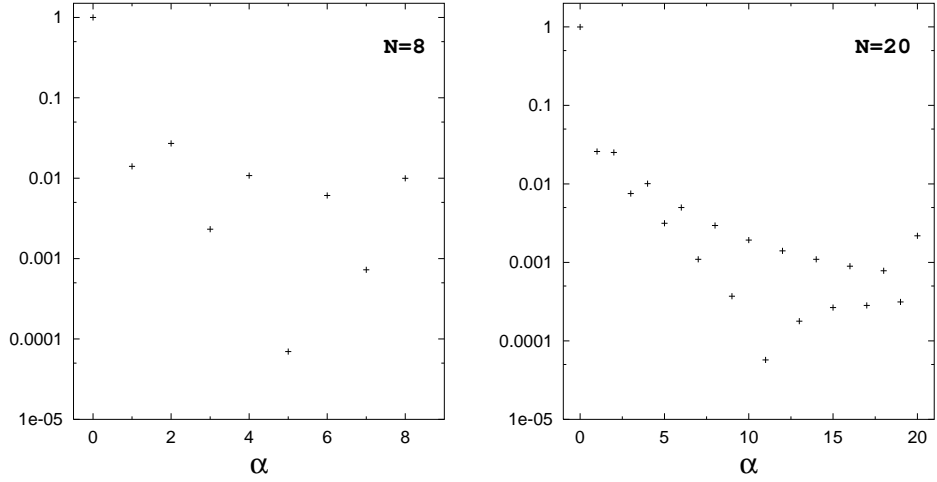}}
\caption[]{\small The coefficients of the  expansion of
$\Phi^q_{N+3}(x_i)/\|\Phi^q_{N+3}\|$
\re{iden} over the eigenstates $\Psi^{q,(0)}_{N,\alpha}$
of the quark Hamiltonian $H_{qq}$.}
\label{Qlow}
\end{figure}
%%%%%%%%%%%%%%%%%%%%%%%%%%%%%%%%%%%%%%%%%%%%%%%%%%%%%%%%%%%%%%%%%%%%%%
%
In Fig.~\ref{Qlow} we
show the coefficients of the expansion of
$\Phi^q_{N+3}(x_i)/\|\Phi^q_{N+3}\|$  over the eigenstates of the
quark Hamiltonian $H_{qq}$ for $N=8$ and $N=20$. It is seen that
the expansion is dominated by the lowest quark state $\alpha=0$.
The corresponding eigenvalue (anomalous dimension) can be
calculated as (see Appendix A for details)
\be{energy-0}
{\cal E}^q_{N,0} \simeq
 \vev{\Phi^q_{N+3} |H_{qq}|\Phi^q_{N+3}}/\|\Phi^q_{N+3}\|^2 =
 N_c E_N^{(0)} +\frac1{N_c} E_N^{(1)}
\ee
where \cite{ABH}
\be{ref-E}
E_N^{(0)} =  2\Psi(N+3) +\frac1{N+3} -\frac12 + 2\gamma_E\,
\ee
and \cite{BKM00}
\be{energy-1}
E_N^{(1)}=-{2}\left[
     \ln(N+3)+\gamma_{_{\rm E}}+\frac34 -
    \frac{\pi^2}{6} + \CO\left(\frac{\ln^2 (N+3)}{(N+3)^2}\right)\right]\,.
\ee
The quality of this approximation is illustrated in Table~\ref{tab:energy}
(compare the numbers in the first two columns).

\subsubsection{The lowest three-gluon state}

\begin{figure}[t]
\centerline{\epsfxsize13.0cm\epsffile{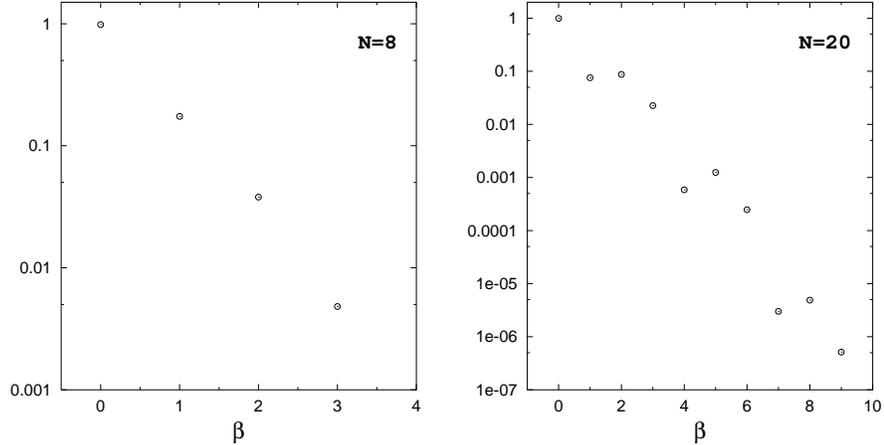}}
\caption[]{\small The coefficients of the expansion of
$\Phi^h_{N+3}(x_i)/\|\Phi^h_{N+3}\|$
\re{lowgluon1} over the eigenstates $\Psi^h_{N-1,\beta}$
of the gluon Hamiltonian $H_{hh}$.}
\label{Ghigh}
\end{figure}

Last but not least, we have to work out  an approximate description
of the lowest three-gluon state. As noticed in \cite{BKM00}, the corresponding
eigenfunction appears to be close to the one-loop gluon coefficient function
$\Phi^g(\xi_i)$ \re{CD} transformed to the helicity representation \re{g-cf-h}
(and continued to arbitrary values of $\xi_i$ using
the conformal symmetry):
\be{lowgluon1}
\Psi^{h,(0)}_{N-1,\beta=0} \simeq \frac{\Phi^h_{N+3}(x_i)}{\|\Phi^h_{N+3}\|}
=\frac1{2\|\Phi^h_{N+3}\|}\lr{\Phi^g_{N+3}(x_1,x_2,x_3)-\Phi^g_{N+3}(x_1,x_3,x_2)}
\,.
\ee
In difference to the above, this approximation cannot be justified in any
formal limit, but is no less accurate, as illustrated in Fig.~\ref{Ghigh}
where we show the corresponding (normalized) expansion coefficients in
the basis of exact three-gluon eigenstates.%
%\footnote{Since the eigenstates of the gluon Hamiltonian are orthogonal to each
%other only in the helicity basis, we apply the transformation \re{Psi-h} to the
%both sides of \re{lowgluon1}, expand the state $\Phi^h_{N+3}$ over the
%orthonormal basis of the eigenstates $\Psi^h_{N-1,\beta}$ and calculate the
%expansion coefficients shown in Fig.~\ref{Ghigh} as
%$\vev{\Psi^h_{N-1,\beta}|\Phi^h_{N+3}}/\|\Phi^h_{N+3}\|$.}

%%%%%%%%%%%%%%%%%%     FIGURE 5          %%%%%%%%%%%%%%%%%%%%%%%%%%%%
%%%%%%%%%%%%%%%%%%%%%%%%%%%%%%%%%%%%%%%%%%%%%%%%%%%%%%%%%%%%%%%%%%%%%%

The corresponding approximation for the energy can be calculated as
(see Appendix A for details)
\be{estim2}
{\cal E}^{g}_{N,0} \simeq \frac{\vev{ \Phi^h_{N+3} |H_{hh}|
\Phi^h_{N+3}}}{\|\Phi^h_{N+3}\|^2}
 = N_c E_N^g
% \left[
%    4 \ln(N\!+\!3) - 0.5843-\frac{3.4333}{(N\!+\!3)}
%+ {\cal O}\lr{\frac1{(N\!+\!3)^2}}\right]
+ \frac23 n_f
\ee
with
\ba{E-n-g}
E_N^g&=&4\ln(N+3)+4\gamma_E+\frac13-\frac{\pi^2}3
+\frac1{N+3}\left[(\ln(N+3)+\gamma_E)\lr{\frac{2\pi^2}3-6}
 - \frac{\pi^2}3\right]
\nonumber\\
&&{}
+ {\cal O}\lr{\frac{\ln^2(N+3)}{(N\!+\!3)^2}}\,.
\ea
It is compared with the exact result in Table~\ref{tab:energy} (the two last
columns).
Note that the energies of the lowest gluon and the
highest quark states are very close to each other, as seen also from
Fig.~\ref{sp-diag}:
\be{degeneracy}
 [{\cal E}^{q}_{N,N}-{\cal E}^{g}_{N,0}]/{\cal E}^{q}_{N,N} \sim
\frac{10^{-2}}{\ln(N+3)}.
 \ee
In fact, the difference is so small that one can treat these two
levels as degenerate for most purposes.
\subsection{The quark-gluon mixing}
The mixing of quark-antiquark gluon and three-gluon operators is governed
by the off-diagonal part ${\cal V}$ of the Hamiltonian \re{split}.
It turns out that the mixing is generally rather weak and for most
practical purposes can be taken into account  using the
standard perturbation theory. To the leading order, the mixing-induced
corrections to the `pure' quark or gluon eigenstates \re{LO}, \re{LO-f}
are equal to
\ba{PT-states}
\Psi^{(1)}_{N,\alpha}(x_i)&= &
\sum_{\beta'}\bin{0}{\Psi^{h,(0)}_{N-1,\beta'}(x_i)}
\frac{{\cal V}_{\beta'\alpha}}
{{\cal E}^{q}_{N,\alpha}-{\cal E}^{g}_{N,\beta'}}\,,
\\
\Psi^{(1)}_{N,\beta+N+1}(x_i)&=& -
\sum_{\alpha'}\bin{\Psi^{q,(0)}_{N,\alpha'}(x_i)}{0}
\frac{{\cal V}_{\beta\alpha'}}
{{\cal E}^{q}_{N,\alpha'}-{\cal E}^{g}_{N,\beta}} \,,
\ea
where the notation was introduced for the mixing matrix
\be{mix-mat}
{\cal V}_{\beta\alpha} =
\vev{\Psi^{h,(0)}_{N-1,\beta}|H_{hq}|\Psi^{q,(0)}_{N,\alpha}}\,.
\ee
It is easy to see that $\vev{\Psi^{(0)}_{N}|{\cal V}|\Psi^{(0)}_{N}}=0$
and, therefore, the energy eigenvalues do not receive any
corrections to the first order of the perturbative expansion.
This explains why the mixing-induced corrections to the anomalous
dimensions are very small (see Table \ref{tab:energy-ex}).
\begin{table}[t]
\begin{center}
\begin{tabular}{|c||c|c||c|c||c|c|}
\hline
  &  ${\cal E}_{N,0}$ &  ${\cal E}_{N,0}-{\cal E}^q_{N,0}$ &
     ${\cal E}_{N,N}$ &  ${\cal E}_{N,N}\!-{\cal E}^q_{N,N}$ &
     ${\cal E}_{N,N+1}$ &  ${\cal E}_{N,N+1}\!-{\cal E}^g_{N,0}$
\\
\hline ${N=2}$ & 10.739 & -0.193 & 17.679  & -0.428 &  18.171 & 0.821
\\
${N=4}$ & 12.571 & -0.058 & 22.324  & -0.262 &  22.648 & 0.561
\\
${N=6}$ & 13.912 & -0.027 & 25.637  & -0.148 &  25.872 & 0.397
\\
${N=8}$ & 14.980 & -0.015 & 28.195  & -0.090 &  28.420 & 0.312
\\
${N=10}$ & 15.867 & -0.009 & 30.278  & -0.065 & 30.530 & 0.270
\\
${N=30}$ & 20.823 & -0.001 & 41.600  & -0.034 & 42.056 & 0.209
\\
\hline
\end{tabular}
\end{center}
\caption{The exact energy of the three `special' levels (see
text) with the mixing-induced corrections taken into account compared with the
energy of the corresponding eigenstates with the mixing effects neglected. The
results for the lowest quark-antiquark-gluon, the highest quark-antiquark-gluon
and the lowest three-gluon eigenstates are shown in the two left, two middle and
two right columns, respectively.}
\label{tab:energy-ex}
\end{table}
The numerical results presented below are in all cases based on the
exact diagonalization of the mixing matrix, and the first-order expressions
in \re{PT-states} only serve to illustrate the picture. Note that
the perturbation theory cannot be used to describe the mixing betwen the
highest quark-antiquark gluon and the lowest three-gluon states because
of the vanishing energy denominators; in this case the explicit
diagonalization is mandatory.

%%%%%%%%%%%%%%%%%%     FIGURE 6          %%%%%%%%%%%%%%%%%%%%%%%%%%%%
\begin{figure}[t]
\centerline{\epsfxsize10.0cm\epsffile{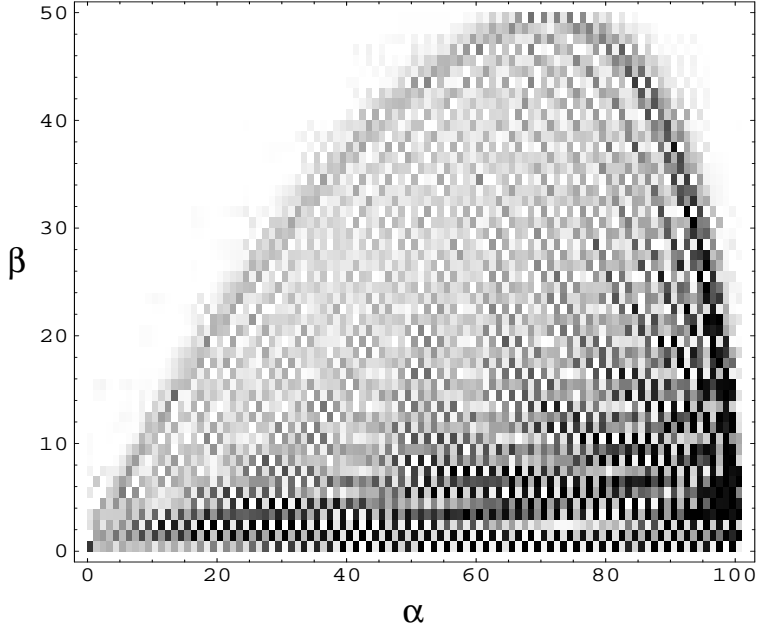}}
\caption[]{\small A density plot for the matrix $|{\cal
V}_{\beta\alpha}|$ for $N=100$.}
\label{Dens100}
\end{figure}
%%%%%%%%%%%%%%%%%%%%%%%%%%%%%%%%%%%%%%%%%%%%%%%%%%%%%%%%%%%%%%%%%%%%%%

The mixing matrix ${\cal V}_{\beta\alpha}$ has a rather peculiar wing-like
shape as illustrated in Fig.~\ref{Dens100}. We remind that
the index $\alpha$ along the horizontal axis
serves to numerate
different quark-antiquark-gluon states, starting with the one with the lowest
anomalous dimension at $\alpha=0$, and the index $\beta$ enumerates
the three-gluon eigenstates in the similar fashion. Dark regions in
Fig.~\ref{Dens100} correspond to large absolute values of
${\cal V}_{\beta\alpha}$ and light regions indicate small matrix elements.
Note the complicated `chess-board' pattern  with alternating large and
small entries.

\begin{figure}[t]
\centerline{\epsfxsize11.0cm\epsffile{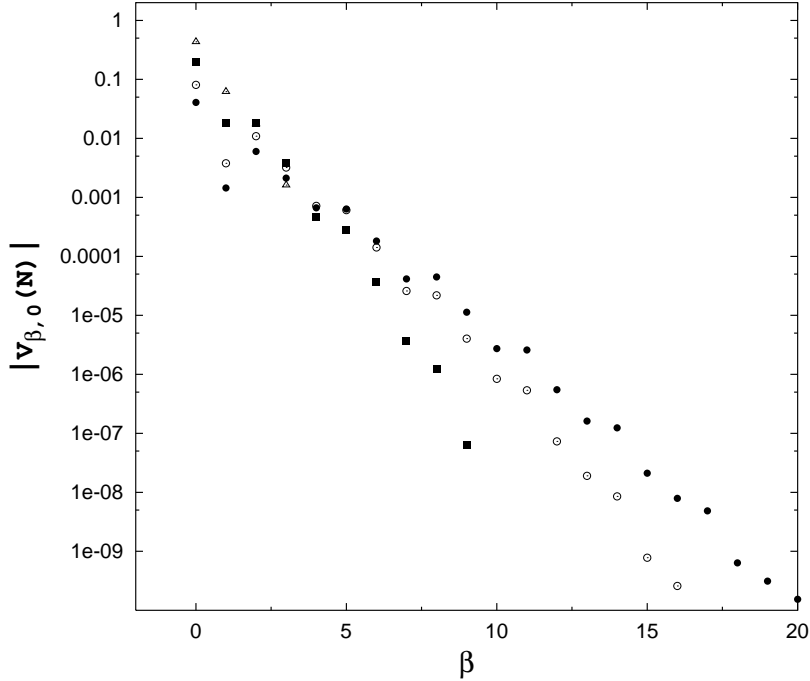}}
\caption[]{\small The mixing coefficients ${\cal V}_{\beta 0}$ of
the different three-gluon states $\beta =0,1,\ldots, [N/2]-1$
with the lowest quark-antiquark-gluon state, for four
different values of $N$: $N=8$ (triangles), $N=20$ (squares),
$N=50$ (open circles) and $N=100$ (full circles).}
\label{hierarchy}
\end{figure}

The most important feature that is seen in Fig.~\ref{Dens100} is that
the lowest quark eigenstate $\alpha=0$ mixes significantly only with
the lowest gluon eigenstate $\beta=0$. In fact, we find
a (roughly) exponential hierarchy of the matrix elements
\be{hierar}
|{\cal V}_{0,0}| \gg |{\cal V}_{1,0}| \gg ...\gg |{\cal V}_{[N/2]-1,0}|\,,
\ee
that is valid for all $N$, see Fig.~\ref{hierarchy}.
%
%%%%%%%%%%%%%%%%%%     FIGURE 7          %%%%%%%%%%%%%%%%%%%%%%%%%%%%
%%%%%%%%%%%%%%%%%%%%%%%%%%%%%%%%%%%%%%%%%%%%%%%%%%%%%%%%%%%%%%%%%%%%%%
%
Higher quark-antiquark-gluon states get mixed with gluons more heavily
and involve many three-gluon states in an essential way. For a few highest
quark states, $\alpha\sim N$ , the mixing is again simplified somewhat, but still involves
several (in general $\sim \ln N$) gluon states, see Fig.~\ref{Dens100}.
It can be shown that the mixing coefficient of the highest quark-antiquark
gluon state \re{highest} with the lowest three-gluon state \re{lowgluon1}
${\cal V}_{0,N}$ vanishes in the large-$N$ limit and this smallness
overcomes the enhancement due the small energy denominator in \re{PT-states}
in the same limit, cf. \re{degeneracy}.

Since the Hamiltonian \re{split} is hermitian, the same mixing matrix
describes the mixing of the gluon states with the quark states.
It is seen that the lowest gluon states $\beta=0,1,\ldots$ get mixed with
many quark states while the highest gluon states $\beta \sim N/2$ only
mix with quark states with $\alpha\sim 2/3 N$.

\begin{figure}[t]
%\centerline{\epsfxsize10.0cm\epsffile{admix.eps}}
\centerline{\epsfxsize16.0cm\epsffile{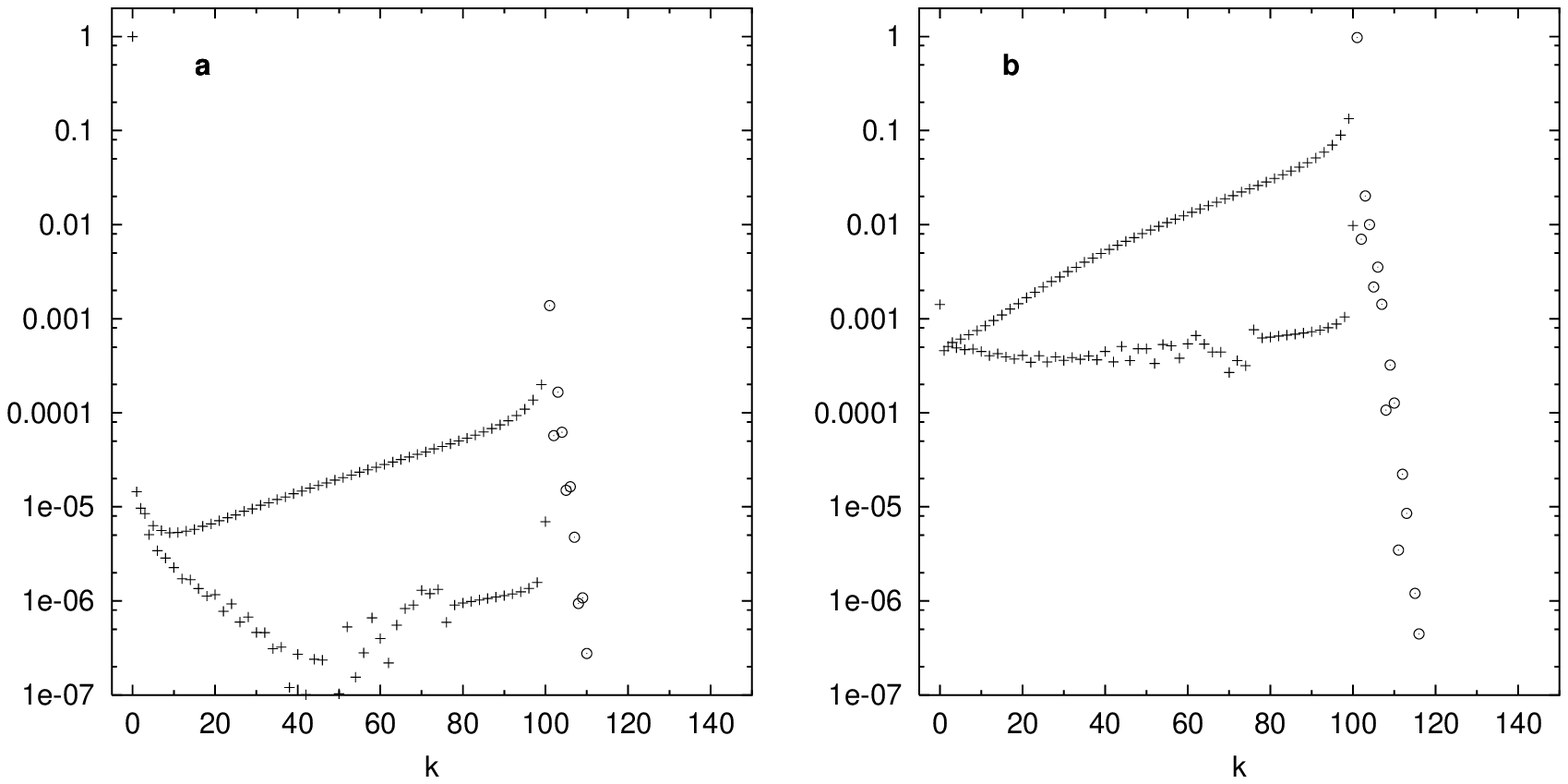}}
\caption[]{Exact numerical results for the expansion coefficients
of the exact eigenstates of the Hamiltonian ${\cal H}^h$ over the
eigenvectors of the diagonal blocks: {\rm \bf a)} - the exact lowest eigenstate,
$\vev{\Psi_{N,0}|\Psi^{(0)}_{N,k}}$;
{\rm \bf b)} - the exact lowest eigenstate in `gluon' sector,
$\vev{\Psi_{N,N+1}|\Psi^{(0)}_{N,k}}$.
The index $k$ enumerates the operators ordered according to their anomalous
dimension; $0\le k=\alpha\le N$ corresponds to quark operators,
$k=N+1+\beta \ge N+1$ to gluons, cf.~\re{LO-f}.
The `quark' eigenstates are shown by crosses, the `gluon' eigenstates by
open circles.
}
\label{fig:admix}
\end{figure}

As a yet another illustration of the mixing pattern, we show
in Fig.~\ref{fig:admix} the exact numerical results for the expansion
coefficients of two different exact eigenstates of the `full'
Hamiltonian \re{split}, $\Psi_{N,0}(x_i)$ and $\Psi_{N,N+1}(x_i)$,
over the `pure' quark and gluon
eigenstates. The chosen states are those that reduce to the lowest quark
and the lowest gluon states if the mixing is `switched off'.
%
%%%%%%%%%%%%%%%%%%     FIGURE 8         %%%%%%%%%%%%%%%%%%%%%%%%%%%%
%%%%%%%%%%%%%%%%%%%%%%%%%%%%%%%%%%%%%%%%%%%%%%%%%%%%%%%%%%%%%%%%%%%%%%
%
For the lowest quark state (which is the lowest eigenstate in the
whole spectrum) the single significant contribution comes from the
lowest gluon state ($\sim 10^{-3}$) while all other contributions are
suppressed by another order of magnitude. For the gluon state
the mixing is much larger ( $\leq 10^{-1}$) and involves many quark states.
This is the same structure that we observed earlier in Fig.~\ref{Dens100}.
Another conclusion is that since the mixing-induced corrections are at
most of the order of 10\% (for large $N\sim 100$), the perturbative
expansion in \re{PT-states} should  be rather accurate and this indeed
can be verified by the direct numerical calculation.
\subsection{Reduction of the mixing matrix: The two-channel DGLAP equation}
We have defined the transverse spin quark and gluon distributions
$\Delta q_T(x)$  and $\Delta g_T(x)$ as the
specific projections of the generic quark-antiquark-gluon and three-gluon
operators, Eqs.~\re{Tq} and \re{Tg}, respectively. They correspond to
the leading order quark and gluon contributions to the operator product
expansion of the T-product of the two electromagnetic currents \re{OPE-tw3}
and determine the odd moments of the structure function \re{otvet} at
a high scale $Q^2$
\be{g2-Delta}
[g_2(n,Q^2)]^{\rm tw-3}= \frac12\sum_{q=u,d,s,\ldots}\!\!\!e^2_q\,\,
\left[\frac{1}{n}\,\Delta q^+_T(n,Q^2)
+ \frac{\alpha_s}{\pi} C_g^T(n)\,\Delta g_T(n,Q^2)
\right].
\ee
Here, we took into account the relation between the gluon coefficient functions
\re{close} and introduced a notation for the overall
gluon coefficient function
$C_g^T$
\be{C-t}
C_g^T(n) = \int_0^1 dx\, x^{n-1} C_g^T(x) =
\frac1{n(n+1)}\lr{1-c(n)[\psi\left(n\right)+\!\gamma_E\!+1]}
\ee
with $c(n)$ given by \re{c-large}.
It is therefore natural to exercise a model for the structure function
in which $g_2(x,Q^2)$ can be expressed in terms of
these two distributions defined at a low scale $\mu^2$.
Such a model cannot be exact, since it is not theoretically
consistent: Other partonic degrees of freedom are generated through the
QCD evolution of $\Delta q_T(x,Q^2)$ and $\Delta g_T(x,Q^2)$ down to a low
scale.
Assuming the absence of their contribution at two different
scales  $Q^2$ and $\mu^2$ simultaneously is, therefore, not possible.
We shall argue, however, that the admixture of the additional degrees of
freedom %to \re{g2-Delta} at low scale $\mu$
turns out to be small numerically, at least for large $N$.
This means that the two-component model for the structure function
$g_2(x,Q^2)$ based on the parton distributions $\Delta q_T(x,Q^2)$ and
$\Delta g_T(x,Q^2)$ has a good numerical accuracy and can
eventually be made more sophisticated (involve more states) when and if
high accuracy experimental data become available.

In order to unravel the qualitative structure of the evolution it is
necessary to go over to large moments $N$ so that the number of
contributing operators becomes large and possible systematic effects
more transparent. To this end we take $N=100$ and, as a first step,
 examine the coefficients in the expansion of $\Delta q_T^+(N+3) =
\int_0^1 \!dx\, x^{N+2}\Delta q_T^+(\xi,Q^2)$ and
$\Delta g_T(N+3) = \int_0^1 \!dx\, x^{N+2}\Delta g_T(\xi,Q^2)$
in contributions of the multiplicatively renormalizable operators
\re{Red-O}. We recall that the moments $\Delta q_T^+(n)$ and
$\Delta g_T(n)$ are given by reduced matrix elements of the local
composite quark-antiquark-gluon and three-gluon operators defined as
\be{mom-pro}
\Delta q_T^+(n)= 2\langle\!\langle \Phi_n^q(\partial_i)\, S_\mu(z_i)
\rangle\!\rangle\bigg|_{z_i=0}
\,,
\qquad
\Delta g_T(n) = \langle\!\langle \Phi_n^g(\partial_i)\,
\widetilde O_\mu(z_i) \rangle\!\rangle\bigg|_{z_i=0}\,.
%=\langle\!\langle \Phi_n^h(\partial_i)
%T_\mu(z_i) \rangle\!\rangle\,.
\ee
Replacing the nonlocal operators by the expansion over the multiplicatively
renormalizable operators \re{short-dist} one gets
\ba{mom-exp}
\Delta q_T^+(N+3,Q^2)&=& 2\sum_{\alpha=0}^{3N/2} \vev{\Phi_{N+3}^q|\Psi^q_{N,\alpha}}
\langle\!\langle{\cal O}_{N,\alpha}(\mu^2)
\rangle\!\rangle
%\left(\frac{\alpha_s(Q^2)}{\alpha_s(\mu^2)}\right)
\,L^{E_{N,\alpha}/b}\,,
\nonumber
\\
\qquad
\Delta g_T(N+3,Q^2)&=&\sum_{\alpha=0}^{3N/2} \vev{\Phi_{N+3}^h|\Psi^h_{N-1,\alpha}}
\langle\!\langle{\cal O}_{N,\alpha}(\mu^2)
\rangle\!\rangle
%\left(\frac{\alpha_s(Q^2)}{\alpha_s(\mu^2)}\right)
\,L^{E_{N,\alpha}/b}\,,
\ea
where $\Psi^q_{N,\alpha}$ and $\Psi^h_{N-1,\alpha}$ are `quark' and `gluon'
components of the exact eigenstate of the evolution kernel in the helicity
basis.

\begin{figure}[t]
\centerline{\epsfxsize15.0cm\epsffile{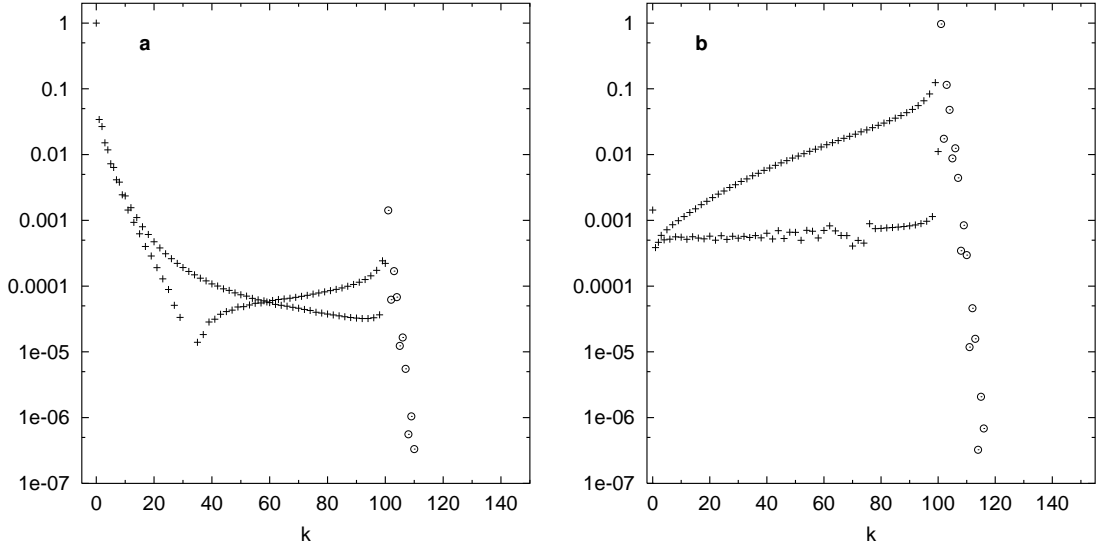}}
\caption[]{Exact numerical results for the coefficients:
a) --- $|\vev{\Psi_{N,k}^q|\Phi_{N+3}^q}|/\|\Phi_{N+3}^q\|$
and
b) --- $|\vev{\Psi_{N-1,k}^h|\Phi_{N+3}^h}|/\|\Phi_{N+3}^h\|$
of the expansion
of the $N=100$th moment of quark and gluon transverse spin distributions,
respectively,
in the contributions of multiplicatively renormalized operators.
The index $k$ enumerates the operators ordered according to their anomalous
dimension; $0\le k=\alpha\le N$ corresponds to quark operators,
$k=N+1+\beta \ge N+1$ to gluons, cf.~\re{LO-f}.
The `quark' eigenstates are shown by crosses, the `gluon' eigenstates by
open circles.}
\label{fig:Dq-expand}
\end{figure}

The expansion coefficients $\vev{\Phi_{N+3}^q|\Psi^q_{N,\alpha}}$
for $N=100$ are shown in Fig.~\ref{fig:Dq-expand}a.
%%%%%%%%%%%%%%%%%%     FIGURE 9          %%%%%%%%%%%%%%%%%%%%%%%%%%%%
%%%%%%%%%%%%%%%%%%%%%%%%%%%%%%%%%%%%%%%%%%%%%%%%%%%%%%%%%%%%%%%%%%%%%%
It is seen that $\Delta q_T^+(N+3)$ is grossly dominated by the contribution
of the local operator with the lowest anomalous dimension, which means that
all other contributions to the evolution are small. Among the
other contributions, noticeable corrections come from a few quark operators
with the low anomalous dimensions, and a single gluon operator with
the lowest anomalous dimension in the gluon sector. The contributions
of the quark operators with the anomalous dimensions close the the lowest
one are less interesting than the gluon contribution for the following two
reasons:
\begin{itemize}
\item{} Since the anomalous dimensions of important quark operators
   are all close to that of the leading quark operator,
   the corresponding contributions are barely distinguishable by the evolution.
\item{} The admixture of quark operators with the next-to-the-lowest etc.
   anomalous  dimensions is roughly the same in the flavor-nonsiglet and the
   flavor-singlet channels. Because of this, one does not expect any
   qualitative effects. On the other hand, the appearance of the gluon
   operator is a new feature of the singlet evolution.
\end{itemize}
Neglecting the higher quark and gluon states in
\re{mom-exp} we find that for arbitrary (large) $N$ the moments of the
quark distribution receive the dominant contribution from
only two states -- the lowest quark ($\alpha=0$)  and the lowest gluon
($\alpha=N+1$) levels and, therefore, depend on only two
nonperturbative parameters.
As a consequence, thus defined
moments $\Delta q^+(n,Q^2)$ satisfy the two-channel evolution equation
which can be written in the standard DGLAP form
\be{DGLAPq}
Q^2\frac{d}{d Q^2}\Delta q_T^+(x;Q^2) = \frac{\alpha_s}{4\pi}\int_{x}^1
\frac{dy}{y} \left[P^T_{qq}(x/y)\Delta q^+_T(y;Q^2)+
P^T_{qg}(x/y)\Delta g_T(y;Q^2)
\right].
\ee
In order to determine the splitting functions one can take moments so
that the convolution in \re{DGLAPq} gets replaced by a product
and identify the moments of the splitting functions with the
corresponding anomalous dimensions (with sign minus).

Since
moments of the quark distributions are obtained from the nonlocal
quark-antiquark-gluon operators by projecting onto the quark coefficient
function \re{mom-pro}, one can calculate the anomalous dimensions
by applying $\Phi^q_n(\partial_i)$ to the evolution equation \re{EQ-nonlocal}.
Since the coefficient functions $\Phi^q_{N+3}$
and $\Phi^g_{N+3}$ of the quark and gluon transverse spin distributions
$\Delta q_T(N+3)$ and $\Delta g_T(N+3)$ \re{partonmoments}
turn out to be very close to the lowest eigenstates of the `pure' quark
and `pure' gluon Hamiltonian
${\cal H}_0$ \re{split}, see Eqs.~\re{iden}, \re{lowgluon1},
we may invert \re{mom-pro} and expand the nonlocal operators $S_\mu(z_i)$ and
$\widetilde O_\mu(z_i)$ as
\ba{SO-exp}
\langle\!\langle S_\mu(z_i) \rangle\!\rangle &=& \sum_{n}
\frac{\widehat\Phi^q_n(z_i)}{2\|\Phi^q_n\|^2} \Delta q^+_T(n,Q^2) + \ldots\ ,
\nonumber
\\
\langle\!\langle \widetilde O_\mu(z_i) \rangle\!\rangle &=&\sum_n
\frac{\widehat\Phi^g_n(z_i)}{\|\Phi^h_n\|^2} \Delta g_T(n,Q^2) + \ldots\ ,
\ea
where $\widehat\Phi^q_n$ and $\widehat\Phi^g_n$ are related to the coefficient
functions through the transformation \re{map-ex} and the
dots denote the contribution of higher quark and gluon eigenstates.
Then, substituting \re{SO-exp} into the evolution equation \re{EQ-nonlocal},
one calculates the corresponding anomalous dimensions (for even $N$) as
\ba{Pq}
\int_{0}^1 dy\, y^{N+2} P_{qq}^T(y)&=& -
\frac{\vev{\Phi^q_{N+3} |H_{qq}|\Phi^q_{N+3}}}{\|\Phi^q_{N+3}\|^2}
  = -{\cal E}^q_{N,0}\,,
\nonumber\\
\int_{0}^1 dy\, y^{N+2} P_{qg}^T(y)&=& -2
\frac{\vev{\Phi^q_{N+3}|{H_{qh}|\Phi^h_{N+3}}}}
 {\| \Phi^q_{N+3}\|\|\Phi^h_{N+3}\|}
 \cdot \frac{\| \Phi^q_{N+3}\|}{\|\Phi^h_{N+3}\|},
\ea
where the last factor in the second expression serves to
correct for the different
normalization of the quark and gluon coefficient functions.
The first matrix element has been calculated in \cite{ABH,BKM00} and the
answer for the large-$N$ expansion of ${\cal E}^q_{N,0}$ is given in
\re{ref-E}, \re{energy-1}.
\be{Pq-exp}
\int_{0}^1 dy\, y^{n-1} P_{qq}^T(y)
= -4C_F\left[\psi(n)+\frac1{2n}+\gamma_E\right]+ C_F+\frac1{N_c}\lr{2-\frac{\pi^2}3}
+\CO\lr{\frac{\ln^2n}{n^2}}.
\ee
For the second matrix element we have
\be{inter-V}
\vev{\Phi^q_{N+3}|H_{qh}|\Phi^h_{N+3}}
= -n_f\sum_{k=1}^N\frac{(k+1)(k+2)+2(-1)^k}{(k+1)(k+2)\sqrt{k(k+3)}}
\phi_{N,k}^q \phi_{N-1,k-1}^h\,.
\ee
Using the explicit expressions for the coefficients
$\phi_{N,k}^q$ \re{Phi-coef} and $\phi_{N-1,k-1}^h$ \re{coeff-even}
one gets
\be{fin-V}
\vev{\Phi^q_n|H_{qh}|\Phi^h_n}=
 n_f\frac{\Gamma^4(n)n}{2\Gamma(2n)}\!\left[
{\frac{(n\!-\!1)(n^3\!-\!2n^2\!-\!6n\!-\!12)}{4(n+1)(n-2)}}+
 {\frac{(n^2\!-\!2n\!+\!4)(\psi(n)\!+\!\gamma_E)}{2(n-1)}}
\right]\!.
\ee
Combining together Eqs.~\re{fin-V}, \re{norm-g-cf} and \re{norm-q-cf}
and expanding at large $n=N+3$ we obtain
\be{P-qg}
\int_{0}^1 dy\, y^{n-1} P_{qg}^T(y) =
 -4n_f\left[\frac1{n}-\frac1{n^2} +\frac{-5+4\gamma_E+4\ln n}{n^3}
+{\cal O}\left(\frac1{n^4}\right)
\right]\,,
\ee
which agrees well with the exact expression for all $n > 3$.

In order to get a closed system of equations, we have to consider the
evolution of $\Delta g_T(x)$ as well. The coefficients of the expansion
of the corresponding coefficient function over the basis of multiplicately
renormalizable operators, Eq.~\re{mom-exp},
are shown in Fig.~\ref{fig:Dq-expand}b for $N=100$.
%%%%%%%%%%%%%%%%%%     FIGURE 10          %%%%%%%%%%%%%%%%%%%%%%%%%%%%
\begin{figure}[t]
\centerline{\epsfxsize15.0cm\epsffile{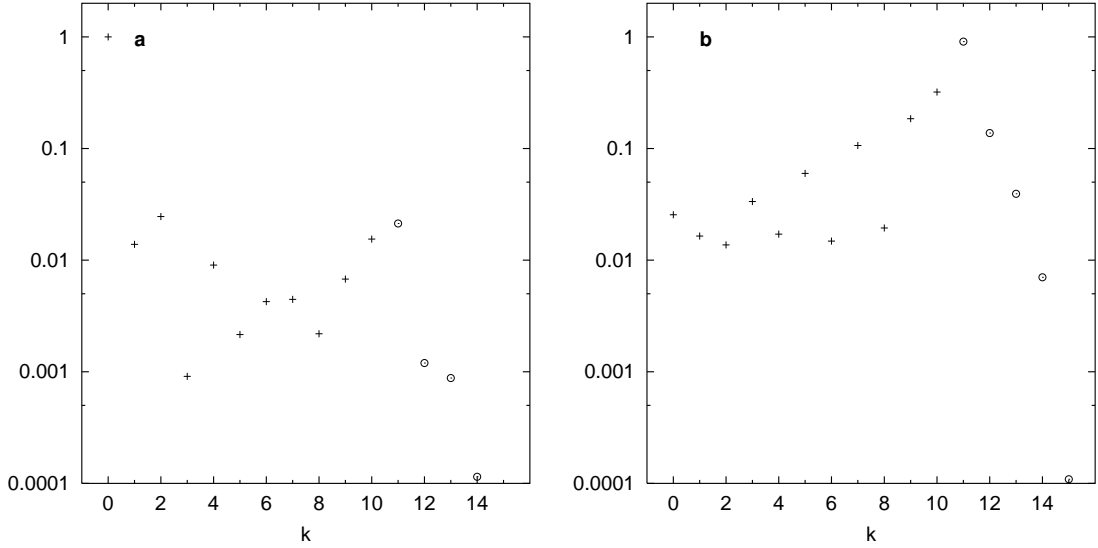}}
\caption[]{Same as in Fig.~\ref{fig:Dq-expand}, but for $N=10$.}
\label{fig:10-expand}
\end{figure}
%%%%%%%%%%%%%%%%%%%%%%%%%%%%%%%%%%%%%%%%%%%%%%%%%%%%%%%%%%%%%%%%%%%%%%
We see that the gluon distribution is dominated by the contribution of the
gluon operator with the lowest anomalous dimension in the gluon sector,
and contributions of all other gluon operators with higher anomalous
dimensions is strongly suppressed. Notice, however, that the contribution
of the quark operator with the lowest anomalous dimension is small
compared to the contributions of a large number $\sim N$
of other quark operators with larger anomalous dimensions. This means that
although we can formally write
 \be{DGLAPg}
Q^2\frac{d}{d Q^2}\Delta g_T(x;Q^2) = \frac{\alpha_s}{4\pi}\int_{x}^1
\frac{dy}{y} \left[P^T_{gg}(x/y)\Delta g_T(y;Q^2)+
P^T_{gq}(x/y)\Delta q^+_T(y;Q^2)+\ldots
\right],
\ee
with
\ba{Pg}
\int_{0}^1 dy\, y^{N+2} P_{gg}^T(y)&=& -
\frac{\vev{\Phi^h_{N+3} |H_{hh}|\Phi^h_{N+3}}}{\|\Phi^h_{N+3}\|^2}
  = -{\cal E}^g_{N,0}\,,
\nonumber\\
\int_{0}^1 dy\, y^{N+2} P_{gq}^T(y)&=&
-\frac12\frac{\vev{\Phi^h_{N+3}|{H_{hq}|\Phi^q_{N+3}}}}
 {\| \Phi^h_{N+3}\|\|\Phi^q_{N+3}\|}
 \cdot \frac{\| \Phi^h_{N+3}\|}{\|\Phi^q_{N+3}\|},
\ea
where ${\cal E}^g_{N,0}$ is given in \re{estim2},
taking into account the mixing term $\sim P^T_{gq}(x/y)$ is not justified
to the accuracy that the mixing with other quark degrees of freedom is omitted.
We suggest, therefore, to neglect the mixing of quarks to the gluon
distribution altogether. It is this mixing that sets the limitations
for the accuracy the two-component transverse spin model. This accuracy
is in fact rather high since the sum of squares of the coefficients of
all quark operators in Fig.~\ref{fig:Dq-expand}b is less than 2\%.

For smaller values of $N$, the hierarchy of different contributions does
not look so convincing, see Fig.~\ref{fig:10-expand}, but qualitatively
remains the same. Most importantly, contributions of all gluon operators
other than the one with the lowest anomalous dimension remain negligible.

The DGLAP equations \re{DGLAPq} and \re{DGLAPg} can easily be solved
going over to moments. Neglecting the quark mixing in \re{DGLAPg}
we find using \re{g2-Delta}
\ba{DGLAP-g2}
g_2^{\rm LL}(n,Q^2)&=&\frac{\vev{e_q^2}}{2n}\Bigg[
\Delta q_T^{+}(n,\mu^2)\,L^{\gamma^T_{qq}(n)/b}
\nonumber
\\&&{}\hspace*{1cm}
+\Delta g_T(n,\mu^2)\, \lr{L^{\gamma^T_{gg}(n)/b}-L^{\gamma^T_{qq}(n)/b}}
\frac{\gamma^T_{qg}(n)}{\gamma^T_{gg}(n)-\gamma^T_{qq}(n)}
\Bigg],
\ea
where $\gamma^T_{ab}(n) = -\int_0^1 dx\, x^{n-1} P_{ab}^T(x)$ and
$L=\alpha_s(Q^2)/\alpha_s(\mu^2)$.
For the two lowest moments $n=3$ and $n=5$, expanding
moments of the parton distributions at a low normalization scale
$\Delta q_T^{+}(n,\mu^2)$ and $\Delta g_T(n,\mu^2)$
over the reduced matrix elements of local operators
$\langle\!\langle [S]^k_N \rangle\!\rangle$ and
$\langle\!\langle [G]^k_N \rangle\!\rangle$,
Eqs.~\re{Explicit-q} -- \re{Explicit-g},
and using the explicit expressions for the evolutions kernels, one obtains
\ba{N=0,2-approx}
\frac{3}{2}\langle e^2_q\rangle^{-1} g_2^{\rm LL}(3,Q^2) &=& L^{10.556/b}
  \langle\!\langle S^0_0\rangle\!\rangle\,,
\nonumber\\
\frac{5}{2}\langle e^2_q\rangle^{-1} g_2^{\rm LL}(5,Q^2) &=&
   L^{10.987/b}
  \Big[
     \langle\!\langle S^2_2 \rangle\!\rangle
  - 2\, \langle\!\langle S^1_2 \rangle\!\rangle
  + 3\,\langle\!\langle S^0_2 \rangle\!\rangle
  + 0.974\, \langle\!\langle G^0_2 \rangle\!\rangle
 \Big]
\nonumber\\
&+&%{}+
 L^{17.350/b}
  \Big[
  -0.974\,\langle\!\langle G^0_2 \rangle\!\rangle
 \Big].
\ea
These results have to be compared with the exact expressions in
\re{N=0,2-sing}.
We see that the approximation advocated in this work corresponds to
taking into account (for $n=5$) the two multiplicatively renormalizable
operators  with a large gluon contribution
(the first and the third lines in \re{N=0,2-sing}) and neglecting the
other terms. The coefficients in front of the three quark local operators
in the first line in \re{N=0,2-sing} are reproduced reasonably well.
%and one can prove \cite{ABH} that the deviations become much smaller
%(decrease as $\sim 1/n$) for higher moments,
%so that the  case considered here, $n=5$, is in fact the worst one.
Note that values of the anomalous dimensions come out to be very close to
the exact results.

\setcounter{equation}{0}
\section{Summary and conclusions}
Based on our systematic analysis of the evolution pattern of twist-three
operators we arrive at
the following approximate two-channel evolution
equation for the flavor-singlet quark and gluon transverse spin distributions:
\ba{stand}
&&Q^2\frac{d}{d Q^2}\Delta q^{+,\,\rm S}_T(x;Q^2) = \frac{\alpha_s}{4\pi}\int_{x}^1
\frac{dy}{y} \left[P^T_{qq}(x/y)\Delta q^{+,\,\rm S}_T(y;Q^2)+
P^T_{qg}(x/y)\Delta g_T(y;Q^2)
\right],
\nonumber\\
&&Q^2\frac{d}{d Q^2}\Delta g_T(x;Q^2)=\frac{\alpha_s}{4\pi}\int_{x}^1
\frac{dy}{y} P^T_{gg}(x/y)\Delta g_T(y;Q^2)\,,
\ea
where the flavor-singlet quark distribution is defined similar to \re{singlet}
\be{Delta-q-s}
\Delta q^{+,\,\rm S}_T(x)=
%\sum_{q=u,d,s,...}\Delta q_T(x)+ \Delta \bar q_T(x)
\Delta u_T(x)+ \Delta \bar u_T(x)+\Delta d_T(x)+ \Delta \bar d_T(x)+\ldots
\,.
\ee
The Hamiltonian approach developed in this paper can be used to
determine the asymptotic expansion of the moments of the splitting functions
at large $N$ that translates to the expansion in powers of $1-x$
at large momentum fractions $x$. Since quark and gluon distributions in the
nucleon appear to be decreasing strongly at large $x$, this contribution
to the splitting functions is the most important one numerically.
Using the  results for the moments given in the text,
we derive the following expressions for the splitting functions:
\ba{otvet2}
 P^T_{qq}(x) &=&
\left[\frac{4C_F}{1-x}\right]_+
      +\delta(1-x)\left[C_F+\frac{1}{N_c}\left(2-\frac{\pi^2}{3}\right)\right]
      - 2C_F\,,
\nonumber\\
 P^T_{gg}(x) &=& \left[\frac{4 N_c}{1-x}\right]_+
 + \delta(1-x) \left[ N_c\lr{\frac{\pi^2}3-\frac13}\,-\frac23{n_f}\right]
\nonumber\\
 &&{}+
  N_c\lr{\frac{\pi^2}3-2}+ N_c\ln\frac{1-x}{x}\lr{\frac{2\pi^2}{3}-6}\,,
\nonumber\\
 P^T_{qg}(x) &=& %2n_f \left[1 + x^2 + (1-x)^2\right]\,.
      -4 n_f \left[ x - 2(1-x)^2\ln(1-x)\right]\,.
\ea
Here, the first two expressions are accurate up to corrections of order
$\CO(1-x)$ for $x\to 1$ and the third expression has the
accuracy $\CO((1-x)^3)$.
Note that the quark splitting function $P^T_{qq}(x)$ is the
same in the flavor singlet and flavor-nonsinglet channels, cf.~\re{AP:NS}.
Remarkably enough, the obtained the twist-3 evolution kernels turn out to
be very similar to the well-known expressions for the twist-2 DGLAP kernels.
Moreover, the leading $x\to 1$ asymptotics of the diagonal kernels is the same
for twist-2 and twist-3 and the difference occurs at the level of subleading
$\sim\delta(1-x)$ corrections. One can argue following \cite{K89} that this
property is rather general and it holds to all orders of perturbation theory.

To the leading logarithmic accuracy, the structure function $g_2(x,Q^2)$
is expressed through the quark distribution
\ba{g2final}
 g_2^{\rm LL}(x,Q^2) &=& g^{WW}_2(x,Q^2)+
\frac{1}{2}\sum_q e^2_q
      \int_{x}^1 \frac{dy}{y} \,\Delta q^+_T(y,Q^2)\,,
\ea
where $g^{WW}_2(x,Q^2)$ is the Wandzura-Wilczek contribution \re{gWW}
and the gluon contribution arises entirely through the evolution equations
\re{stand} (after the separation of the flavor-singlet part).
Note that both the gluon and the quark distributions are defined by analytic
continuation from the odd moments $n=1,3,\ldots$ and must satisfy
the constraints
\be{BC1}
  \int_0^1 \!dx\, \Delta q^+_T(x,Q^2) = \int_0^1\! dx\, \Delta g_T(x,Q^2) =0
\ee
that follow from properties of the coefficient functions
$\Phi^q$, $\Phi^g$ in \re{partonmoments}. The Burkhardt-Cottingham
sum rule \cite{BC70} $\int_0^1 dx \,g_2(x,Q^2) =0$ then follows
from \re{g2final} which derivation involves an additional implicit
assumption about the absence of subtraction constants in the
dispersion relation for the spin-dependent Compton amplitude, see
\cite{Book}.

The gluon distribution is subject to the additional constraint
\be{BC2}
  \int_0^1\! dx\, x^2 \Delta g_T(x,Q^2) =0\,,
\ee
which ensures that the gluon contribution vanishes for the third
moment.

%Beyond the leading logarithms,
To the next-to-leading logarithmic accuracy,
%(NL),
the twist-3 gluon contribution at large scales
can be calculated as a finite part of the box diagram in the background
gluon field and the result \cite{I} projected onto the gluon transverse
spin distribution has the form
\ba{glueQ}
  g_2^{\rm NL}(x,Q^2) &=& \ldots +
\frac{1}{2}{\sum_q e^2_q}\ \frac{\alpha_s}{\pi}
      \int_{x}^1 \frac{dy}{y} \,C_g^T(x/y)\,\Delta g_T(y,Q^2)\,,
\nonumber\\
 C_g^T(x) &=& -(1-x)\left[1+\ln\frac{x}{1-x}\right]
\\\nonumber
 && -(1-x)^3\left[\frac23\ln^2(1-x)-\frac{19}{9}\ln(1-x)+\frac{83}{54}-\frac{\pi^2}{9}\right]
+\CO((1-x)^4)
\,,
\ea
where the dots stand for
the leading-order quark contribution and the $\CO(\alpha_s)$ quark corrections.
We have argued in \cite{I} and in this paper that this projection has a high
accuracy for all integer moments (i.e. all contributions left out by this
projection are very small).

To summarize, the set of formulas given in this Section presents a
theoretically motivated approximation for the QCD description of
deep inelastic scattering from a transversely
polarized nucleon. Importance of this approximation is not so much in the
possibility to calculate the scale dependence, but in the identification
of important transverse spin degrees of freedom that are preserved by
QCD interaction.
The formalism developed in this paper is very general and can be applied
to the study of other higher-twist distributions. It is based on the
operator product expansion and conformal symmetry of the QCD Lagrangian.
This technique turns out to be very effective in dealing with the operator
renormalization for large $N$ and in many cases a WKB-type expansion
in $1/N$ can be constructed analytically. Going over from the moments
to the momentum fraction space involves analytic continuation and
in general may be quite complicated, see \cite{BKM00} for a discussion.
Because of this, the small$-x$ behavior of twist-3 (and higher-twist)
parton distributions presents a nontrivial problem and deserves
further study.

\subsection*{Acknowledgements}

The work by A.M. was supported by the DFG, project Nr. 920585,
by the grant of the Spanish Ministry of Science,
and by the grant 00-01-00500 of the Russian Foundation for
Fundamental Research. The work by G.K. was supported by
the EU network `Training and Mobility of Researchers',
 FMRX--CT98--0194.
G.K. is grateful to L.~Frankfurt and M.~Karliner for
useful discussions and warm hospitality at the Tel-Aviv University.

\appendix
\renewcommand{\theequation}{\Alph{section}.\arabic{equation}}
\setcounter{table}{0}
\renewcommand{\thetable}{\Alph{table}}

\section*{Appendices}

\setcounter{equation}{0}
\section{Appendix: Calculation of the anomalous dimensions}
In this Appendix we describe the calculation of the energy of the
lowest quark and gluon levels, Eqs.~\re{energy-0} and \re{E-n-g},
respectively, and the energy of the highest quark level, Eq.~\re{quark-high}.

The energy of the lowest gluon level is defined by Eq.~\re{estim2}
and involves the expectation value of the Hamiltonian $H_{hh}$ over
the gluon state $\Phi^h_n(x_i)$ and its norm.
To calculate the both, we shall use the expansion
of the wave function over the conformal basis.
An arbitrary three-gluon state can be characterized by
total conformal spin  $J=n+1/2=N+7/2$ and the
conformal spin in a certain two-particle channel. Depending on the particular
choice of the two-particle channel one can define three different
sets of  basis functions $Y^{(12)3}_{n+1/2,k+3}(x_i)$,
$Y^{(23)1}_{n+1/2,k+3}(x_i)$ and
$Y^{(31)2}_{n+1/2,k+3}(x_i)$. The functions belonging to each basis are linear
independent whereas the ones belonging to two different sets are
related to each other through a linear transformation
\be{AppA-Racah}
Y^{(31)2}_{n+1/2,k+3}(x_i)=\sum_{m}\Omega^{(h)}_{km}\,Y^{(12)3}_{n+1/2,m+3}(x_i)
=\sum_{m}(-1)^{m+k}\Omega^{(h)}_{km}\,Y^{(23)1}_{n+1/2,m+3}(x_i)\,,
\ee
with $\Omega^{(h)}_{km}=\vev{Y^{(31)2}_{n+1/2,k+3}|Y^{(12)3}_{n+1/2,m+3}}$
being the Racah $6j-$symbols.

The gluon coefficient function in the helicity representation
can be expanded in any one of the three conformal basis so that one gets
three sets of the expansion coefficients:
\be{all-g}
\Phi^h_n(x_i) = \sum _{k=0}^{n-4}\phi^{(31)2}_{n,k}\, Y^{(31)2}_{n+1/2,k+3}(x_i)
= \sum _{k=0}^{n-4}\phi^{(12)3}_{n,k}\, Y^{(12)3}_{n+1/2,k+3}(x_i)
= \sum _{k=0}^{n-4}\phi^{(23)1}_{n,k}\, Y^{(23)1}_{n+1/2,k+3}(x_i)\,.
\ee
These coefficients can be calculated similar to \re{cf-J} and are given by
\be{2}
\phi^{(12)3}_{n,k} = -\phi^{(31)2}_{n,k} =
h_{n,k}\left[
{\frac {\left (n+2+k\right )\left (n+4+{k}^{2}/2+5\,k/2\right )}{
\left (k+1\right )\left (n-2-k\right )}}
\right]\,,\quad \phi^{(23)1}_{n,k}=0
\ee
for {\it even\/} $k$, and
\be{3}
\phi^{(12)3}_{n,k} = \phi^{(31)2}_{n,k} = -\frac12\phi^{(23)1}_{n,k}=h_{n,k}
{\frac {\left (k+4\right )\left (n+3+k\right )}{2(n-3-k)}}
\ee
for {\it odd\/} $k$.
Here, the normalization factor $h_{n,k}$ is defined as
\be{h-nk}
h^2_{n,k}=
{\frac {\left (k+1\right )\left (-n+3+k\right )\left (-n+2+k
\right )\left (2\,k+5\right ){n}^{2}}{8
\left (k+2\right )\left (k+3\right )\left (k+4\right )\left (n+2+k
\right )\left (n+3+k\right )}}
\frac{\Gamma ^{4}(n)}{\Gamma (2\,n)}
\,.
\ee
The relations between the coefficients respect the symmetry
properties of
$\Phi_n^h(x_i)$. The coefficients $\phi^{(23)1}_{n,k}$ vanish for even $k$
to ensure the antisymmetry of the state under the interchange of gluons
with the same helicity, $x_2 \leftrightarrow x_3$. The sum of the
coefficients in the three different channels vanishes for arbitrary $k$
since the three-gluon state
is annihilated by the operator $1+P+P^2=(P^3-1)/(P-1)$ with $P$ being
the operator of cyclic permutations.

The `spherical harmonics' $Y^{(31)2}_{n,k}(x_i)$ form an orthonormal
basis on the space
of the coefficient functions endowed with the scalar product \re{int-rep},
$\vev{Y^{(31)2}_{n,k}|Y^{(31)2}_{n,m}}=\delta_{k,m}$. The
same is true for the states $Y^{(12)3}_{n,k}(x_i)$ and
$Y^{(23)1}_{n,k}(x_i)$ . Using this property we may calculate the
norm of the gluon state in one of the three equivalent forms
\be{norm-gluon}
\|\Phi^h_n\|^2 =\sum_{k=0}^{n-4}\left[\phi^{(12)3}_{n,k}\right]^2
=\sum_{k=0}^{n-4}\left[\phi^{(23)1}_{n,k}\right]^2
=\sum_{k=0}^{n-4}\left[\phi^{(31)2}_{n,k}\right]^2\,.
\ee
Substituting the expressions for the expansion coefficients \re{2} and \re{3},
one gets a finite sum over two-particle conformal spin $k$ whose evaluation
leads to Eq.~\re{norm-g-cf}. Expanding the result at large $n$ one obtains
\be{norm-gluon-as}
\|\Phi^h_n\|^2=\frac{\Gamma^4(n)}{\Gamma(2n)}\frac{n^4}{16}\left[
1+\frac1{n}\lr{2\ln n +2\gamma_E -1}+\CO(1/n^2)
\right]\,,
\ee
where we factored out the ratio of the $\Gamma-$functions for
later convenience.

In order to calculate  the expectation value of the evolution kernel
defined in \re{on-g}, we notice that $H_{hh}$ has a two-particle
structure and can be split in three contributions each of which only
depends on the operator of the conformal spin $J_{ab}$ in a given
two-particle channel
\ba{re-arr}
\vev{\Phi_n^h|H_{hh}|\Phi_n^h} &=& N_c\, \vev{\Phi_n^h|2U_{gg}(J_{12})
-V_{gg}(J_{12}) |\Phi_n^h}
\nonumber
\\
&+&N_c \,\vev{\Phi_n^h|2U_{gg}(J_{31})
-V_{gg}(J_{31}) |\Phi_n^h}
\nonumber
\\
&+&2N_c\, \vev{\Phi_n^h|U_{gg}(J_{23})|\Phi_n^h}
-b\, \|\Phi^h_n\|^2\,.
\ea
Since the operator $J_{12}$ is diagonal in the conformal basis
$Y^{(12)3}_{n,k}(x_i)$ (by construction),
it is natural to evaluate the first term in
\re{re-arr} by expanding the gluon state $\Phi_n^h$
over this particular basis. By the same token,
the second and the third terms in \re{re-arr} are most easily calculated
using the expansion over
$Y^{(31)2}_{n,k}(x_i)$ and $Y^{(23)1}_{n,k}(x_i)$, respectively. In this
way one arrives at
\ba{App-Energy}
\vev{\Phi_n^h|H_{hh}|\Phi_n^h} &=& 2N_c\sum_{k=0}^{n-4}
U_{gg}(k+3)\left([\phi^{(12)3}_{n,k}]^2+[\phi^{(31)2}_{n,k}]^2
+[\phi^{(23)1}_{n,k}]^2\right)
\nonumber
\\&&
-N_c\sum_{k=0}^{n-4} V_{gg}(k+3)\lr{[\phi^{(12)3}_{n,k}]^2+[\phi^{(31)2}_{n,k}]^2}
-b\, \|\Phi^h_n\|^2,
\ea
where we have used that the conformal spin in the two-gluon channel is
equal to $2j_g+k=k+3$. Using explicit expressions for the kernels
$U_{gg}$ and $V_{gg}$ defined in \re{U,V} and  the expansion coefficients
\re{2} and \re{3}, one can rewrite \re{App-Energy} as
a finite sum over the two-particle spin $k$. The sum involving
$V_{gg}$ can be calculated analytically while the sum involving $U_{gg}$
can be expanded in inverse powers of $1/n$ at large $n$.
After some algebra one arrives at
\be{Sum-V}
\sum_{k=0}^{n-4} V_{gg}(k+3)\lr{[\phi^{(12)3}_{n,k}]^2+[\phi^{(31)2}_{n,k}]^2}
=\frac{\Gamma^4(n)}{\Gamma(2n)}\frac{n^4}4
\left[\frac{\pi^2}6-\frac32+\frac1{n}+\CO\lr{\frac{\ln n}{n^2}}\right]
\ee
and
\ba{Sum-U}
\sum_{k=0}^{n-4}
&&
U_{gg}(k+3)\left([\phi^{(12)3}_{n,k}]^2+[\phi^{(31)2}_{n,k}]^2
+[\phi^{(23)1}_{n,k}]^2\right)
\\
&&
=\frac{\Gamma^4(n)}{\Gamma(2n)}\frac{n^4}8
\left[\ln n+\gamma_E+\frac{\pi^2}{12}-\frac12
+\frac1{n}\lr{2(\ln n+\gamma_E)^2-\frac12(\ln n+\gamma_E)}
+\CO\lr{\frac{\ln^2\! n}{n^2}}\right]\!.
\nonumber
\ea
Substituting these expressions into \re{App-Energy} and
combining them with \re{norm-gluon-as} one arrives at the expression for the
energy of the lowest gluon state given in the text, Eq~\re{estim2}.

Calculation of the energy of the lowest quark level, Eq.~\re{energy-0},
goes along the same lines. The quark coefficient function
$\Phi_n^q$ is defined in Eq.~\re{psi-n} on the hyperplane
$x_1+x_2+x_3=0$ corresponding to the kinematics of the forward scattering,
and can be continued to arbitrary values of the momentum fractions $x_i$
using the conformal symmetry.
Its expansion over the conformal `spherical harmonics' in the three
different two-particle channels looks like
\be{Phi-q-exp}
\Phi_n^q(x_i)
= \sum_{k=0}^{n-3}\varphi^{(12)3}_{n,k} Y^{(12)3}_{n+1/2,k+5/2}(x_i)
= \sum_{k=0}^{n-3}\varphi^{(23)1}_{n,k} Y^{(23)1}_{n+1/2,k+5/2}(x_i)
= \sum_{k=0}^{n-3}\varphi^{(31)2}_{n,k} Y^{(31)2}_{n+1/2,k+2}(x_i)\,,
\ee
where e.g. $Y^{(12)3}_{n+1/2,k+5/2}(x_i)$ is defined by the general expression
\re{Y-q} with $j_1=j_3=1$, $j_2=3/2$, the total conformal spin
equal to $J=n=N+3$ and the two-particle
conformal spin in the $(12)$ channel given by $j_{12}=j_1+j_2+k=3+k$.

The explicit expressions for the expansion coefficients can be obtained
using the representation similar to \re{Geg}.
The result is:
\ba{Phi-q-coef}
\varphi^{(12)3}_{n,k}&=&
    2\left (-1\right )^{k}\,{\frac {n\left (n+2+k\right )}{k+1}}\, {q_{n,k}}\,,
\nonumber
\\
\varphi^{(23)1}_{n,k}&=&
 -2\,\left (-1\right )^{n}\left (k+3\right )\left(n+2+k\right){q_{n,k}}\,,
\\
\varphi^{(31)2}_{n,k}&=& -\left (3+2\,k\right )\left (k+2\right ){\tilde q_{n,k}}
\nonumber\\&\times&
\left [
\frac{\left(-1\right )^{n-k}-1}2{\frac{n+2+k}{n-2-k}}
+\frac{\left (-1\right )^{n-k}+1}2{\frac {n+1+k} {n-1-k}}\right ],
\nonumber
\ea
with the normalization factors
\ba{Phi-q-q}
q_{n,k}^2&=&{\frac {\left (k+1\right )\left (n-2-k\right )}{4\left (n+2+k
\right )\left (k+2\right )\left (k+3\right )}}\frac{\Gamma^4(n)}{\Gamma(2n)},
\nonumber\\
\widetilde q_{n,k}^2&=&
2\,{\frac {\left (n-1-k\right )\left (k+3\right )}{\left (3+2\,k
\right )\left (n+1+k\right )}}q_{n,k}^2\,.
\ea
Using the expansion in  \re{Phi-q-exp} one finds three equivalent
representations for the norm of the quark state
\be{Phi-q-norm-sum}
\|\Phi_n^q\|^2=\sum_{k=0}^{n-3}[\varphi^{(12)3}_{n,k}]^2
=\sum_{k=0}^{n-3}[\varphi^{(23)1}_{n,k}]^2
=\sum_{k=0}^{n-3}[\varphi^{(31)2}_{n,k}]^2\,.
\ee
Substituting explicit expressions for the expansion coefficients and
performing the summation one arrives at Eq.~\re{norm-q-cf}.
The expansion of the norm at large $n$ reads
\be{Phi-q-norm}
\|\Phi_n^q\|^2=\frac{\Gamma^4(n)}{\Gamma(2n)}\frac{n^4}4
\left[1+\frac1{n^2}\lr{1-4\ln n -4\gamma_E}+\CO(1/n^2)\right]\,.
\ee
The expectation value of the diagonal quark evolution kernel
$H_{qq}$ defined in \re{on-q} can be written in the form analogous
to \re{re-arr}
\ba{App-H-qq}
\vev{\Phi_n^q|H_{qq}|\Phi_n^q} &=& \vev{\Phi_n^q|N_c V_{qg}^{(0)}(J_{12})
-\frac2{N_c}  V_{qg}^{(1)}(J_{12}) |\Phi_n^q}
\nonumber
\\
&+&\vev{\Phi_n^q|N_c U_{qg}^{(0)}(J_{23})
-\frac2{N_c} U_{qg}^{(1)}(J_{23}) |\Phi_n^q}
\\
&+&\vev{\Phi_n^q|\frac{2n_f}3\delta_{J_{31},2}-\frac2{N_c} U_{qq}^{(1)}(J_{31}) |\Phi_n^q}
\,.\nonumber
\ea
Expanding the quark state \re{Phi-q-exp} over the suitable conformal
basis one obtains
\ba{App-H-qq-sum}
\vev{\Phi_n^q|H_{qq}|\Phi_n^q} &=&
\sum_{k=0}^{n-3}[\varphi^{(12)3}_{n,k}]^2
\lr{N_c V_{qg}^{(0)}(k+5/2)-\frac2{N_c}  V_{qg}^{(1)}(k+5/2)}
\nonumber
\\
&+&
\sum_{k=0}^{n-3}[\varphi^{(23)1}_{n,k}]^2
\lr{N_c U_{qg}^{(0)}(k+5/2)-\frac2{N_c} U_{qg}^{(1)}(k+5/2)}
\\
&+&
\sum_{k=0}^{n-3}[\varphi^{(31)2}_{n,k}]^2
\lr{ \frac{2n_f}3\delta_{k,0}-\frac2{N_c} U_{qq}^{(1)}(k+2)}
\,,
\nonumber
\ea
where we have replaced the operators of the two-particle conformal spins by
the corresponding eigenvalues $J_{ab}=j_a+j_b+(n-3)$.
The necessary expansion coefficients
and explicit expressions for the evolution kernels
entering this expression are given  in \re{Phi-q-coef}
and \re{nonplanar}, respectively.

It turns out that the part of the sum in \re{App-H-qq-sum}
proportional to $N_c$ can be calculated exactly
\be{App-planar}
\sum_{k=0}^{n-3}[\varphi^{(12)3}_{n,k}]^2 V_{qg}^{(0)}(k+5/2)+
[\varphi^{(23)1}_{n,k}]^2 U_{qg}^{(0)}(k+5/2)
=\|\Phi_n^q\|^2 \lr{\psi(n)+\frac1{n}-\frac12+2\gamma_E}.
\ee
%and it coincides with the ground state energy of the open spin chain
%studied in \cite{?}.
The $1/N_c$ correction to the sum
\re{App-H-qq-sum} is given by a ratio of rather complicated sums
and can easily be expanded  at large $n$ leading to
\ba{App-nonplanar}
&& \sum_{k=0}^{n-3}[\varphi^{(12)3}_{n,k}]^2 V_{qg}^{(1)}(k+5/2)
 =\|\Phi_n^q\|^2 \lr{\frac74-\frac{\pi^2}{6}+\CO(\ln  n/n^2)},
\nonumber
\\
&& \sum_{k=0}^{n-3}[\varphi^{(23)1}_{n,k}]^2 U_{qg}^{(1)}(k+5/2)
 =\|\Phi_n^q\|^2\lr{0+ \CO(1/n^2)},
\\
&& \sum_{k=0}^{n-3}[\varphi^{(31)2}_{n,k}]^2 U_{qq}^{(1)}(k+2)
 =\|\Phi_n^q\|^2 \lr{\ln n +\gamma_E -1+\CO(\ln^2 n/n^2)}\,.
\nonumber
\ea
Finally, the $n_f-$dependent contribution to the \re{App-H-qq-sum}
is given for odd $n$ by
\be{App-nf}
[\varphi^{(31)2}_{n,0}]^2
= 3\frac{(n-1)(n+2)}{(n+1)(n-2)}\frac{\Gamma^4(n)}{\Gamma(2n)}
= \|\Phi_n^q\|^2\lr{0+ \CO(1/n^4)}\,.
\ee
Combining \re{App-planar}, \re{App-nonplanar}, \re{App-nf} and
\re{App-H-qq} we obtain the expression for the energy of the lowest quark level
given in Eq.~\re{energy-0}.% \cite{BKM00}.

The eigenstate of the highest quark level can be approximated by
the expression in \re{exp1}. Calculating the corresponding energy
\re{quark-high} and using \re{on-q}, one gets
\be{App-E}
{\cal E}^{q}_{N,N} = \frac23 n_f + \vev{Y^q_{N,k=0} |H_{S^+}| Y^q_{N,k=0}}\,.
\ee
Using the explicit expression \re{HSp} for the Hamiltonian $H_{S^+}$
and taking into account that  $Y^q_{N,k=0}$ is symmetric with respect
to the interchange of the quarks, $(P_{13}-1)Y^q_{N,k=0}=0$, one obtains
\be{App-sym}
{\cal E}^{q}_{N,N} = \frac23 n_f-\frac2{N_c} U_{qq}^{(1)}(J_{13}=2)
+ \vev{Y^q_{N,0} |V(J_{12})| Y^q_{N,0}}\,,
\ee
where the notation was introduced for the linear combination of the
kernels \re{nonplanar}
\be{App-not}
V(J)=
N_c\,\lr{V_{qg}^{(0)}(J)+ U_{qg}^{(0)}(J)}
 -\frac{2}{N_c}\, \lr{V_{qg}^{(1)}(J) + U_{qg}^{(1)}(J)}.
\ee
In order to calculate the matrix element entering \re{App-sym} we use the
Racah decomposition \re{tr-M}
to expand $Y^q_{N,k=0}$ over the conformal basis in the quark-gluon channel
\be{App-reex}
Y^q_{N,0}(x_i)=\sum_m \Omega_{2,m+5/2}(N+7/2)\,Y^{(12)3}_{N+7/2,m+5/2}(x_i)\,.
\ee
Taking into account that $U_{qq}^{(1)}(2)=0$ one finds %from \re{App-sym}
\be{App-fin}
{\cal E}^{q}_{N,N}=\frac23 n_f+\sum_{m=0}^NV(m+5/2)
  \,[\Omega_{2,m+5/2}(N+7/2)]^2\,.
\ee
Finally, using the explicit expressions for  the Racah symbols \re{App-Racah}
and the two-particle evolution kernels \re{nonplanar} and
and performing the summation in \re{App-fin} one arrives at
the result given in Eq.~\re{quark-high}.

\setcounter{equation}{0}
\section{Appendix: Racah symbols for the $SL(2,R)$ group}

In this Appendix we derive the explicit expression for the Racah $6j-$symbols
$\Omega_{jj'}$ defined in \re{tr-matr}
\be{tr-M}
{Y}_{Jj}^{(31)2}(x_i)=\sum_{j_1+j_2\le j'\le J-j_3}\Omega_{jj'}(J)\,
{Y}_{Jj'}^{(12)3}(x_i)\,,
\ee
where the basis functions ${Y}_{Jj}^{(31)2}$ are given in
Eq.~(\ref{Y-q}). In order to find the coefficients $\Omega_{jj'}(J)$
it is sufficient to set the three variables $x_i$ in the Eq.~(\ref{tr-M})
to the following values: $x_1=-x$, $x_2=x$ and
$x_3=1$. Using the explicit expressions for the
basis functions~(\ref{Y-q}) one gets for~(\ref{tr-M}):
\ba{gen-id}
&& (-1)^{J-j-j_2}
r_{Jj}\frac{\Gamma(j\!+\!j_3\!-\!j_1)\,\Gamma(J\!-\!j\!+\!j_2)}{
\Gamma(2j_3)\,\Gamma(2j_2)\,(j\!-\!j_1\!-\!j_3)!\,(J\!-\!j\!-\!j_2)!}
\times\\
&&\times\,\, {}_2F_1(j_1\!+\!j_3\!-\!j,1\!+\!j_3\!-\!j_1\!-\!j,2j_3;x)
\, {}_2F_1(\!-\!J\!+\!j\!+\!j_2,J\!+\!j\!+\!j_2\!-\!1,2j_2;x)\nonumber \\
&=&
%(\!-\!1)^{J\!-\!j_1\!-\!j_2\!-\!j_3}
\sum_{j'=j_1\!+\!j_2}^{J\!-\!j_3}
 \Omega_{jj'}(J) \,
\frac{r_{Jj'}\, (-x)^{j'\!-\!j_1\!-\!j_2}}{(j'\!-\!j_1\!-\!j_2)!\,(J\!-\!j'\!-\!j_3)!}
\frac{\Gamma(J\!+\!j'\!-\!j_3)}{(2j'\!-\!1)\,\Gamma(j'\!+\!j_1\!+\!j_2\!-\!1)}\,\nonumber.
\ea
The product of the two hypergeometric functions in the l.h.s. of this
indentity is nothing else but the generating
function for the Racah polynomials (see \cite{A-W} for details).
Expanding the hypergeometric functions and comparing the terms in the
l.h.s. and the r.h.s. of~(\ref{gen-id})
with the same power of $x$ one derives after some algebra the following
explicit expression:
\ba{App-Racah}
\Omega_{jj'}(J)&=&(-1)^{J\!-\!j\!-\!j_2}
\frac{\Gamma(J\!-\!j_1\!-\!j_2\!-\!j_3\!+\!1)}{\Gamma(2j_1)\Gamma(J\!+\!j_1\!+\!j_2\!+\!j_3\!-\!1)}\,
f(J,j',j_1,j_2,j_3)\, f(J,j,j_1,j_3,j_2)\nonumber \\
&&\times\,\,
{}_4F_3\left ({-j'\!+\!j_1\!+\!j_2,\,j'\!+\!j_1\!+\!j_2\!-\!1,\,\!-\!j\!+\!j_1\!+\!j_3,\,j\!+\!j_1\!+\!j_3\!-\!1\atop
2j_1,\,\!-\!J\!+\!j_1\!+\!j_2\!+\!j_3,\,J\!+\!j_1\!+\!j_2\!+\!j_3\!-\!1} \bigg |\,1
\right),
\ea
where
\be{fk}
f(J,j',j_1,j_2,j_3)=
\!\left[
(2j'\!-\!1)\frac{\Gamma(j'\!+\!j_1\!-\!j_2)\Gamma(j'\!+\!j_1\!+\!j_2\!-\!1)
  \Gamma(J\!-\!j'\!+\!j_3)
\Gamma(J\!+\!j'\!+\!j_3\!-\!1)}{
\Gamma(j'\!+\!j_2\!-\!j_1)\Gamma(j'\!-\!j_1\!-\!j_2\!+\!1)\Gamma(J\!-\!j'\!-\!j_3\!+\!1)\Gamma(J\!+\!j'\!-\!j_3)}
\right]^{1/2}\!\!\!.
\ee

\setcounter{equation}{0}
\section{Appendix: Evolution kernels in the conformal basis}

To solve the Schr\"odinger equation \re{Sch-eq-h} we expand the
eigenstates over the conformal basis in the quark-antiquark-gluon and
three-gluon sectors, Eq.~\re{ex-exp}.
In this representation the evolution kernels
are given by real and symmetric matrices acting on the vector of the expansion
coefficients $(u^q_{N,k},\, u^h_{N,k})$ defined in \re{ex-exp}.
In this appendix we work out the explicit form of these matrices and
discuss their properties.

The diagonal quark evolution kernel \re{on-q} is described in the conformal
basis $Y^q_{N,k}$ by a square matrix of dimension $\ell_q=N+1$
\be{App-on-q}
[H_{qq}]_{km} = \vev{Y^q_{N,k}| H_{qq}|Y^q_{N,m}} =
\vev{Y^q_{N,k}| H_{S^+}|Y^q_{N,m}}+ \frac{2n_f}3 \delta_{k,m}\delta_{k,0}
\ee
with $0\le k,m \le N$.
In turn, the Hamiltonian $H_{S^+}$ is given by the sum of pair-wise kernels
\re{nonplanar} depending on the conformal spins in the
different two-particle channels. We recall
that the operators $J_{31}$ are diagonal by definition on the space of the
states $Y^q_{N,k}$ so that
\be{App-J31}
 \vev{Y^q_{N,k}|V(J_{31})|Y^q_{N,m}} = V(k+2)\,\delta_{km}
\ee
for an arbitrary $V$. On the other hand, contributions to the Hamiltonian
that are functions of $J_{12}$ can be expanded over the conformal basis in the
two-particle $(12)-$channel
\be{App-J}
 \vev{Y^q_{N,k}|V(J_{12})|Y^q_{N,m}} = \sum_{l=0}^N V(l+5/2)
 \vev{Y^q_{N,k}|Y^{(12)3}_{N+7/2,l+5/2}}\vev{Y^{(12)3}_{N+7/2,l+5/2}|Y^q_{N,m}}\,.
\ee
The similar representation exists for $V(J_{23})$. The scalar
product of the $Y$-functions belonging to different conformal basis
is given by the Racah $6j-$symbols defined in \re{tr-M} and \re{App-Racah}.
In particular
\ba{App-sc-prod}
\vev{Y^{(12)3}_{N+7/2,l+5/2}|Y^q_{N,m}}&=& \Omega_{m+2,l+5/2}(N+7/2)\equiv
 \Omega_{ml}^{(q)}\,,
\\
\vev{Y^{(23)1}_{N+7/2,l+5/2}|Y^q_{N,m}}&=& (-1)^{l+m} \Omega_{m+2,l+5/2}(N+7/2)
\equiv (-1)^{l+m} \Omega_{ml}^{(q)}\,.
\ea
Taking into account these relations one finds
\ba{App-sample}
\vev{Y^q_{N,k}|V(J_{12})|Y^q_{N,m}} &=& \sum_{l=0}^N V(l+5/2)\Omega_{ml}^{(q)}
\Omega_{kl}^{(q)},
\\
\vev{Y^q_{N,k}|V(J_{23})|Y^q_{N,m}} &=& \sum_{l=0}^N (-1)^{l+m} V(l+5/2)\Omega_{ml}^{(q)}
\Omega_{kl}^{(q)}\,.
\ea
Using these identities and the explicit expression for $H_{S^+}$ given in
Sect.~3.5.1 %Eqs.~\re{HSp} and \re{H_NS}
%and applying the identities \re{App-J31} and \re{App-sample}
one finds the following matrix representation of the diagonal quark kernel
\ba{App-quark-mat}
&&[H_{qq}]_{km} =
\left[-\frac2{N_c}U^{(1)}_{qq}(k+2)+\frac{2n_f}3\delta_{k,0}\right] \delta_{k,m}+
\sum_{l=0}^N \Omega_{ml}^{(q)}\, \Omega_{kl}^{(q)}
\\
%&&\times
%\left[ N_c(V_{qg}^{(0)}(l+5/2) + (-1)^{k+m} U_{gq}^{(0)}(l+5/2))
%-\frac2{N_c}(V_{qg}^{(1)}(l+5/2) + (-1)^{k+m} U_{gq}^{(1)}(l+5/2))
%\right]\,.
%\nonumber
%\\
&&\times
\left( N_c V_{qg}^{(0)}(l\!+\!5/2) -\frac2{N_c}V_{qg}^{(1)}(l\!+\!5/2)
+(-1)^{k+m}\left[N_c U_{gq}^{(0)}(l\!+\!5/2)
 -\frac2{N_c}U_{gq}^{(1)}(l\!+\!5/2)\right]
\right).
\nonumber
\ea
The explicit expressions for the Racah symbols and the evolution kernels
are given in Eqs.~\re{App-Racah} and \re{nonplanar}, respectively.

The gluon diagonal kernel in the helicity representation \re{on-g} is given in
the conformal basis $Y^g_{N-1,k}$ by a square matrix of dimension
$\ell_g=N/2$
\be{App-on-h}
[H_{hh}]_{km} = \vev{Y^h_{N-1,k}| H_{hh}|Y^h_{N-1,m}} = N_c
\vev{Y^h_{N-1,k}| H_{3/2}-V_{3/2}|Y^h_{N-1,m}}
\ee
with $0\le k,m \le [N/2]-1$. To obtain the explicit expression one has
to follow the steps similar to that in the quark case.
Introducing the notation for the relevant Racah symbols
\be{Racah-h}
\vev{Y^{(12)3}_{N+7/2,l+3}|Y^g_{N-1,m}} = \Omega_{m+3,l+3}(N+7/2)\equiv
\Omega_{ml}^{(h)}
\ee
and using the explicit expressions for the gluon kernels \re{h32} one finds
\ba{App-gluon-mat}
[H_{hh}]_{km} &=&
 \delta_{k,m}\big[N_c \,\lr{2U_{gg}(k+3)-V_{gg}(k+3)}-b\big]
\\
&+&
N_c \sum_{l=0}^{[N/2]} \Omega_{ml}^{(h)}\, \Omega_{kl}^{(h)}
\big[2 U_{gg}(l+3)(1+(-1)^{k+m}) - V_{gg}(l+3)\big]\,.
\nonumber
\ea
The explicit expressions for the Racah symbols and the functions
$U_{gg}$, $V_{gg}$
are given in Eqs.~\re{App-Racah} and \re{U,V}, respectively.

Finally, the explicit expressions for the off-diagonal kernels $H_{qh}$ and
$H_{hq}$ are given in \re{herm-off}.

% It follows from their definition \re{off-diag-ex}
%that
%\ba{App-off}
%\vev{Y^q_{N,k}|H_{qh}|Y^h_{N-1,m}}
%&=&-n_f\, V_{\rm off}(k+2) \vev{Y^q_{N,k}|\frac{1-P_{23}}2|Y^q_{N,m}}
%\\
%\vev{Y^h_{N-1,m}|H_{hq}|Y^q_{N,k}}
%&=&-2N_c\, V_{\rm off}(k+2) \vev{Y^h_{N-1,k}|\frac{1-P_{23}}2|Y^h_{N-1,m}}
%\ea

%\input{references}


\addcontentsline{toc}{section}{References}
\begin{thebibliography}{99}

\bibitem{E143}
K.~Abe {\it et al.}  [E143 Collaboration],
Phys.\ Rev.\ Lett.\  {\bf 76} (1996) 587.
%%CITATION = HEP-EX 9511013;%%

\bibitem{E154}
K.~Abe {\it et al.}  [E154 Collaboration],
Phys.\ Lett.\  {\bf B404} (1997) 377.
%%CITATION = HEP-EX 9705017;%%

\bibitem{E155}
P.~L.~Anthony {\it et al.}  [E155 Collaboration],
Phys.\ Lett.\  {\bf B458} (1999) 529;
%%CITATION = HEP-EX 9901006;%%
G.~S.~Mitchell  [E155 Collaboration], hep-ex/9903055.
%%CITATION = HEP-EX 9903055;%%

\bibitem{Book}
B.~L.~Ioffe, V.~A.~Khoze and L.~N.~Lipatov,
``Hard Processes. Vol. 1: Phenomenology, Quark Parton Model'',
{\it  Amsterdam, Netherlands: North-Holland ( 1984)}.

\bibitem{AELreview}
M.~Anselmino, A.~Efremov and E.~Leader,
%``The theory and phenomenology of polarized deep inelastic scattering,''
Phys.\ Rept.\  {\bf 261} (1995) 1.
%%CITATION = HEP-PH 9501369;%%

\bibitem{KTreview}
J.~Kodaira and K.~Tanaka,
%``Polarized structure functions in {QCD},''
Prog.\ Theor.\ Phys.\  {\bf 101} (1999) 191.
%%CITATION = HEP-PH 9812449;%%


\bibitem{SV82}
E.~V.~Shuryak and A.~I.~Vainshtein,
%``Theory Of Power Corrections To Deep Inelastic Scattering In
%Quantum Chromodynamics. 2. Q**4 Effects: Polarized Target,''
Nucl.\ Phys.\  {\bf B201} (1982) 141.
%%CITATION = NUPHA,B201,141;%%

\bibitem{BKL84}
A.P.~Bukhvostov, E.A.~Kuraev and L.N.~Lipatov,
%``Deep Inelastic Electron Scattering By A Polarized Target In
%Quantum Chromodynamics,''
Sov.\ Phys.\ JETP\  {\bf 60} (1984) 22;
%%CITATION = JTPLA,37,482;%%

\bibitem{BB89}
I.~I.~Balitsky and V.~M.~Braun,
%``Evolution Equations For QCD String Operators,''
Nucl.\ Phys.\  {\bf B311} (1989) 541.
%%CITATION = NUPHA,B311,541;%%

\bibitem{ABH}
A.~Ali, V.M.~Braun and G.~Hiller,
%``Asymptotic solutions of the evolution equation for the polarized
%nucleon structure function g-2 (x, Q**2),''
Phys.\ Lett.\ {\bf B266} (1991) 117.
%%CITATION = PHLTA,B266,117;%%

\bibitem{Kodaira}
J.~Kodaira, Y.~Yasui and T.~Uematsu,
Phys.\ Lett.\  {\bf B344} (1995) 348;
J.~Kodaira {\em et al.}, Phys.\ Lett.\  {\bf B387} (1996) 855;
%%CITATION = HEP-PH 9603377;%%
Prog.\ Theor.\ Phys.\  {\bf 99} (1998) 315.
%%CITATION = HEP-PH 9712395;%%

\bibitem{BKM00}
V.~M.~Braun, G.~P.~Korchemsky and A.~N.~Manashov,
Phys.\ Lett.\  {\bf B476} (2000) 455.
%%CITATION = HEP-PH 0001130;%%

\bibitem{Ji1}
X.~Ji, W.~Lu, J.~Osborne and X.~Song,
%``One-loop factorization of the nucleon g2 structure function in the  non-singlet case,''
Phys.\ Rev.\  {\bf D62}, 094016 (2000).

\bibitem{Ji2}
A.~Belitsky, X.~Ji, W.~Lu and J.~Osborne,
%``The singlet g2 structure function in the next-to-leading order,''
hep-ph/0007305.

\bibitem{I}
V.~M.~Braun, G.~P.~Korchemsky and A.~N.~Manashov,
%``Gluon contribution to the structure function g2(x,Q**2),''
hep-ph/0010128.
%%CITATION = HEP-PH 0010128;%%

\bibitem{WW}
S.~Wandzura and F.~Wilczek,
%``Sum Rules For Spin Dependent Electroproduction: Test Of Relativistic Constituent Quarks,''
Phys.\ Lett.\  {\bf B72} (1977) 195.
%%CITATION = PHLTA,B72,195;%%

\bibitem{Muller}
B.~Geyer, D.~Muller and D.~Robaschik,
%``Evolution kernels of twist-3 light-ray operators in polarized deep  inelastic scattering,''
Nucl.\ Phys.\ Proc.\ Suppl.\  {\bf 51C} (1996) 106;
D.~Muller,
%``Calculation of higher-twist evolution kernels for polarized deep  inelastic scattering,''
Phys.\ Lett.\  {\bf B407} (1997) 314.

\bibitem{Jaffe83}
R.L.~Jaffe,
%``Parton Distribution Functions For Twist Four,''
Nucl.\ Phys.\ {\bf B229} (1983) 205.
%%CITATION = NUPHA,B229,205;%%

\bibitem{Okun}
L.B.~Okun, ``Leptons and Quarks'',
{\it  Amsterdam, Netherlands: North-Holland (1982)}.

\bibitem{BFLK}
A.P.~Bukhvostov, G.V.~Frolov, L.N.~Lipatov and E.A.~Kuraev,
%``Evolution Equations For Quasi - Partonic Operators,''
Nucl.\ Phys.\  {\bf B258} (1985) 601.
%%CITATION = NUPHA,B258,601;%%

\bibitem{BDKM}
V.M.~Braun, S.E.~Derkachov, G.P.~Korchemsky and A.N.~Manashov,
%``Baryon distribution amplitudes in {QCD},''
Nucl.\ Phys.\ {\bf B553} (1999) 355.
%%CITATION = NUPHA,B553,355;%%

\bibitem{BDM98}
V.M.~Braun, S.E.~Derkachov and A.N.~Manashov,
%``Integrability of three-particle evolution equations in {QCD},''
Phys.\ Rev.\ Lett.\ {\bf 81} (1998) 2020.
%%CITATION = PRLTA,81,2020;%%

\bibitem{Belitsky99}
A.~V.~Belitsky,
%``Integrability and WKB solution of twist-three evolution equations,''
Nucl.\ Phys.\ {\bf B558} (1999) 259;
%%CITATION = HEP-PH 9903512;%%
%``Renormalization of twist-three operators and integrable lattice models,''
Nucl.\ Phys.\ {\bf B574} (2000) 407.
%%CITATION = HEP-PH 9907420;%%

\bibitem{DKM99}
S.E.~Derkachov, G.P.~Korchemsky and A.N.~Manashov,
Nucl.\ Phys.\ {\bf B566} (2000) 203.
%``Evolution equations for quark-gluon distributions in multi-color {QCD}
%and open spin chains,''
%%CITATION = HEP-PH 9909539;%%

\bibitem{K89}
G.~P.~Korchemsky,
%``Asymptotics Of The Altarelli-Parisi-Lipatov Evolution Kernels Of Parton Distributions,''
Mod.\ Phys.\ Lett.\ {\bf A4} (1989) 1257.
%%CITATION = MPLAE,A4,1257;%%

\bibitem{BC70}
H.~Burkhardt and W.~N.~Cottingham, Annals Phys.\  {\bf 56} (1970) 453.
%%CITATION = APNYA,56,453;%%

\bibitem{A-W}
R.~Koekoek, R.~Swarttouw, The Askey scheme of hypergeometric orthogonal
polynomials and its q-analogue, Report 98-17, Delft University of Technology,
Faculty TWI, 1998.

\end{thebibliography}
\end{document}